\documentclass[a4paper,11pt]{article}
\usepackage{amsmath}
\usepackage{tikz}
\usepackage{graphicx}
\graphicspath{figs/}
\usepackage{amsfonts}
\usepackage{physics}
\usepackage{xcolor}
\usepackage{graphbox}
\usepackage{jheppub}
\usepackage{tcolorbox}

\definecolor{Red}{RGB}{214, 39, 40}
\definecolor{Blue}{RGB} {31, 119, 180}
\definecolor{Orange}{RGB}{255, 153, 51}
\definecolor{Green}{RGB}{44, 160, 44}
\newcommand{\drawA}[4]{\draw[line width=1.5pt,->] (#1,#2)--(#1+0.6*#3-0.6*#1,#2+0.6*#4-0.6*#2);
	\draw[line width=1.5pt] (#1+0.5*#3-0.5*#1,#2+0.5*#4-0.5*#2)--(#3,#4);}

\newcommand{\twositechain}{\raisebox{-11pt}{\begin{tikzpicture}
		\coordinate (X1) at (0,0);
		\coordinate (X2) at (1.5,0);
		\node[below] at (X1) {\small{$\omega_1,\ q_1$}};
		\node[below] at (X2) {\small{$\omega_2,\ q_2$}};
		\draw[line width=1.5pt,->] (X1)--(0.85,0);
        \draw[line width=1.5pt] (0.75,0)--(X2);
		\path[fill=black] (X1) circle[radius=0.1];
		\path[fill=black] (X2) circle[radius=0.1];
\end{tikzpicture}}}

\newcommand{\onesiteT}{\raisebox{-11pt}{\begin{tikzpicture}
		\coordinate (X1) at (0,0);
		\node[below] at (X1) {$\omega_{1,2},\ q_{1,2}$};
		\path[fill=black] (X1) circle[radius=0.1];
\end{tikzpicture}}}

\newcommand{\rmd}{{\rm d}}
\newcommand{\ii}{\text{i}}

\title{Differential equations and recursive solutions for cosmological amplitudes}

\author[a,b,c]{Song He,}\emailAdd{songhe@itp.ac.cn}
\author[a]{Xuhang Jiang,}\emailAdd{xhjiang@itp.ac.cn}
\author[a,d]{Jiahao Liu,}\emailAdd{liujiahao@itp.ac.cn}
\author[a]{Qinglin Yang,}\emailAdd{yangqinglin@itp.ac.cn}
\author[a,d]{Yao-Qi Zhang}\emailAdd{zhangyaoqi@itp.ac.cn}
\affiliation[a]{CAS Key Laboratory of Theoretical Physics, Institute of Theoretical Physics, Chinese Academy of Sciences, Beijing 100190, China}
\affiliation[b]{School of Fundamental Physics and Mathematical Sciences, Hangzhou Institute for Advanced Study, UCAS \& ICTP-AP, Hangzhou, 310024, China}
\affiliation[c]{Peng Huanwu Center for Fundamental Theory, Hefei, Anhui 230026, P. R. China}
\affiliation[d]{School of Physical Sciences, University of Chinese Academy of Sciences, No.19A Yuquan Road, Beijing 100049, China}

\abstract{
Recently considerable efforts have been devoted to computing cosmological correlators and the corresponding wavefunction coefficients, as well as understanding their analytical structures. In this note, we revisit the computation of these ``cosmological amplitudes" associated with any tree or loop graph for conformal scalars with time-dependent interactions in the power-law FRW universe, directly in terms of iterated time integrals. We start by decomposing any such cosmological amplitude (for loop graph, the ``integrand" prior to loop integrations) as a linear combination of {\it basic time integrals}, one for each 
{\it directed graph}. We derive remarkably simple first-order differential equations involving such time integrals with edges ``contracted" one at a time, which can be solved recursively and the solution takes the form of Euler-Mellin integrals/generalized hypergeometric functions. By combining such equations, we then derive a complete system of differential equations for all time integrals needed for a given graph. 
Our method works for any graph: for a tree graph with $n$ nodes, this system can be transformed into the {\it canonical differential equations} of size $4^{n{-}1}$ equivalent to the graphic rules derived recently
, and we also derive the system of differential equations for loop integrands {\it e.g.} of all-loop two-site graphs and one-loop $n$-gon graphs. Finally, we show how the differential equations truncate for the de Sitter (dS) case (in a way similar to differential equations for Feynman integrals truncate for integer dimensions), which immediately yields the complete symbol for the dS amplitude with interesting structures {\it e.g.} for $n$-site chains and $n$-gon cases. 
}

\begin{document}
\maketitle

\section{Introduction and Review}
A great deal of information about the history of our universe can be extracted from the measurement of cosmic perturbations, which are generated in a primordial phase before the hot big bang~\cite{Achucarro:2022qrl,Maldacena:2002vr,Chen:2011zf,Chen:2011tu,Chen:2014cwa,Chen:2014joa,Chen:2015lza,Chen:2018cgg}.
In recent years, a new paradigm dubbed cosmological collider physics has received significant attention~\cite{Chen:2009we,Chen:2009zp,Chen:2012ge,Noumi:2012vr,Arkani-Hamed:2015bza,Chen:2016hrz,Chen:2016nrs,Chen:2016uwp,Lee:2016vti,An:2017hlx}, which has aimed at extracting from such measurement invaluable information about the high energy physics associated with primordial universe. In inflationary cosmology~\cite{Guth:1980zm, Linde:1981mu, Albrecht:1982wi, Baumann:2009ds}, such perturbations are generated by quantum fluctuations at the end of inflation, where they can be viewed as correlation functions live at the “boundary” of an approximate de Sitter (dS) spacetime. Thus, the study of cosmological correlation functions, which can be viewed as the analog of ``scattering amplitudes" in dS spacetime, is not only important for cosmology, but also of great interests for understanding QFT in curved spacetime in general~\cite{Anninos:2014lwa,Chen:2017ryl,Sleight:2021plv,Stefanyszyn:2023qov,Penedones:2023uqc,Loparco:2023rug,Loparco:2023akg,Marolf:2010nz,Cespedes:2023aal}. Notably at least two strategies have been developed to study these cosmological correlation functions. One is the traditional approach which evolves the cosmological correlations from past infinity to now, which necessitates the time integrals for each interaction vertex~\cite{Maldacena:2002vr,Weinberg:2005vy,Chen:2009we,Chen:2009zp,Chen:2012ge,Noumi:2012vr}. A second approach is the so-called cosmological bootstrap, which directly reconstructs the correlation functions based on principles like unitarity, locality, and symmetries~\cite{Arkani-Hamed:2018kmz,Baumann:2019oyu,Baumann:2020dch,Pajer:2020wnj,Hogervorst:2021uvp,Baumann:2022jpr,Pimentel:2022fsc,Jazayeri:2022kjy,Wang:2022eop,DuasoPueyo:2023kyh,Goodhew:2020hob,Jazayeri:2021fvk,Melville:2021lst,Goodhew:2021oqg,Baumann:2021fxj,Meltzer:2021zin,DiPietro:2021sjt,Tong:2021wai}. 

In this setting, a particularly simple and important class of toy models are given by conformal scalar fields with time-dependent self-intersections in power-law FRW universes. Given the conformal invariance of the scalar coupling in any FRW universe, the kinetic term of our scalars can be transformed to a massless scalar in Minkowski spacetime; note that the correlator differs from the flat space result due to the non-conformal nature of the self-interactions, but it still takes relatively simple form, which makes such models an ideal laboratory for studying cosmological correlators and corresponding wavefunction coefficients~\cite{Arkani-Hamed:2017fdk,Arkani-Hamed:2018bjr,Hillman:2019wgh,Arkani-Hamed:2023kig,Arkani-Hamed:2023bsv,De:2023xue,Benincasa:2024leu,Benincasa:2024lxe}. For any ``Feynman diagram" contributing to the cosmological correlator, one can express the corresponding wavefunction coefficient as a nice twisted energy integral~\cite{Arkani-Hamed:2023kig,Arkani-Hamed:2023bsv} (with twists that depend on the self-coupling and FRW background), whose integrands in energy space is given by a simple logarithmic form associated with the so-called ``cosmological polytope"~\cite{Arkani-Hamed:2017fdk,Arkani-Hamed:2018bjr,Benincasa:2019vqr,Benincasa:2018ssx,Benincasa:2020aoj,Juhnke-Kubitzke:2023nrj,Benincasa:2024leu}. Recently in~\cite{Arkani-Hamed:2023kig,Arkani-Hamed:2023bsv}, the authors have derived a set of graphic rules dubbed ``kinematic flow" which govern the change of wavefunction for any {\it tree} graph with respect to the kinematics; this amounts to a system of differential equations for a basis of $4^{n{-}1}$ twisted integrals (for $n$-site tree), which resemble the canonical differential equations for Feynman integrals (with the dimensional regulator $\epsilon$ replaced by the twist). Note that in the limit when the twist vanishes, one must recover the special case of conformal scalars in de Sitter (dS) spacetime, where the wavefunction evaluates to weight-$n$ multiple polylogarithmic functions (MPL), thus the differential equations must {\it truncate}. Given the difficulty in solving such differential equations analytically, very recently in~\cite{Fan:2024iek} a different approach has been proposed which allows one to obtain analytic results in terms of series expansions for tree-level wavefunction and correlator based on their definitions: they can be decomposed into linear combinations of simple, iterated {\it time integrals}. In this way, one can avoid solving any complicated system of differential equations and in principle obtain analytic results for such ``cosmological amplitudes"~\footnote{Following~\cite{Fan:2024iek}, we will refer to cosmological correlators and wavefunction coefficients for such conformal scalars as ``cosmological amplitudes".}

In this note, we will revisit the problem of deriving differential equations and solving such equations  for cosmological amplitudes of conformal scalars with time-dependent interactions in the power-law FRW universe. There are various motivations for considering our method, and let us quickly summarize the key points and advantages here. 
\begin{itemize}
    \item We derive differential equations for wavefunctions/correlators without referring to twisted energy integrals but directly in terms of the original time integrals: as we will see the building blocks called {\it basic time integrals} satisfy extremely simple differential equations, and the complete system of differential equations for any tree amplitude/loop integrand can be obtained simply by adding them up. 
    \item We can solve such differential equations recursively and obtain solutions for any basic time integral of tree/loop graph in terms of Euler-Mellin integrals ({\it c.f.}~\cite{Matsubara-Heo:2023ylc,berkesch2014euler}) or equivalently series expansions (similar to those in~\cite{Fan:2024iek}), which provides a link between the approach using twisted energy integrals and the one based on time integrals. 
    \item Our method generalizes previous results to differential equations directly for correlators and also to any loop graph (as a byproduct one can easily obtain a basis for any graph, which might prove difficult for twisted integrals of complicated loop graphs), and there are various potential new applications: for example, we will see that it is straightforward to go back to dS case and obtain the symbol for tree amplitudes and loop integrands. 

\end{itemize}
Let us elaborate on these points and give a preview of the rest of the paper. Essentially by definition, the wavefunction coefficient or correlator for any graph can be decomposed as a simple linear combination of building blocks we call ``basic time integrals", where each edge of the graph can be decomposed according to time ordering: one can either associate an arrow of each direction (for advanced/retarded time ordering) or remove the edge for the case without time ordering. Thus we have one such basic time integral for each {\it directed graph}: there are exactly $3^e$ such building blocks for a $n$-site tree graph with $e=n{-}1$ edges, while at loop level the number of basic time integrals is smaller than $3^e$ since only a subset of orientations are allowed. We will review such a decomposition in Sec.~\ref{subsec:decomp}. 

Given the decomposition, as we have mentioned, for any tree graph by introducing the so-called ``family tree/chain" decomposition, one can obtain analytic expressions for basic time integrals in terms of series expansions as achieved in~\cite{Fan:2024iek}. Here we adopt a different approach and derive differential equations for such building blocks: as we will show in Sec.~\ref{subsec:de1}, each basic time integral satisfies a first-order differential equation which involves graphs where one edge is ``contracted"; this procedure can be trivially iterated which express derivatives of the basic time integral we start with in terms of those for directed graphs with less and less edges (until we reach ``vacuum" graph with no edges at all). Moreover, such differential equations can be solved recursively: the basic time integral for any $n$-site directed graph can be expressed as a one-fold integral of that for an $(n{-}1)$-site directed graph, as we will see in Sec.~\ref{sec:sol}. For any tree graph, such a recursion relation can also lead to the series solutions as in~\cite{Fan:2024iek} (which sometimes can be recognized as known generalized hypergeometric functions). In this way, our alternative approach via differential equations produce analytic results for basic time integral of any directed graph, which can then be assembled to give cosmological tree amplitudes or loop integrands, which take the form of Euler-Mellin integrals and we will initiate studies on their analytic properties in Sec.~\ref{sec:sol}. As already emphasized in~\cite{Fan:2024iek}, these results for basic time integrals apply equally well for wavefunction coefficients and for correlators: once we obtain results for these building blocks, both of them can be obtained via respective linear combinations. 

In Sec.~\ref{sec:DE} we will illustrate how to combine these differential equations to obtain the complete system of differential equations for any graph. Note that our idea of studying differential equations and solutions for basic time integrals differs significantly from the approach of~\cite{Arkani-Hamed:2023bsv, Arkani-Hamed:2023kig}, where one considers the full wavefunction coefficient of a graph, written as twisted energy integrals of the canonical function of the so-called cosmological polytopes; then at least for any tree graph, the complete system of differential equations is obtained by finding a basis of such twisted integrals, which then nicely organized according to the graphic rules given in~\cite{Arkani-Hamed:2023bsv, Arkani-Hamed:2023kig}. For our case, while the differential equations are almost trivial for individual basic time integral, it still takes significant work to combine them together to obtain the ``kinematic flow" (with a slight generalization that each site has its own twist $q_i$ which can be chosen independently). However, as we will see there are certain advantages to considering basic time integrals: they are building blocks for both wavefunction coefficients and correlators and there are no differences between tree and loop graphs (prior to loop integrations), thus we can derive the complete system of differential equations (including a basis in terms of time integrals) for any graph! We will present examples for trees as well as loop integrands such as $n$-gon and two-site all-loop graphs (``banana"), both for wavefunction and correlators. 

As another application, we will show in Sec.~\ref{sec:dS} how to truncate the differential equations and the resulting recursion relations to go back to the de Sitter space with twist $q=0$. Although amplitudes for an individual directed graph diverge in the $q\to 0$ limit, the combination for wavefunction coefficients is finite and it is straightforward to obtain the truncated differential equations for any graph. By recursively applying the truncated differential equations we directly obtain the {\it symbol} of the MPL functions for any dS tree amplitude or loop integrand. We will illustrate our results with some closed formula for the symbol of dS amplitudes for the $n$-site chain as well as the $n$-gon graphs. We end with conclusions and some directions for further investigations, and provide more material in the appendices. 

\subsection{Basics of cosmological amplitudes}
Let us quickly review basic facts about self-interacting conformal scalar theory in power-law FRW background and associated cosmological amplitudes, which are the main interests of this note (more details can be found in {\it e.g.} \cite{Fan:2024iek} and references therein). We will begin with the following action for scalar field $\phi$ in $(d{+}1)$-dimensional spacetime
\begin{equation}\label{eq:action}
    S[\phi]=-\int{\rm d}^dx \, {\rm d}\tau\sqrt{-g}\left(\frac12(\partial_\mu\phi)^2+\frac12\xi R \phi^2+\sum_{k>2}\frac{\tilde\lambda_k}{k!}\phi^k\right),
\end{equation}
with the conformally flat FRW metric 
\begin{equation}
{\rm d}^2s=a(\tau)^2(-{\rm d}\tau^2+{\rm d}x^2)=\left(\frac{\tau}{\tau_0}\right)^{2p}(-{\rm d}\tau^2+{\rm d}x^2).
\end{equation}
Here $\tau_0$ is the renormalized time scale, and we consider non-derivative self-interacting of the field to arbitrary powers. Different choices of power $p$ in the metric cover many interesting cases such as de Sitter spacetime for $p=-1$, inflation for $p=-1+\epsilon$, Minkowski spacetime for $p=0$, {\it etc.}. As mentioned above, we will focus on the special case where $\xi$ is chosen as $\frac{d{-}1}{4d}$: after a field redefinition $\phi\to (a(\tau))^{\frac{1{-}d}{2}}\varphi$, the theory is conformally equivalent to a massless scalar theory living in Minkowski spacetime as
\begin{equation}\label{eq:actionp}
    S[\varphi]=-\int {\rm d}^dx \, {\rm d}\tau \left(\frac12 (\partial \varphi)^2+\sum_{k>2}\frac{\lambda_k(\tau)}{k!}\varphi^k\right),
\end{equation}
with time-dependent coupling constants $\lambda_k(\tau)=\tilde\lambda_k (-\tau)^{q_k{-}1}$, and twist exponents $q_k$ here read $q_k=\left(d{+}1-\frac{k(d{-}1)}2 \right)p{+}1$, which vary according to different $\varphi^k$ self-interaction and spacetime background. Specially when $d{=}3$ and $k{=}4$, $q_k\equiv1$. Another important case is when $d{=}3$ and $k{=}3$, under which $q_k=p{+}1$. Therefore the inflationary background results in $q_k=\epsilon$, while de Sitter spacetime corresponds to the limit $q_k\to 0$. This is an important case we will consider in Sec.\ref{sec:dS}. 

In this way, ``cosmological amplitudes" associated with action \eqref{eq:action} can be computed effectively by correlation functions from flat spacetime action \eqref{eq:actionp}. We will consider two types of such amplitudes, namely {\it wavefunction coefficients} and {\it correlators}. 

\paragraph{Wavefunction coefficients}
Wavefunction of the universe $\Phi[\hat\varphi]$ is defined as the overlapping of vacuum state at the past infinity $\tau=-\infty$ with eigenstate $|\hat\varphi(x)\rangle$ at $\tau=0$, {\it i.e.} $\Phi[\hat\varphi]=\langle\hat\varphi(x)|0\rangle$, which can be formally presented by the path integral
\begin{equation}
   \Phi[\hat\varphi]=\int_{\varphi(\tau=-\infty)=|0\rangle}^{\varphi(\tau=0)=|\hat\varphi\rangle}\mathcal{D}\,\varphi e^{\ii \,  S[\varphi]} .
\end{equation}
Working in momentum space, the wavefunction can be perturbatively expanded as 
\begin{equation}
   \Phi[\hat\varphi]\equiv \exp{-\ii\, \sum_{n=2}^\infty\frac{1}{n!}\int\prod_{i=1}^n\left(\frac{{\rm d}^d\mathbf{k}_i}{(2\pi)^d}\hat\varphi(\mathbf{k}_i)\right)\tilde{\psi}_n(\mathbf{k}_1,\cdots,\mathbf{k}_n)\ \delta^{(d)}\left(\sum_i\mathbf{k}_i\right)} .
\end{equation}
All non-trivial information of the functional is encoded in the quantities $\tilde{\psi}_n(\mathbf{k}_1,\cdots,\mathbf{k}_n)$, which are called $n$-site wavefunction coefficients. Pretty similar to the method in studying scattering amplitudes, computation for these functions can also be carried out through ``Feynman graphs". At tree level, corresponding integrals read 
\begin{equation}    \psi_n(\mathbf{k}_1,\cdots,\mathbf{k}_n;\mathcal{E})=\int_{-\infty}^0\prod_{v=1}^n{\rm d}\tau_v\ \ii\,\lambda_v(\tau_v) B(X_v,\tau_v)\prod_{e\in\mathcal{E}}G_e(Y_e,\tau_{v_e},\tau_{v_{e^\prime}}).
\end{equation}
Here $\mathcal{E}$ are edges from topology of the graph. $G_e(Y_e,\tau_{v_e},\tau_{v_{e^\prime}})$ are {\it bulk-to-bulk} propagators following the definition
\begin{equation}\label{eq:wavefuncprop}
\begin{aligned}
    G_e(Y_e,\tau_{v_e},\tau_{v_{e^\prime}}){=}\frac{1}{2Y_e}\!\left[ e^{-\ii \, Y_e(\tau_{v_e}{-}\tau_{v_{e^\prime}})}\theta(\tau_{v_e}{-}\tau_{v_{e^\prime}}){+}e^{+\ii \, Y_e(\tau_{v_e}{-}\tau_{v_{e^\prime}})}\theta(\tau_{v_{e^\prime}}{-}\tau_{v_e}){-}e^{+\ii \, Y_e(\tau_{v_e}+\tau_{v_{e^\prime}})} \right]\!,
\end{aligned}
\end{equation}
which is between two bulk points on the ends of edge $e$ at conformal time $\tau_{v_e}$ and $\tau_{v_{e^\prime}}$, carrying energy $Y_e=|\mathbf{K}_e|$, and $\mathbf{K}_e$ are certain sum over external $\mathbf{k}_v$. 
They are derived by solving the inhomogeneous equation of motion of the theory with delta-function source and boundary condition $G\to0$ when $\tau_{v_e}$ or $\tau_{v_{e^\prime}}$ goes to $0$ or $-\infty$. $B(X_v,\tau_v)$, on the other hand,  are {\it bulk-to-boundary} propagators between a bulk point at $\tau_v$ and the boundary $\tau=0$, carrying energy  $X_v=|\mathbf{k}_v|$, whose definition is
\begin{equation}\label{eq:vertexwf}
    B(X_v,\tau_v)=e^{\ii X_v\tau_v},
\end{equation}
which are solutions for homogeneous equation of motion satisfying $B(X_v,-\infty)=0$ and $B(X_v,0)=1$. Finally, $\lambda_v(\tau_v)$ is coupling constant at vertex $v$ in the diagram, which yields an extra factor $\lambda_v(\tau_v)\propto (-\tau)^{q_v{-}1}$ in the integrand. We emphasize that here and in the following sections, we will always use the notation $q_v$ as the twist exponent on vertex $v$ instead of the power of $(-\tau)$ in \eqref{eq:actionp} at $\varphi^v$, and simply fix all $\tilde\lambda_k\equiv1$ since they only contribute overall factors. To make the discussion more general, we will always assume $q_i$ on each vertex to be free parameters. In the whole paper, we will use $\psi_n$ to denote an $n$-site wavefunction coefficient integral, and add a description for the topologies, {\it i.e.} $\psi_{\text{2-chain}}$, if necessary.

Finally, once the diagrams contain loops, $\mathbf{K}_e$ in the above expression sums over some external $\mathbf{k}_i$ and loop momenta $\mathbf{l}_i$. Therefore, besides integrating conformal time,  we should also perform the loop integration for the wavefunction coefficient as
\begin{equation}
\int\prod_{i=1}^L\frac{{\rm d}^d\mathbf{l}_i}{(2\pi)^d} \psi_n(\mathbf{k}_1,\cdots,\mathbf{k}_n;\mathbf{l}_1,\cdots,\mathbf{l}_L;\mathcal{E}).
\end{equation}
Generally speaking, performing this ${\rm d}^d\mathbf{l}_i$ integration can be pretty complicated, and in this note we will mainly focus on the computation for the integrand $\psi_n(\mathbf{k}_1,\cdots,\mathbf{k}_n;\mathbf{l}_1,\cdots,\mathbf{l}_L;\mathcal{E})$, and we leave the exploration for the loop integration for future study.

\paragraph{Correlators}
Correlators are direct observables in cosmological study, and are closedly related to wavefunction of the universe formally as
\begin{equation}    \langle\varphi(x_1)\cdots\varphi(x_n)\rangle=\frac{\int\mathcal{D}\varphi\prod_{i=1}^n\varphi(x_i)|\Phi[\varphi]|^2}{\int\mathcal{D}\varphi|\Phi[\varphi]|^2}.
\end{equation}
They can also be perturbatively computed by Feynman graphs, and graphical rules for these observables can be directly extracted from wavefunction. In practice, the quantity can be instead studied through in-in formalism \cite{Chen:2017ryl} after a Fourier transformation as
\begin{align}   \langle\varphi(\mathbf{k}_1)\cdots\varphi(\mathbf{k}_n)\rangle&(2\pi)^d\delta^{(d)}(\sum_i\mathbf{k}_i)\\&{=}\int\mathcal{D}\varphi_{+}\mathcal{D}\varphi_{-}\prod_{i=1}^n\delta\left[\varphi_+(\mathbf{k}_i){-}\varphi_-(\mathbf{k}_i)\right]
\varphi_+(\mathbf{k}_1)\cdots\varphi_+(\mathbf{k}_n)e^{i(S[\varphi_+]{-}S[\varphi_-])}\nonumber,
\end{align}
and perturbatively computed by corresponding ``Feynman integrals" 
\begin{align}    \mathcal{T}_n(\mathbf{k}_1,\cdots,\mathbf{k}_n;&\mathbf{l}_1,\cdots,\mathbf{l}_L;\mathcal{E})\\
&=\sum_{a_v=\pm} \int_{-\infty}^0\prod_{v=1}^n  {\rm d}\tau_v\ \ii \,  a_v\lambda_v(\tau_v) D_{a_v}(X_v,\tau_v)\prod_{e\in\mathcal{E}}D_{a_{v_e}a_{v_{e^\prime}}}(Y_e,\tau_{v_e},\tau_{v_{e^\prime}}),\nonumber
\end{align}
{\it i.e.} we have two types of bulk points, and the corresponding bulk-to-bulk and bulk-to-boundary propagators read
\begin{equation}\label{eq:correprop}
\begin{aligned}
    & D_{\pm \pm}(Y;\tau_1,\tau_2)=\frac{1}{2Y}\left[ e^{\mp \ii \, Y (\tau_1-\tau_2)}\theta(\tau_1-\tau_2)+e^{\pm \ii \, Y (\tau_1-\tau_2)} \theta(\tau_2-\tau_1) \right],\\
    & D_{\pm \mp}(Y;\tau_1,\tau_2)=\frac{1}{2Y} e^{\pm \ii \, Y (\tau_1-\tau_2)},
\end{aligned}
\end{equation}
as well as 
\begin{equation}
    D_\pm(X,\tau)=\frac{1}{2X}e^{\pm \ii \, X \tau}.
\end{equation}
We will only focus on the integrand for correlators as well, and use $\mathcal{T}_n$ or, {\it e.g.} $\mathcal{T}_{\text{2-chain}}$ to denote correlators in this note.

\section{Basic time integrals and their differentials}\label{sec:decomp}
In this section, we derive simple differential equations satisfied by the so-called basic time integrals for directed graphs. 
Recall that to compute cosmological amplitudes for a graph with $n$ sites amounts to carrying out $n$-fold iterated time integrals. As we have mentioned, differential equations (DEs) for corresponding twisted energy integrals~\cite{Arkani-Hamed:2023bsv} and analytic results in terms of series expansions~\cite{Fan:2024iek} have been obtained at least for tree amplitudes.  Here we adopt the time integral representation as in \cite{Fan:2024iek} for both tree and loop graphs and propose to study DEs for simpler building blocks. We will first review a natural decomposition for any cosmological amplitudes and then take their differentials with respect to the kinematics (energies). 

\subsection{Decomposing cosmological amplitudes into basic time integrals}\label{subsec:decomp}

Following the basic definition of wavefunction coefficients and correlators, we have already seen that both of their integrands enjoy very similar structures. Therefore, after a simple expansion of propagators \eqref{eq:wavefuncprop} or \eqref{eq:correprop}, without loss of generality, it is very natural to see that these iterated time integrals can be decomposed into basic building blocks \cite{Fan:2024iek}, which are called {\it basic time integrals} in this note, and defined as the following:
\begin{equation}\label{eq:Tdef}
	\mathbf{T}^{q_1,q_2,\ldots,q_n}_{\mathcal{N}}(\omega_1,\omega_2,\ldots,\omega_n)=(-\ii)^n\int_{-\infty}^{0}\prod_{\ell=1}^n\left[\mathrm{d\tau}_{\ell}(-\tau_{\ell})^{q_\ell-1}e^{\ii \, \omega_\ell \tau_\ell}\right]\prod_{(j,k)\in\mathcal{N}}\theta_{j,k}.
\end{equation}
Here $\mathcal{N}$ denotes the time order structure of the integral, $\theta_{j,k}\equiv\theta(\tau_j-\tau_k)$ is the theta function, $\omega_i$ is the energy of vertex $i$\footnote{Note that this energy $\omega_i$ is a linear combination of  energies $X_i$ and $Y_{i,j}$ carried by propagators attached to this vertex.}, and $q_i$ represents the twist exponent of vertex $i$. Note that this definition includes time integrals of disconnected type, {\it i.e.}, we have decomposition for $\mathcal{N}=\mathcal{N}_1\cup\mathcal{N}_2$, such that there is no shared vertex between any time structures in $\mathcal{N}_1$ and $\mathcal{N}_2$, and the corresponding integrals turn out to be trivially a product of two basic time integrals.  Graphically, we have a natural representation by assigning a directed edge from vertex $k$ to $j$ for every $(j,k) \in \mathcal{N}$ in the time structure. Then each basic time integral one-to-one corresponds to a directed graph, with all its edges oriented and all its vertices assigned the energy. 

For tree-level wavefunction coefficients, there are 3 terms for each propagator \eqref{eq:wavefuncprop}, graphically represented as two kinds of directed edges and one dash line edge. Thus there will be $3^e$ terms if we expand the wavefunction integral in terms of basic time integrals, with $e$ the number of edges. Taking the three-site chain as an example, it turns out to be a sum over 9 terms: 
\begin{equation}
	\begin{aligned}
	\raisebox{-1em}{\begin{tikzpicture}
		\coordinate (X1) at (0,0);
		\coordinate (X2) at (1,0);
		\coordinate (X3) at (2,0);
		\node[below] at (X1) {\small{$1$}};
		\node[below] at (X2) {\small{$2$}};
		\node[below] at (X3) {\small{$3$}};
		\draw[line width=1.5pt] (X2)--(X3);
		\draw[line width=1.5pt] (X1)--(X2);
		\path[fill=black] (X1) circle[radius=0.1];
		\path[fill=black] (X2) circle[radius=0.1];
		\path[fill=black] (X3) circle[radius=0.1];
\end{tikzpicture}} 
 = & \raisebox{-1em}{\begin{tikzpicture}
				\coordinate (X1) at (0,0);
				\coordinate (X2) at (1,0);
				\coordinate (X3) at (2,0);;
				\node[below] at (X1) {\small{$1$}};
				\node[below] at (X2) {\small{$2$}};
				\node[below] at (X3) {\small{$3$}};
				\draw[line width=1.5pt,->] (X1)--(0.6,0);
				\draw[line width=1.5pt] (0.5,0)--(X2);
				\draw[line width=1.5pt,->] (X2)--(1.6,0);
				\draw[line width=1.5pt] (1.5,0)--(X3);
				\path[fill=black] (X1) circle[radius=0.1];
				\path[fill=black] (X2) circle[radius=0.1];
				\path[fill=black] (X3) circle[radius=0.1];
		\end{tikzpicture}} + 
		\raisebox{-1em}{\begin{tikzpicture}
		\coordinate (X1) at (0,0);
		\coordinate (X2) at (1,0);
		\coordinate (X3) at (2,0);;
		\node[below] at (X1) {\small{$1$}};
		\node[below] at (X2) {\small{$2$}};
		\node[below] at (X3) {\small{$3$}};
		\draw[line width=1.5pt,->] (X1)--(0.6,0);
		\draw[line width=1.5pt] (0.5,0)--(X2);
		\draw[line width=1.5pt,->] (X3)--(1.4,0);
		\draw[line width=1.5pt] (1.5,0)--(X2);
		\path[fill=black] (X1) circle[radius=0.1];
		\path[fill=black] (X2) circle[radius=0.1];
		\path[fill=black] (X3) circle[radius=0.1];
	\end{tikzpicture}} - 
	\raisebox{-1em}{\begin{tikzpicture}
	\coordinate (X1) at (0,0);
	\coordinate (X2) at (1,0);
	\coordinate (X3) at (2,0);
	\node[below] at (X1) {\small{$1$}};
	\node[below] at (X2) {\small{$2$}};
	\node[below] at (X3) {\small{$3$}};
	\draw[line width=1.5pt,->] (X1)--(0.6,0);
	\draw[line width=1.5pt] (0.5,0)--(X2);
	\draw[line width=1.5pt,dashed] (X2)--(X3);
	\path[fill=black] (X1) circle[radius=0.1];
	\path[fill=black] (X2) circle[radius=0.1];
	\path[fill=black] (X3) circle[radius=0.1];
\end{tikzpicture}}\\
+ & \raisebox{-1em}{\begin{tikzpicture}
		\coordinate (X1) at (0,0);
		\coordinate (X2) at (1,0);
		\coordinate (X3) at (2,0);;
		\node[below] at (X1) {\small{$1$}};
		\node[below] at (X2) {\small{$2$}};
		\node[below] at (X3) {\small{$3$}};
\draw[line width=1.5pt,->] (X2)--(0.4,0);
\draw[line width=1.5pt] (0.5,0)--(X1);
		\draw[line width=1.5pt,->] (X2)--(1.6,0);
		\draw[line width=1.5pt] (1.5,0)--(X3);
		\path[fill=black] (X1) circle[radius=0.1];
		\path[fill=black] (X2) circle[radius=0.1];
		\path[fill=black] (X3) circle[radius=0.1];
\end{tikzpicture}} + 
\raisebox{-1em}{\begin{tikzpicture}
		\coordinate (X1) at (0,0);
		\coordinate (X2) at (1,0);
		\coordinate (X3) at (2,0);;
		\node[below] at (X1) {\small{$1$}};
		\node[below] at (X2) {\small{$2$}};
		\node[below] at (X3) {\small{$3$}};
\draw[line width=1.5pt,->] (X2)--(0.4,0);
\draw[line width=1.5pt] (0.5,0)--(X1);
		\draw[line width=1.5pt,->] (X3)--(1.4,0);
		\draw[line width=1.5pt] (1.5,0)--(X2);
		\path[fill=black] (X1) circle[radius=0.1];
		\path[fill=black] (X2) circle[radius=0.1];
		\path[fill=black] (X3) circle[radius=0.1];
\end{tikzpicture}} - 
\raisebox{-1em}{\begin{tikzpicture}
		\coordinate (X1) at (0,0);
		\coordinate (X2) at (1,0);
		\coordinate (X3) at (2,0);;
		\node[below] at (X1) {\small{$1$}};
		\node[below] at (X2) {\small{$2$}};
		\node[below] at (X3) {\small{$3$}};
\draw[line width=1.5pt,->] (X2)--(0.4,0);
\draw[line width=1.5pt] (0.5,0)--(X1);
\draw[line width=1.5pt,dashed] (X2)--(X3);
		\path[fill=black] (X1) circle[radius=0.1];
		\path[fill=black] (X2) circle[radius=0.1];
		\path[fill=black] (X3) circle[radius=0.1];
\end{tikzpicture}}\\
- & \raisebox{-1em}{\begin{tikzpicture}
		\coordinate (X1) at (0,0);
		\coordinate (X2) at (1,0);
		\coordinate (X3) at (2,0);;
		\node[below] at (X1) {\small{$1$}};
		\node[below] at (X2) {\small{$2$}};
		\node[below] at (X3) {\small{$3$}};
		\draw[line width=1.5pt,dashed] (X1)--(X2);
		\draw[line width=1.5pt,->] (X2)--(1.6,0);
		\draw[line width=1.5pt] (1.5,0)--(X3);
		\path[fill=black] (X1) circle[radius=0.1];
		\path[fill=black] (X2) circle[radius=0.1];
		\path[fill=black] (X3) circle[radius=0.1];
\end{tikzpicture}} - 
\raisebox{-1em}{\begin{tikzpicture}
		\coordinate (X1) at (0,0);
		\coordinate (X2) at (1,0);
		\coordinate (X3) at (2,0);;
		\node[below] at (X1) {\small{$1$}};
		\node[below] at (X2) {\small{$2$}};
		\node[below] at (X3) {\small{$3$}};
		\draw[line width=1.5pt,dashed] (X1)--(X2);
		\draw[line width=1.5pt,dashed] (X2)--(X3);
		\draw[line width=1.5pt,->] (X3)--(1.4,0);
		\draw[line width=1.5pt] (1.5,0)--(X2);
		\path[fill=black] (X1) circle[radius=0.1];
		\path[fill=black] (X2) circle[radius=0.1];
		\path[fill=black] (X3) circle[radius=0.1];
\end{tikzpicture}} + 
\raisebox{-1em}{\begin{tikzpicture}
		\coordinate (X1) at (0,0);
		\coordinate (X2) at (1,0);
		\coordinate (X3) at (2,0);
		\node[below] at (X1) {\small{$1$}};
		\node[below] at (X2) {\small{$2$}};
		\node[below] at (X3) {\small{$3$}};
		\draw[line width=1.5pt,dashed] (X2)--(X3);
		\draw[line width=1.5pt,dashed] (X1)--(X2);
		\path[fill=black] (X1) circle[radius=0.1];
		\path[fill=black] (X2) circle[radius=0.1];
		\path[fill=black] (X3) circle[radius=0.1];
\end{tikzpicture}}.
	\end{aligned}
\end{equation}
Note that when drawing the diagrams, we omit possible bulk-to-boundary propagators. It is also worth mentioning that for each directed graph in this decomposition, energy $\omega_i$ at each vertex is defined differently according to the time order structure of each integral. For instance, for the first graph in the first line, we have $\omega_1=X_1{+}Y_{1,2}$, $\omega_2=X_2{-}Y_{1,2}{+}Y_{2,3}$ and $\omega_3=X_3{-}Y_{2,3}$, while for the second graph,  $\omega_2=X_2{-}Y_{1,2}{-}Y_{2,3}$ and $\omega_3=X_3{+}Y_{2,3}$ instead. In the following two sections we only focus on an individual directed graph, so we stick to these $\omega$ variables. One should, however, be very careful about definition of $\omega$s when adding them together and considering the full wavefunction/correlator integrals. 

For diagrams with loops, naively we still have $3^e$ terms after expansion. However, due to the topologies of the diagrams, naive time structures of certain terms in the decomposition may be redundant and can be simplified furthermore, resulting in simpler directed graphs. For instance, the decomposition of a two-site bubble diagram is as follows:
\begin{equation}\label{eq:bubbleexp}
	 \begin{aligned}
		\raisebox{-0.5em}{\begin{tikzpicture}
		\node[left] at (0,0) {\small{$1$}};
		\node[right] at (1,0) {\small{$2$}};
		\draw [line width=1.5pt] (0,0) to[out=60,in=-180] (0.6,0.25);
		\draw [line width=1.5pt] (0.5,0.25) to[out=0,in=120] (1,0);
		\draw [line width=1.5pt] (0,0) to[out=-60,in=-180] (0.5,-0.25);
		\draw [line width=1.5pt] (1,0) to[out=-120,in=0] (0.4,-0.25);
		\path[fill=black] (0,0) circle[radius=0.1];
		\path[fill=black] (1,0) circle[radius=0.1];
\end{tikzpicture}} = & \raisebox{-0.5em}{\begin{tikzpicture}
			\node[left] at (0,0) {\small{$1$}};
			\node[right] at (1,0) {\small{$2$}};
			\draw [line width=1.5pt,->] (0,0) to[out=60,in=-180] (0.6,0.25);
			\draw [line width=1.5pt] (0.5,0.25) to[out=0,in=120] (1,0);
			\draw [line width=1.5pt,->,black!30] (0,0) to[out=-60,in=-180] (0.6,-0.25);
			\draw [line width=1.5pt,black!30] (0.5,-0.25) to[out=0,in=-120] (1,0);
			\path[fill=black] (0,0) circle[radius=0.1];
			\path[fill=black] (1,0) circle[radius=0.1];
		\end{tikzpicture}} +
	\raisebox{-0.5em}{\begin{tikzpicture}
		\node[left] at (0,0) {\small{$1$}};
		\node[right] at (1,0) {\small{$2$}};
		\draw [line width=1.5pt,->] (0,0) to[out=60,in=-180] (0.6,0.25);
		\draw [line width=1.5pt] (0.5,0.25) to[out=0,in=120] (1,0);
		\draw [line width=1.5pt] (0,0) to[out=-60,in=-180] (0.5,-0.25);
		\draw [line width=1.5pt,->] (1,0) to[out=-120,in=0] (0.4,-0.25);
        \draw [line width=1pt,red] (0,-0.25)--(1,0.25);
		\path[fill=black] (0,0) circle[radius=0.1];
		\path[fill=black] (1,0) circle[radius=0.1];
	\end{tikzpicture}} -
\raisebox{-0.5em}{\begin{tikzpicture}
		\node[left] at (0,0) {\small{$1$}};
		\node[right] at (1,0) {\small{$2$}};
		\draw [line width=1.5pt,->] (0,0) to[out=60,in=-180] (0.6,0.25);
		\draw [line width=1.5pt] (0.5,0.25) to[out=0,in=120] (1,0);
		\draw [line width=1.5pt,dashed] (0,0) to[out=-60,in=-180] (0.5,-0.25);
		\draw [line width=1.5pt,dashed] (1,0) to[out=-120,in=0] (0.4,-0.25);
		\path[fill=black] (0,0) circle[radius=0.1];
		\path[fill=black] (1,0) circle[radius=0.1];
\end{tikzpicture}}\\[3pt]
+ & \raisebox{-0.5em}{\begin{tikzpicture}
		\node[left] at (0,0) {\small{$1$}};
		\node[right] at (1,0) {\small{$2$}};
		\draw [line width=1.5pt] (0,0) to[out=60,in=-180] (0.5,0.25);
		\draw [line width=1.5pt,->] (1,0) to[out=120,in=0] (0.4,0.25);
		\draw [line width=1.5pt,->] (0,0) to[out=-60,in=-180] (0.6,-0.25);
		\draw [line width=1.5pt] (0.5,-0.25) to[out=0,in=-120] (1,0);
        \draw [line width=1pt,red] (0,-0.25)--(1,0.25);
		\path[fill=black] (0,0) circle[radius=0.1];
		\path[fill=black] (1,0) circle[radius=0.1];
\end{tikzpicture}} +
\raisebox{-0.5em}{\begin{tikzpicture}
		\node[left] at (0,0) {\small{$1$}};
		\node[right] at (1,0) {\small{$2$}};
		\draw [line width=1.5pt] (0,0) to[out=60,in=-180] (0.5,0.25);
		\draw [line width=1.5pt,->] (1,0) to[out=120,in=0] (0.4,0.25);
		\draw [line width=1.5pt,black!30] (0,0) to[out=-60,in=-180] (0.5,-0.25);
		\draw [line width=1.5pt,->,black!30] (1,0) to[out=-120,in=0] (0.4,-0.25);
		\path[fill=black] (0,0) circle[radius=0.1];
		\path[fill=black] (1,0) circle[radius=0.1];
\end{tikzpicture}} -
\raisebox{-0.5em}{\begin{tikzpicture}
		\node[left] at (0,0) {\small{$1$}};
		\node[right] at (1,0) {\small{$2$}};
		\draw [line width=1.5pt] (0,0) to[out=60,in=-180] (0.5,0.25);
		\draw [line width=1.5pt,->] (1,0) to[out=120,in=0] (0.4,0.25);
		\draw [line width=1.5pt,dashed] (0,0) to[out=-60,in=-180] (0.5,-0.25);
		\draw [line width=1.5pt,dashed] (1,0) to[out=-120,in=0] (0.4,-0.25);
		\path[fill=black] (0,0) circle[radius=0.1];
		\path[fill=black] (1,0) circle[radius=0.1];
\end{tikzpicture}}\\[3pt]
- & \raisebox{-0.5em}{\begin{tikzpicture}
		\node[left] at (0,0) {\small{$1$}};
		\node[right] at (1,0) {\small{$2$}};
		\draw [line width=1.5pt,dashed] (0,0) to[out=60,in=-180] (0.6,0.25);
		\draw [line width=1.5pt,dashed] (0.5,0.25) to[out=0,in=120] (1,0);
		\draw [line width=1.5pt,->] (0,0) to[out=-60,in=-180] (0.6,-0.25);
		\draw [line width=1.5pt] (0.5,-0.25) to[out=0,in=-120] (1,0);
		\path[fill=black] (0,0) circle[radius=0.1];
		\path[fill=black] (1,0) circle[radius=0.1];
\end{tikzpicture}} -
\raisebox{-0.5em}{\begin{tikzpicture}
		\node[left] at (0,0) {\small{$1$}};
		\node[right] at (1,0) {\small{$2$}};
		\draw [line width=1.5pt,dashed] (0,0) to[out=60,in=-180] (0.6,0.25);
		\draw [line width=1.5pt,dashed] (0.5,0.25) to[out=0,in=120] (1,0);
		\draw [line width=1.5pt] (0,0) to[out=-60,in=-180] (0.5,-0.25);
		\draw [line width=1.5pt,->] (1,0) to[out=-120,in=0] (0.4,-0.25);
		\path[fill=black] (0,0) circle[radius=0.1];
		\path[fill=black] (1,0) circle[radius=0.1];
\end{tikzpicture}} +
\raisebox{-0.5em}{\begin{tikzpicture}
		\node[left] at (0,0) {\small{$1$}};
		\node[right] at (1,0) {\small{$2$}};
		\draw [line width=1.5pt,dashed] (0,0) to[out=60,in=-180] (0.6,0.25);
		\draw [line width=1.5pt,dashed] (0.5,0.25) to[out=0,in=120] (1,0);
		\draw [line width=1.5pt,dashed] (0,0) to[out=-60,in=-180] (0.5,-0.25);
		\draw [line width=1.5pt,dashed] (1,0) to[out=-120,in=0] (0.4,-0.25);
		\path[fill=black] (0,0) circle[radius=0.1];
		\path[fill=black] (1,0) circle[radius=0.1];
\end{tikzpicture}}.\\[3pt]
	\end{aligned}
\end{equation}
One can see that due to the basic relations of $\theta$ functions $\theta_{i,j}^2=\theta_{i,j}$ and $\theta_{i,j}\theta_{j,i}=0$, the grey arrows can be removed, then the final decomposition of this wavefunction coefficient integral contains no loop structures actually, and ends up with a combination of 7 (instead of 9) terms. Similar thing also happens for triangle diagrams since we have $\theta_{i,j}\theta_{j,k}\theta_{i,k}=\theta_{i,j}\theta_{j,k}$ and $\theta_{i,j}\theta_{j,k}\theta_{k,i}=0$. Graphically, this is depicted as
\begin{equation}
	\raisebox{-2em}{\begin{tikzpicture}
		\drawA{0}{0}{0.5}{0.866025}
		\drawA{0.5}{0.866025}{1}{0}
		\draw[line width=1.5pt,black!30,->] (0,0)--(0.6,0);
		\draw[line width=1.5pt,black!30] (0.5,0)--(1,0);
		\path[fill=black] (0,0) circle[radius=0.1];
		\path[fill=black] (1,0) circle[radius=0.1];
		\path[fill=black] (0.5,0.866025) circle[radius=0.1];
		\node[below left] at (0,0) {\small{$k$}};
		\node[above] at (0.5,0.866025) {\small{$j$}};
		\node[below right] at (1,0) {\small{$i$}};
	\end{tikzpicture}} =
	\raisebox{-1em}{\begin{tikzpicture}
		\coordinate (X1) at (0,0);
		\coordinate (X2) at (1,0);
		\coordinate (X3) at (2,0);
		\node[below] at (X1) {\small{$k$}};
		\node[below] at (X2) {\small{$j$}};
		\node[below] at (X3) {\small{$i$}};
		\draw[line width=1.5pt,->] (X1)--(0.6,0);
		\draw[line width=1.5pt] (0.5,0)--(X2);
		\draw[line width=1.5pt,->] (X2)--(1.6,0);
		\draw[line width=1.5pt] (1.5,0)--(X3);
		\path[fill=black] (X1) circle[radius=0.1];
		\path[fill=black] (X2) circle[radius=0.1];
		\path[fill=black] (X3) circle[radius=0.1];
	\end{tikzpicture}} \quad \text{and} 
 \raisebox{-2em}{\begin{tikzpicture}
		\drawA{0}{0}{0.5}{0.866025}
		\drawA{0.5}{0.866025}{1}{0}
        \drawA{1}{0}{0}{0}
		\path[fill=black] (0,0) circle[radius=0.1];
		\path[fill=black] (1,0) circle[radius=0.1];
		\path[fill=black] (0.5,0.866025) circle[radius=0.1];
		\node[below left] at (0,0) {\small{$k$}};
		\node[above] at (0.5,0.866025) {\small{$j$}};
		\node[below right] at (1,0) {\small{$i$}};
        \draw[line width=1.5pt,red] (0,-0.3)--(1,0.85);
	\end{tikzpicture}}.
\end{equation}
The first non-trivial loop structure for basic time integral, which cannot be further reduced to a tree, only appears at four-site with $\theta_{i,j}\theta_{j,k}\theta_{i,l}\theta_{l,k}$, which can be shown as
\begin{equation}
    \raisebox{-2em}{\begin{tikzpicture}
        \coordinate (X1) at (0,1);
		\coordinate (X2) at (0,0);
        \coordinate (X3) at (1,1);
        \coordinate (X4) at (1,0);
		\node[above left] at (X1) {\small{$k$}};
        \node[below left] at (X2) {\small{$l$}};
        \node[above right] at (X3) {\small{$j$}};
        \node[below right] at (X4) {\small{$i$}};
		\draw[line width=1.5pt,->] (X1)--(0,0.4);
        \draw[line width=1.5pt] (0,0.5)--(X2);
        \draw[line width=1.5pt,->] (X3)--(1,0.4);
        \draw[line width=1.5pt] (1,0.5)--(X4);
        \draw[line width=1.5pt,->] (X1)--(0.6,1);
        \draw[line width=1.5pt] (0.5,1)--(X3);
        \draw[line width=1.5pt,->] (X2)--(0.6,0);
        \draw[line width=1.5pt] (0.5,0)--(X4);
		\path[fill=black] (X1) circle[radius=0.1];
		\path[fill=black] (X2) circle[radius=0.1];
        \path[fill=black] (X3) circle[radius=0.1];
		\path[fill=black] (X4) circle[radius=0.1];
    \end{tikzpicture}}.
\end{equation}
After deleting all vanishing graphs, we have 25 and 79 terms in the decomposition of the triangle and box respectively. Generally, for an $n$-gon graph, there are two vanishing terms among those naively $3^n$ terms in the decomposition: the clockwise cycle and the anti-clockwise cycle. So there are $(3^n-2)$ terms in the decomposition:
\begin{equation}\label{eq:ngon}
	   \raisebox{-3.5em}{\begin{tikzpicture}[scale=0.8]
		\draw[line width=1.5pt] (0,0) circle [radius=1.25];
		\draw[fill=black] (0.323524, 1.20741) circle [radius=0.1];
		\draw[fill=black] (0.883883, 0.883883) circle [radius=0.1];
		\draw[fill=black] (-0.323524, 1.20741) circle [radius=0.1];
		\draw[fill=black] (-0.883883, 0.883883) circle [radius=0.1];
		\draw[fill=black] (0,-1.25) circle [radius=0.1];
		\draw[fill=black] (0.535757, -1.29343) circle [radius=0.03];
		\draw[fill=black] (0.989949, -0.989949) circle [radius=0.03];
		\draw[fill=black] (1.29343, -0.535757) circle [radius=0.03];
		\draw[fill=black] (1.4, 0) circle [radius=0.03];
		\draw[fill=black] (1.29343, 0.535757) circle [radius=0.03];
		\draw[fill=black] (-0.535757, -1.29343) circle [radius=0.03];
		\draw[fill=black] (-0.989949, -0.989949) circle [radius=0.03];
		\draw[fill=black] (-1.29343, -0.535757) circle [radius=0.03];
		\draw[fill=black] (-1.4, 0) circle [radius=0.03];
		\draw[fill=black] (-1.29343, 0.535757) circle [radius=0.03];
		\node[above right] at (0.323524, 1.20741) {\tiny{$n$}};
		\node[above right] at (0.883883, 0.883883) {\tiny{$n-1$}};
		\node[above left] at (-0.323524, 1.20741) {\tiny{$1$}};
		\node[above left] at (-0.883883, 0.883883) {\tiny{$2$}};
		\node[below] at (0, -1.4) {\tiny{$j$}};
	\end{tikzpicture}}\!\!\!\!\!\!=
	\raisebox{-3.5em}{\begin{tikzpicture}[scale=0.8]
		\draw[line width=1.5pt] (0,0) circle [radius=1.25];
		\draw[fill=black] (0.323524, 1.20741) circle [radius=0.1];
		\draw[fill=black] (0.883883, 0.883883) circle [radius=0.1];
		\draw[fill=black] (-0.323524, 1.20741) circle [radius=0.1];
		\draw[fill=black] (-0.883883, 0.883883) circle [radius=0.1];
		\draw[fill=black] (0,-1.25) circle [radius=0.1];
		\draw[fill=black] (0.535757, -1.29343) circle [radius=0.03];
		\draw[fill=black] (0.989949, -0.989949) circle [radius=0.03];
		\draw[fill=black] (1.29343, -0.535757) circle [radius=0.03];
		\draw[fill=black] (1.4, 0) circle [radius=0.03];
		\draw[fill=black] (1.29343, 0.535757) circle [radius=0.03];
		\draw[fill=black] (-0.535757, -1.29343) circle [radius=0.03];
		\draw[fill=black] (-0.989949, -0.989949) circle [radius=0.03];
		\draw[fill=black] (-1.29343, -0.535757) circle [radius=0.03];
		\draw[fill=black] (-1.4, 0) circle [radius=0.03];
		\draw[fill=black] (-1.29343, 0.535757) circle [radius=0.03];
		\node[above right] at (0.323524, 1.20741) {\tiny{$n$}};
		\node[above right] at (0.883883, 0.883883) {\tiny{$n-1$}};
		\node[above left] at (-0.323524, 1.20741) {\tiny{$1$}};
		\node[above left] at (-0.883883, 0.883883) {\tiny{$2$}};
		\node[below] at (0, -1.4) {\tiny{$j$}};
		\draw[line width=1.3pt,->] (-0.603704-0.01, 1.04565+0.045)--(-0.603704+0.01, 1.04565+0.06);
		\draw[line width=1.3pt,->] (0.603704+0.045, 1.04565+0.01)--(0.603704+0.07, 1.04565-0.01);
		\draw[line width=1.3pt,->] (0.04,1.25)--(0.05,1.25);
		\draw[line width=1.3pt,->] (-0.883883-0.2+0.04, 0.883883-0.2)--(-0.883883-0.2+0.08, 0.883883-0.2+0.05);
		\draw[line width=1.3pt,->] (0.883883+0.2-0.01+0.003, 0.883883-0.25+0.01)--(0.883883+0.2, 0.883883-0.25);
		\draw[line width=1.3pt,->] (-0.3+0.01-0.04-0.003, -1.25+0.02-0.001+0.02)--(-0.3-0.04, -1.25+0.02+0.02);
		\draw[line width=1.3pt,->] (0.3+0.01-0.05-0.06, -1.25+0.023-0.01)--(0.3-0.05-0.06, -1.25+0.02-0.01);
		\draw[line width=1.5pt,red] (1.3,1.3)--(-1.3,-1.3);
	\end{tikzpicture}}\!\!\!\!\!\!+
	\raisebox{-3.5em}{\begin{tikzpicture}[scale=0.8]
		\draw[line width=1.5pt] (0,0) circle [radius=1.25];
		\draw[fill=black] (-0.323524, 1.20741) circle [radius=0.1];
		\draw[fill=black] (-0.883883, 0.883883) circle [radius=0.1];
		\draw[fill=black] (0.323524, 1.20741) circle [radius=0.1];
		\draw[fill=black] (0.883883, 0.883883) circle [radius=0.1];
		\draw[fill=black] (0,-1.25) circle [radius=0.1];
		\draw[fill=black] (-0.535757, -1.29343) circle [radius=0.03];
		\draw[fill=black] (-0.989949, -0.989949) circle [radius=0.03];
		\draw[fill=black] (-1.29343, -0.535757) circle [radius=0.03];
		\draw[fill=black] (-1.4, 0) circle [radius=0.03];
		\draw[fill=black] (-1.29343, 0.535757) circle [radius=0.03];
		\draw[fill=black] (0.535757, -1.29343) circle [radius=0.03];
		\draw[fill=black] (0.989949, -0.989949) circle [radius=0.03];
		\draw[fill=black] (1.29343, -0.535757) circle [radius=0.03];
		\draw[fill=black] (1.4, 0) circle [radius=0.03];
		\draw[fill=black] (1.29343, 0.535757) circle [radius=0.03];
		\node[above left] at (-0.323524, 1.20741) {\tiny{$1$}};
		\node[above left] at (-0.883883, 0.883883) {\tiny{$2$}};
		\node[above right] at (0.323524, 1.20741) {\tiny{$n$}};
		\node[above right] at (0.883883, 0.883883) {\tiny{$n-1$}};
		\node[below] at (0, -1.4) {\tiny{$j$}};
		\draw[line width=1.3pt,->] (0.603704+0.01, 1.04565+0.045)--(0.603704-0.01, 1.04565+0.06);
		\draw[line width=1.3pt,->] (-0.603704-0.045, 1.04565+0.01)--(-0.603704-0.07, 1.04565-0.01);
		\draw[line width=1.3pt,->] (-0.04,1.25)--(-0.05,1.25);
		\draw[line width=1.3pt,->] (0.883883+0.2-0.04, 0.883883-0.2)--(0.883883+0.2-0.08, 0.883883-0.2+0.05);
		\draw[line width=1.3pt,->] (-0.883883-0.2+0.01-0.003, 0.883883-0.25+0.01)--(-0.883883-0.2, 0.883883-0.25);
		\draw[line width=1.3pt,->] (0.3-0.01+0.04+0.003, -1.25+0.02-0.001+0.02)--(0.3+0.04, -1.25+0.02+0.02);
		\draw[line width=1.3pt,->] (-0.3-0.01+0.05+0.06, -1.25+0.023-0.01)--(-0.3+0.05+0.06, -1.25+0.02-0.01);
		\draw[line width=1.5pt,red] (1.3,1.3)--(-1.3,-1.3);
	\end{tikzpicture}} \!\!\!\!\!\! + \text{other $(3^n-2)$ terms.}
\end{equation}
Similarly for general $L$-loop two-site graphs, every time integral in the expansion is either a two-site chain or a product of two one-site graphs, which can be labeled by their energies on each vertex. By considering all possible energy assignments and ignoring cases that lead to incompatible time order (like the red-slashed ones in \eqref{eq:bubbleexp}), we conclude that the number of time integrals in the expansion of an $L$-loop two-site graph is $ (2^{L+2}-1) $.

For correlators, things become a little more complicated because of the double types of vertices and more kinds of propagators introduced in \eqref{eq:correprop}. We should consider graphs with all possible bulk point assignments. Take the two-site correlator as an example: following from the definition, there are four terms from the correlator corresponding to four distributions $\{+,+\}$, $\{-,-\}$, $\{+.-\}$ and $\{-,+\}$ for the bulk points, each of which corresponds to different propagator \eqref{eq:correprop} along the edge. After decomposing the correlator to basic time integrals, propagators with the same sign on its vertices yield two terms, while propagators with different signs yield one term.  Moreover, terms from $D_{++}$ or $D_{--}$, as well as terms from $D_{+-}$ or $D_{-+}$ are parity conjugate to each other. So finally the two-site chain is decomposed into $2\times(2+1)^1=6$ basic time integrals. 
\begin{equation}
	\begin{aligned}
		\raisebox{-11pt}{\begin{tikzpicture}
				\coordinate (X1) at (0,0);
				\coordinate (X2) at (1.5,0);
				\node[below] at (X1) {\small{$1$}};
				\node[below] at (X2) {\small{$2$}};
				\draw[line width=.5pt] (0,0.05)--(1.5,0.05);
                \draw[line width=.5pt] (0,-0.05)--(1.5,-0.05);
				\path[fill=black] (X1) circle[radius=0.1];
				\path[fill=black] (X2) circle[radius=0.1];
		\end{tikzpicture}} & = \raisebox{-11pt}{\begin{tikzpicture}
				\coordinate (X1) at (0,0);
				\coordinate (X2) at (1.5,0);
				\node[below] at (X1) {\small{$1,+$}};
				\node[below] at (X2) {\small{$2,+$}};
				\draw[line width=1.5pt] (X1)--(X2);
				\path[fill=black] (X1) circle[radius=0.1];
				\path[fill=black] (X2) circle[radius=0.1];
		\end{tikzpicture}}+\raisebox{-11pt}{\begin{tikzpicture}
		\coordinate (X1) at (0,0);
		\coordinate (X2) at (1.5,0);
		\node[below] at (X1) {\small{$1,-$}};
		\node[below] at (X2) {\small{$2,-$}};
		\draw[line width=1.5pt] (X1)--(X2);
		\path[fill=black] (X1) circle[radius=0.1];
		\path[fill=black] (X2) circle[radius=0.1];
	\end{tikzpicture}}-\raisebox{-11pt}{\begin{tikzpicture}
	\coordinate (X1) at (0,0);
	\coordinate (X2) at (1.5,0);
	\node[below] at (X1) {\small{$1,+$}};
	\node[below] at (X2) {\small{$2,-$}};
	\draw[line width=1.5pt] (X1)--(X2);
	\path[fill=black] (X1) circle[radius=0.1];
	\path[fill=black] (X2) circle[radius=0.1];
\end{tikzpicture}}-\raisebox{-11pt}{\begin{tikzpicture}
\coordinate (X1) at (0,0);
\coordinate (X2) at (1.5,0);
\node[below] at (X1) {\small{$1,-$}};
\node[below] at (X2) {\small{$2,+$}};
\draw[line width=1.5pt] (X1)--(X2);
\path[fill=black] (X1) circle[radius=0.1];
\path[fill=black] (X2) circle[radius=0.1];
\end{tikzpicture}}\\
		& = \raisebox{-11pt}{\begin{tikzpicture}
				\coordinate (X1) at (0,0);
				\coordinate (X2) at (1.5,0);
				\node[below] at (X1) {\small{$1,+$}};
				\node[below] at (X2) {\small{$2,+$}};
				\draw[line width=1.5pt,->] (X1)--(0.85,0);
				\draw[line width=1.5pt] (0.75,0)--(X2);
				\path[fill=black] (X1) circle[radius=0.1];
				\path[fill=black] (X2) circle[radius=0.1];
		\end{tikzpicture}}+\raisebox{-11pt}{\begin{tikzpicture}
		\coordinate (X1) at (0,0);
		\coordinate (X2) at (1.5,0);
		\node[below] at (X1) {\small{$1,+$}};
		\node[below] at (X2) {\small{$2,+$}};
		\draw[line width=1.5pt] (X1)--(0.85,0);
		\draw[line width=1.5pt,->] (X2)--(0.65,0);
		\path[fill=black] (X1) circle[radius=0.1];
		\path[fill=black] (X2) circle[radius=0.1];
	\end{tikzpicture}} + \raisebox{-11pt}{\begin{tikzpicture}
	\coordinate (X1) at (0,0);
	\coordinate (X2) at (1.5,0);
	\node[below] at (X1) {\small{$1,-$}};
	\node[below] at (X2) {\small{$2,-$}};
	\draw[line width=1.5pt,->] (X1)--(0.85,0);
	\draw[line width=1.5pt] (0.75,0)--(X2);
	\path[fill=black] (X1) circle[radius=0.1];
	\path[fill=black] (X2) circle[radius=0.1];
\end{tikzpicture}}+\raisebox{-11pt}{\begin{tikzpicture}
\coordinate (X1) at (0,0);
\coordinate (X2) at (1.5,0);
\node[below] at (X1) {\small{$1,-$}};
\node[below] at (X2) {\small{$2,-$}};
\draw[line width=1.5pt] (X1)--(0.85,0);
\draw[line width=1.5pt,->] (X2)--(0.65,0);
\path[fill=black] (X1) circle[radius=0.1];
\path[fill=black] (X2) circle[radius=0.1];
\end{tikzpicture}}\\
& \hspace{14em}\hspace{9pt} -\raisebox{-11pt}{\begin{tikzpicture}
		\coordinate (X1) at (0,0);
		\coordinate (X2) at (1.5,0);
		\node[below] at (X1) {\small{$1,+$}};
		\node[below] at (X2) {\small{$2,-$}};
		\draw[line width=1.5pt,dashed] (X1)--(X2);
		\path[fill=black] (X1) circle[radius=0.1];
		\path[fill=black] (X2) circle[radius=0.1];
\end{tikzpicture}}-\raisebox{-11pt}{\begin{tikzpicture}
\coordinate (X1) at (0,0);
\coordinate (X2) at (1.5,0);
\node[below] at (X1) {\small{$1,-$}};
\node[below] at (X2) {\small{$2,+$}};
\draw[line width=1.5pt,dashed] (X1)--(X2);
\path[fill=black] (X1) circle[radius=0.1];
\path[fill=black] (X2) circle[radius=0.1];
\end{tikzpicture}}\\
& = \raisebox{-11pt}{\begin{tikzpicture}
				\coordinate (X1) at (0,0);
				\coordinate (X2) at (1.5,0);
				\node[below] at (X1) {\small{$1,+$}};
				\node[below] at (X2) {\small{$2,+$}};
				\draw[line width=1.5pt,->] (X1)--(0.85,0);
				\draw[line width=1.5pt] (0.75,0)--(X2);
				\path[fill=black] (X1) circle[radius=0.1];
				\path[fill=black] (X2) circle[radius=0.1];
		\end{tikzpicture}}+\raisebox{-11pt}{\begin{tikzpicture}
		\coordinate (X1) at (0,0);
		\coordinate (X2) at (1.5,0);
		\node[below] at (X1) {\small{$1,+$}};
		\node[below] at (X2) {\small{$2,+$}};
		\draw[line width=1.5pt] (X1)--(0.85,0);
		\draw[line width=1.5pt,->] (X2)--(0.65,0);
		\path[fill=black] (X1) circle[radius=0.1];
		\path[fill=black] (X2) circle[radius=0.1];
	\end{tikzpicture}}-\raisebox{-11pt}{\begin{tikzpicture}
		\coordinate (X1) at (0,0);
		\coordinate (X2) at (1.5,0);
		\node[below] at (X1) {\small{$1,+$}};
		\node[below] at (X2) {\small{$2,-$}};
		\draw[line width=1.5pt,dashed] (X1)--(X2);
		\path[fill=black] (X1) circle[radius=0.1];
		\path[fill=black] (X2) circle[radius=0.1];
\end{tikzpicture}}+ \text{c.c.}
	\end{aligned}
\end{equation}
Generally speaking, for a tree correlator, this decomposition results in $2\times 3^e$ terms. It is easier to see this by firstly drawing all possible $ 3^e $ basis graphs for correlators and assigning possible bulk point types to each vertex. By adding $\pm$ signs on the vertices in each basic time integral, there are only two allowed configurations of assignment, which are complex conjugate to each other. Thus there are $ 3^e $ independent terms, together with their complex conjugation, totally $ 2 \times 3^e $ terms contributing to tree correlators. 

For correlators with a loop structure, the decomposition is straightforward and similar to the previous discussion. As an example, the decomposition of a bubble diagram reads
\begin{equation}\label{eq:bubbleexp}
	\begin{aligned}
		\raisebox{-0.5em}{\begin{tikzpicture}
				\node[left] at (0,0) {\small{$1$}};
				\node[right] at (1,0) {\small{$2$}};
				\draw [line width=0.5pt] (-0.05,0) to[out=60,in=-180] (0.5,0.29);
				\draw [line width=0.5pt] (0.05,0) to[out=60,in=-180] (0.5,0.21);
				\draw [line width=0.5pt] (0.5,0.29) to[out=0,in=120] (1.05,0);
				\draw [line width=0.5pt] (0.5,0.21) to[out=0,in=120] (0.95,0);
				\draw [line width=0.5pt] (-0.05,0) to[out=-60,in=-180] (0.5,-0.29);
				\draw [line width=0.5pt] (0.05,0) to[out=-60,in=-180] (0.5,-0.21);
				\draw [line width=0.5pt] (0.5,-0.29) to[out=0,in=-120] (1.05,0);
				\draw [line width=0.5pt] (0.5,-0.21) to[out=0,in=-120] (0.95,0);
				\path[fill=black] (0,0) circle[radius=0.1];
				\path[fill=black] (1,0) circle[radius=0.1];
		\end{tikzpicture}} = & \raisebox{-0.5em}{\begin{tikzpicture}
				\node[left] at (0,0) {\tiny{$1,+$}};
				\node[right] at (1,0) {\tiny{$2,+$}};
				\draw [line width=1.5pt,->] (0,0) to[out=60,in=-180] (0.6,0.25);
				\draw [line width=1.5pt] (0.5,0.25) to[out=0,in=120] (1,0);
				\draw [line width=1.5pt,->] (0,0) to[out=-60,in=-180] (0.6,-0.25);
				\draw [line width=1.5pt] (0.5,-0.25) to[out=0,in=-120] (1,0);
				\path[fill=black] (0,0) circle[radius=0.1];
				\path[fill=black] (1,0) circle[radius=0.1];
		\end{tikzpicture}} + \raisebox{-0.5em}{\begin{tikzpicture}
		\node[left] at (0,0) {\tiny{$1,+$}};
		\node[right] at (1,0) {\tiny{$2,+$}};
		\draw [line width=1.5pt] (0,0) to[out=60,in=-180] (0.5,0.25);
		\draw [line width=1.5pt,->] (1,0) to[out=120,in=0] (0.4,0.25);
		\draw [line width=1.5pt] (0,0) to[out=-60,in=-180] (0.5,-0.25);
		\draw [line width=1.5pt,->] (1,0) to[out=-120,in=0] (0.4,-0.25);
		\path[fill=black] (0,0) circle[radius=0.1];
		\path[fill=black] (1,0) circle[radius=0.1];
	\end{tikzpicture}}+ \raisebox{-0.5em}{\begin{tikzpicture}
	\node[left] at (0,0) {\tiny{$1,+$}};
	\node[right] at (1,0) {\tiny{$2,-$}};
	\draw [line width=1.5pt,dashed] (0,0) to[out=60,in=-180] (0.6,0.25);
	\draw [line width=1.5pt,dashed] (0.5,0.25) to[out=0,in=120] (1,0);
	\draw [line width=1.5pt,dashed] (0,0) to[out=-60,in=-180] (0.5,-0.25);
	\draw [line width=1.5pt,dashed] (1,0) to[out=-120,in=0] (0.4,-0.25);
	\path[fill=black] (0,0) circle[radius=0.1];
	\path[fill=black] (1,0) circle[radius=0.1];
\end{tikzpicture}}+ \text{c.c.}
	\end{aligned}
\end{equation}
Interestingly, for two-site $L$-loop graph, since the propagator type depends on vertex type, there are only 6 terms in the decomposition: 2 terms for $\{+,+\}$ type vertex, 1 term for $\{+,-\}$ type vertex and their complex conjugation.
\begin{equation}\label{eq:bubbleexp}
	\begin{aligned}
		\raisebox{-0.5em}{\begin{tikzpicture}
				\node[left] at (0,0) {\small{$1$}};
				\node[right] at (1,0) {\small{$2$}};
				\draw [line width=0.5pt] (-0.05,0) to[out=60,in=-180] (0.5,0.29);
				\draw [line width=0.5pt] (0.05,0) to[out=60,in=-180] (0.5,0.21);
				\draw [line width=0.5pt] (0.5,0.29) to[out=0,in=120] (1.05,0);
				\draw [line width=0.5pt] (0.5,0.21) to[out=0,in=120] (0.95,0);
				\draw [line width=0.5pt] (-0.05,0) to[out=-60,in=-180] (0.5,-0.29);
				\draw [line width=0.5pt] (0.05,0) to[out=-60,in=-180] (0.5,-0.21);
				\draw [line width=0.5pt] (0.5,-0.29) to[out=0,in=-120] (1.05,0);
				\draw [line width=0.5pt] (0.5,-0.21) to[out=0,in=-120] (0.95,0);
				\path[fill=black] (0,0) circle[radius=0.1];
				\path[fill=black] (1,0) circle[radius=0.1];
                \path[fill=black] (0.5,0.1) circle[radius=0.025];
				\path[fill=black] (0.5,0.0) circle[radius=0.025];
				\path[fill=black] (0.5,-0.1) circle[radius=0.025];
		\end{tikzpicture}} = & \raisebox{-0.5em}{\begin{tikzpicture}
				\node[left] at (0,0) {\tiny{$1,+$}};
				\node[right] at (1,0) {\tiny{$2,+$}};
				\draw [line width=1.5pt,->] (0,0) to[out=60,in=-180] (0.6,0.25);
				\draw [line width=1.5pt] (0.5,0.25) to[out=0,in=120] (1,0);
				\draw [line width=1.5pt,->] (0,0) to[out=-60,in=-180] (0.6,-0.25);
				\draw [line width=1.5pt] (0.5,-0.25) to[out=0,in=-120] (1,0);
				\path[fill=black] (0,0) circle[radius=0.1];
				\path[fill=black] (1,0) circle[radius=0.1];
    \path[fill=black] (0.5,0.1) circle[radius=0.025];
				\path[fill=black] (0.5,0.0) circle[radius=0.025];
				\path[fill=black] (0.5,-0.1) circle[radius=0.025];
		\end{tikzpicture}} + \raisebox{-0.5em}{\begin{tikzpicture}
		\node[left] at (0,0) {\tiny{$1,+$}};
		\node[right] at (1,0) {\tiny{$2,+$}};
		\draw [line width=1.5pt] (0,0) to[out=60,in=-180] (0.5,0.25);
		\draw [line width=1.5pt,->] (1,0) to[out=120,in=0] (0.4,0.25);
		\draw [line width=1.5pt] (0,0) to[out=-60,in=-180] (0.5,-0.25);
		\draw [line width=1.5pt,->] (1,0) to[out=-120,in=0] (0.4,-0.25);
		\path[fill=black] (0,0) circle[radius=0.1];
		\path[fill=black] (1,0) circle[radius=0.1];
  \path[fill=black] (0.5,0.1) circle[radius=0.025];
				\path[fill=black] (0.5,0.0) circle[radius=0.025];
				\path[fill=black] (0.5,-0.1) circle[radius=0.025];
	\end{tikzpicture}}+ \raisebox{-0.5em}{\begin{tikzpicture}
	\node[left] at (0,0) {\tiny{$1,+$}};
	\node[right] at (1,0) {\tiny{$2,-$}};
	\draw [line width=1.5pt,dashed] (0,0) to[out=60,in=-180] (0.6,0.25);
	\draw [line width=1.5pt,dashed] (0.5,0.25) to[out=0,in=120] (1,0);
	\draw [line width=1.5pt,dashed] (0,0) to[out=-60,in=-180] (0.5,-0.25);
	\draw [line width=1.5pt,dashed] (1,0) to[out=-120,in=0] (0.4,-0.25);
	\path[fill=black] (0,0) circle[radius=0.1];
	\path[fill=black] (1,0) circle[radius=0.1];
 \path[fill=black] (0.5,0.1) circle[radius=0.025];
				\path[fill=black] (0.5,0.0) circle[radius=0.025];
				\path[fill=black] (0.5,-0.1) circle[radius=0.025];
\end{tikzpicture}}+ \text{c.c.}
	\end{aligned}
\end{equation}



\subsection{Contracting edges: differentials of a time integral}\label{subsec:de1}
After discussing the decomposition, let us turn to the computation of these integrals. In \cite{Fan:2024iek}, the authors furthermore decompose these basic time integrals into family tree integrals and consider series expansion. In this work, we will instead focus on the differential equations these basic time integrals satisfy.

To derive the differential equations, we treat $q_i$ as constants and take the partial derivative with respect to each energy $\omega_m$ for $\mathbf{T}^{q_1,\cdots,q_n}_\mathcal{N}$. Following the definition, it is very straightforward to get the relation
\begin{equation}	\partial_m\mathbf{T}^{q_1,q_2,\ldots,q_n}_{\mathcal{N}}(\omega_1,\omega_2,\ldots,\omega_n)=-\ii \, \mathbf{T}^{q_1,q_2,\ldots,q_m+1,\ldots,q_n}_{\mathcal{N}}(\omega_1,\omega_2,\ldots,\omega_n),
\end{equation}
where $\partial_m\equiv\frac{\partial}{\partial \omega_m}$, {\it i.e.} the operator simply raises the twist exponent for vertex $m$ by $1$. However, since we would like to stick to the integrals with $q_m$ fixed, we need to perform integration by part and bring the twist back to $q_{m}$. To achieve this, we use the differential relation
\begin{equation}\label{eq:qshift}
	\begin{aligned}
		& \hspace{-5em} {\partial}_{\tau_m} \left[ (-\tau_m)^{q_m} e^{\ii\,\omega_m \tau_m} \prod_{(m,b)\in\mathcal{N}}\theta_{m,b}\prod_{(f,m)\in\mathcal{N}}\theta_{f,m}\right]\\
		  &\hspace{5em}  ={-}q_m (-\tau_m)^{q_m-1} e^{\ii\,\omega_m \tau_m}  \prod_{(m,b)\in\mathcal{N}}\theta_{m,b}\prod_{(f,m)\in\mathcal{N}}\theta_{f,m}\\
        & \hspace{6.5em} {+}\ii \, \omega_m ({-}\tau_m)^{q_m} e^{\ii \, \omega_m \tau_m}  \prod_{(m,b)\in\mathcal{N}}\theta_{m,b}\prod_{(f,m)\in\mathcal{N}}\theta_{f,m}\\
		& \hspace{6.5em} {+}\sum_{b_0}(-\tau_m)^{q_m} e^{\ii\,\omega_m \tau_m}  \prod_{\substack{(m,b)\in\mathcal{N}\\b\neq b_0}}\theta_{m,b}\delta_{m,b_0}\prod_{(f,m)\in\mathcal{N}}\theta_{f,m}\\
        & \hspace{6.5em} {-}\sum_{f_0}(-\tau_m)^{q_m} e^{\ii\,\omega_m \tau_m}  \prod_{(m,b)\in\mathcal{N}}\theta_{m,b}\prod_{\substack{(f,m)\in\mathcal{N}\\f\neq f_0}}\theta_{f,m}\delta_{f_0,m},
	\end{aligned}
\end{equation}
where $b$ are all the points that are prior to point $m$ {\it i.e.} $\tau_b<\tau_m$ and $f$ are all the points that are later than $m$ in the time structure $\mathcal{N}$. Let us take a more careful look into these terms and show their graphical meanings. Firstly, the first term on the RHS corresponds to the integrand of $-q_m\mathbf{T}^{q_1,q_2,\ldots,q_n}_{\mathcal{N}} (\omega_1,\omega_2,\ldots,\omega_n)$. Similarly, the second term corresponds to the integrand of $\ii \, \omega_m\mathbf{T}^{q_1,q_2,\ldots,q_m+1,\ldots,q_n}_{\mathcal{N}}(\omega_1,\omega_2,\ldots,\omega_n)$. Finally, due to the existence of the $\delta$ function, $\tau_m$ can be trivially integrated out, leaving each term in the last two lines corresponding to a lower-point basic time integral by contracting the edges between vertices $b_0$ and $m$ or $f_0$ and $m$. Therefore after performing the integration by part using \eqref{eq:qshift}, the following identity can be derived
\begin{equation}\label{eq:pm}
	\begin{aligned}
		\partial_m\mathbf{T}^{q_1,q_2,\ldots,q_n}_{\mathcal{N}}(\omega_1,\omega_2,\ldots,\omega_n)
		&=-\frac{q_m}{\omega_m}\mathbf{T}^{q_1,q_2,\ldots,q_n}_{\mathcal{N}}(\omega_1,\omega_2,\ldots,\omega_n)\\
		&-\frac{\ii}{\omega_m}\sum_{(m,b)\in{\mathcal{N}}}\mathbf{T}^{q_1,q_2,\ldots,q_{m,b},\ldots,q_n}_{\mathcal{N}_{m,b}}(\omega_1,\omega_2,\ldots,\omega_{m,b}\ldots,\omega_n)\\
		&+\frac{\ii}{\omega_m}\sum_{(f,m)\in{\mathcal{N}}}\mathbf{T}^{q_1,q_2,\ldots,q_{m,f},\ldots,q_n}_{\mathcal{N}_{m,f}}(\omega_1,\omega_2,\ldots,\omega_{m,f}\ldots,\omega_n),
	\end{aligned}
\end{equation}
where $q_{a,b}\equiv q_a+q_b$ and similar for $\omega_{a,b}$\footnote{It is also the same for longer indices. For example, $q_{a,\ldots,c}=q_{a}+\cdots+q_{c}$ and sometimes we omit the commas for simplicity: $q_{a\ldots c}=q_{a,\ldots,c}$.}, and $\mathcal{N}_{m,b}$ denotes the time structure for a $(n{-}1)$-site graph with original $m$ and $b$ identified. For convenience, we introduce the notation $\mathrm{sgn}^{i,j}_\mathcal{N}$:
\begin{equation}
	\mathrm{sgn}^{i,j}_\mathcal{N}=\left\{
	\begin{aligned}
		&1,\qquad & \text{if}\quad(i,j)\in\mathcal{N}\\
		&-1,\qquad & \text{if}\quad(j,i)\in\mathcal{N}\\
		&0,\qquad & \text{if} \quad i,j\quad\text{not connected}
	\end{aligned}
\right. .
\end{equation}
Then we can write down the total differential of a general basic time integral as the following
\begin{tcolorbox}
\begin{align}\label{eq:dT} &\mathrm{d}\mathbf{T}^{q_1,q_2,\ldots,q_n}_{\mathcal{N}}(\omega_1,\omega_2,\ldots,\omega_n)=\sum_{m=1}^n \rmd\!\log\omega_m \Bigg ({-}q_m \mathbf{T}^{q_1,q_2,\ldots,q_n}_{\mathcal{N}}(\omega_1,\omega_2,\ldots,\omega_n)\\		&\hspace{10em}\left.{+}\ii\sum_{j\neq m}\mathrm{sgn}^{j,m}_\mathcal{N}\,\mathbf{T}^{q_1,q_2,\ldots,q_{j,m},\ldots,q_n}_{\mathcal{N}_{j,m}}(\omega_1,\omega_2,\ldots,\omega_{j,m}\ldots,\omega_n)\right).\nonumber
\end{align}
\end{tcolorbox}
This is the main result of this section. Eq.~\eqref{eq:d5-site} is an illustrative example of a five-site tree. Here $i{+}j$ denotes the energy on the node to be $\omega_i+\omega_j$, and the twist exponent to be $q_i+q_j$.
\begin{equation}\label{eq:d5-site}
	\begin{aligned}
		\rmd \raisebox{-2.25em}{
} 
	     \right]
	\end{aligned}
\end{equation}

We can also apply \eqref{eq:dT} to basic time integrals beyond tree level. For one-loop, the first non-trivial graph is 4-site. 
Eq.~\eqref{eq:1l4pt} shows the differential of a one-loop 4-site integral, where the grey arrow can be eliminated.
\begin{equation}\label{eq:1l4pt}
    \begin{aligned}
      \rmd  \raisebox{-2em}{%
}
    \right],
    \end{aligned}
\end{equation}

\section{Recursive solutions for basic time integrals}\label{sec:sol}

In this section, we solve the differential equations satisfied by $\mathbf{T}^{q_1,q_2,\ldots,q_n}_{\mathcal{N}}(\omega_1,\omega_2,\ldots,\omega_n)$ recursively. This amounts to choosing a specific integration path in the parameter space of $\omega:= (\omega_1,\omega_2,\ldots,\omega_n)$. Then any basic time integral can be expressed by integrals in \eqref{eq:euler} which can be viewed as a special case of \textit{Euler-Mellin Integrals}\footnote{They appear in different kinds of literature with many different names, such as generalized Euler integrals or Aomoto-Gelfand integrals. See \cite{Matsubara-Heo:2023ylc} for an overview. We also note that the integration domain $[0,1]$ seems to be different from the definition but can be transformed to $[0,\infty]$ by $\alpha\equiv y/(1+y)$, which makes relation with a lot of integrals appearing in physical context such as the stringy integrals defined in \cite{Arkani-Hamed:2019mrd}.}~\cite{Matsubara-Heo:2023ylc,berkesch2014euler}, plus some products of lower-point functions which can be solved in the same way.
\begin{equation}\label{eq:euler}
    I(\mathbf{\omega};\nu(\mathbf{q}))=\Gamma(q_{12\cdots n})\int_{0}^{1}\frac{\mathrm{d}\alpha}{\alpha}\alpha^{\nu(q)} P(\alpha,\mathbf{\omega})^{-q_{12\cdots n}},\, \frac{\mathrm{d}\alpha}{\alpha}=\frac{\mathrm{d}\alpha_{2}}{\alpha_{2}}\ldots \frac{\mathrm{d}\alpha_{n}}{\alpha_{n}},
\end{equation}
$\nu(q)$ is a summation of $q_{i}$ and $P(\alpha,\omega)$ is a polynomial of $\alpha_{i}$ with $\omega$ as coefficients. The explicit expressions of both depend on the specific integration path we choose.
Finally, we will show that series representations in \cite{Fan:2024iek} can be deduced from our results. Therefore, it provides us with a novel representation of basic time integrals and in this representation many analytic tools developed for Euler integrals or GKZ system \cite{gelfand_hypergeometric_1989} can be applied \cite{Matsubara-Heo:2023ylc,Aomoto:2011ggg,yoshida2013hypergeometric}.

\subsection{Solving differential equations recursively}\label{sec:recursivesol}
Let us start from the differential equation \eqref{eq:dT} and rewrite it in the following form
\begin{equation}\label{eq:tildeTdef}
    \begin{aligned}
        &\left(\omega_{m}\frac{\partial}{\partial \omega_{m}}+q_{m}\right)\mathbf{T}^{q_1,q_2,\ldots,q_n}_{\mathcal{N}}(\omega_1,\omega_2,\ldots,\omega_n)=-\ii \, \tilde{\mathbf{T}}_{\mathcal{N}_{c}}, \\
        &\hspace{8em}\tilde{\mathbf{T}}_{\mathcal{N}_{c}}(\omega_m)=-\sum_{j\ne m}\mathrm{sgn}_{\mathcal{N}}^{j,m}\,\mathbf{T}^{q_1,q_2,\ldots,q_{j,m},\ldots,q_n}_{\mathcal{N}_{j,m}}(\omega_1,\omega_2,\ldots,\omega_{j,m}\ldots,\omega_n),
    \end{aligned}
\end{equation}
where $\tilde{\mathbf{T}}_{\mathcal{N}_{c}}(\omega_m)$ is a shorthand for the basic time integrals with some edges contracted and we suppress its dependence on $q_{i}$ and other $\omega_{i}$ for simplicity. The coefficient $(-\ii)$ is due to the normalization for time integrals in \eqref{eq:Tdef}. Defining
\begin{equation}
    \omega_{m}^{-q_{m}}\mathbf{G}^{q_1,q_2,\ldots,q_n}_{\mathcal{N}}(\omega_1,\omega_2,\ldots,\omega_n)\equiv \mathbf{T}^{q_1,q_2,\ldots,q_n}_{\mathcal{N}}(\omega_1,\omega_2,\ldots,\omega_n),
\end{equation}
it is easy to derive that for $\mathbf{G}^{q_1,q_2,\ldots,q_n}_{\mathcal{N}}$,
\begin{equation}
\omega_{m}\partial_{m}\mathbf{G}^{q_1,q_2,\ldots,q_n}_{\mathcal{N}}(\omega_1,\omega_2,\ldots,\omega_n)=(-\ii) \, \omega_{m}^{q_{m}}\,\tilde{\mathbf{T}}_{\mathcal{N}_{c}}(\omega_m).
\end{equation}
Then this first-order partial differential equation can be straightforwardly solved to be
\begin{equation}
    \mathbf{G}^{q_1,\ldots,q_n}_{\mathcal{N}}(\omega_{m};\omega_1,\ldots,\omega_n)=(-\ii)\int_{0}^{\omega_{m}}\mathrm{d}t_{m} \,  t_{m}^{q_{m}-1}\,\tilde{\mathbf{T}}_{\mathcal{N}_{c}}(t_m)+C[\omega_{1},\ldots,\hat{\omega}_{m},\ldots,\omega_{n}],
\end{equation}
where $C$ is an arbitrary function free of $\omega_{m}$. And the expression for $\mathbf{T}^{q_1,q_2,\ldots,q_n}_{\mathcal{N}}$ follows:
\begin{equation}
    \begin{aligned}
        \mathbf{T}^{q_1,q_2,\ldots,q_n}_{\mathcal{N}}(\omega_1,\omega_2,\ldots,\omega_n)=\frac{(-\ii)}{\omega_{m}^{q_{m}}}\int_{0}^{\omega_{m}}\mathrm{d}t_{m}  \,  t_{m}^{q_{m}-1} \, \tilde{\mathbf{T}}_{\mathcal{N}_{c}}(t_{m})+\omega_{m}^{-q_{m}}C[\omega_{1},\ldots,\hat{\omega}_{m},\ldots,\omega_{n}].
    \end{aligned}
\end{equation}
Here we use $\tilde{\mathbf{T}}_{\mathcal{N}_{c}}(t_{m})$ to emphasize $\tilde{\mathbf{T}}_{\mathcal{N}_{c}}$ is a function of $t_{m}$. Now we normalize the integration variables $t_{m}$ with $\omega_{m}$, that is, perform the transformation $t_{m}\equiv \alpha_{m}\omega_{m}$ and arrive at
\begin{equation}\label{eq:Tsols}
    \begin{aligned}
        \mathbf{T}^{q_1,q_2,\ldots,q_n}_{\mathcal{N}}(\omega_1,\omega_2,\ldots,\omega_n)=(-\ii)\int_{0}^{1}\mathrm{d}\alpha_{m} \,  \alpha_{m}^{q_{m}-1} \, \tilde{\mathbf{T}}_{\mathcal{N}_{c}}(\alpha_{m}\omega_{m})\!+\!\omega_{m}^{-q_{m}}C[\omega_{1},\ldots,\hat{\omega}_{m},\ldots,\omega_{n}].
    \end{aligned}
\end{equation}

Then we determine $C$ using boundary conditions. In general, the boundary conditions can be taken as the soft limit $\omega_{m}\to 0$. However, when we properly choose the integration path, that is, choosing which $\omega_{m}$ to be integrated first, this condition will simplify a lot. Supposing $\omega_{m}$ is not a ``source'', that is, there is at least one adjacent vertex to $m$ which is earlier than $\tau_{m}$ and let us suppose it to be $\tau_{j}$, then the soft limit for the left-hand side of \eqref{eq:Tsols} will be an integral like
\begin{equation}\label{eq:leftlimit}
\begin{aligned}
    \lim_{\omega_{m}\to 0}\mathbf{T}^{q_1,q_2,\ldots,q_n}_{\mathcal{N}}(\omega_1,\omega_2,\ldots,\omega_n)=(-\ii) \, \hat{\mathbf{T}}_{\hat{\mathcal{N}}_{e}}\int_{\tau_{j}}^{0}\rmd\tau_{m}(-\tau_{m})^{q_{m}-1}\hat{\mathbf{T}}_{\hat{\mathcal{N}}_{l}},
\end{aligned}
\end{equation}
where $\hat{\mathbf{T}}_{\hat{\mathcal{N}}_{e}}$ means the integration of earlier-than-$m$ part and $\hat{\mathbf{T}}_{\hat{\mathcal{N}}_{l}}$ means the integration of later-than-$m$ part. The not-a-source condition guarantees that the integration of $\tau_{m}$ will be bounded below. Thus no divergence will appear when integrating $\tau_{m}$ and neither for remaining integrations since they all have been regulated by factors like $e^{i\omega_{i}\tau_{i}}$ which is zero when $\tau_{i}\to -\infty$. The expression is finite. Meanwhile, the first term in the right-hand side of \eqref{eq:Tsols} under the same limit will become
\begin{equation}\label{eq:rightlimit}
    \lim_{\omega_{m}\to 0}(-\ii)\int_{0}^{1}\mathrm{d}\alpha_{m} \, \alpha_{m}^{q_{m}-1}\tilde{\mathbf{T}}_{\mathcal{N}_{c}}(\alpha_{m}\omega_{m})=\frac{(-\ii)}{q_{m}}\tilde{\mathbf{T}}_{\mathcal{N}_{c}}(0),
\end{equation}
which is also finite. We note that in \eqref{eq:rightlimit} $\alpha_{m}\omega_{m}$ is always combined with some non-zero $\omega_{j}$ because vertex $m$ will always be contracted with some adjacent one $j$. So $\omega_{m}$ can be taken to 0 smoothly without singularities. Therefore, $C[\omega_{1},\ldots,\hat{\omega}_{m},\ldots,\omega_{n}]=0$ since otherwise the RHS will diverge when $q_{m}>0$. Actually, we can directly calculate the soft limit $\omega_{m}\to 0$ of LHS of \eqref{eq:Tsols} and prove that it indeed matches \eqref{eq:rightlimit} by induction which is presented in Appendix~\ref{app:boundary}. We finally have the following recursion relations for basic time integrals:
\begin{tcolorbox}
\begin{equation}\label{eq:Trecursion}
    \begin{aligned}        \mathbf{T}^{q_1,q_2,\ldots,q_n}_{\mathcal{N}}(\omega_1,\omega_2,\ldots,\omega_n)=(-\ii)\int_{0}^{1}\mathrm{d}\alpha_{m} \, \alpha_{m}^{q_{m}-1} \, \tilde{\mathbf{T}}_{\mathcal{N}_{c}}(\alpha_{m}\omega_{m}), \quad \text{for $m$ not a ``source''}.
    \end{aligned}
\end{equation}
\end{tcolorbox}
When there is only one vertex adjacent to $m$, that is, $m$ is a ``leaf'', this can be represented diagrammatically as
\begin{equation}\label{eq:leafrecursion}
\raisebox{-11pt}{\begin{tikzpicture}
		\node[below] at (0.55,0) {\,\, \small{$\omega_{j}$}};
		\node[below] at (1.5,0) {\small{$\omega_{m}$}};
		\draw[fill=black!10] (0,0) circle [radius=0.5];
		\path[fill=black] (0.5,0) circle[radius=0.1];
		\draw[line width=1.5pt,->] (0.5,0)--(1.1,0);
		\draw[line width=1.5pt] (0.9,0)--(1.5,0);
		\path[fill=black] (1.5,0) circle[radius=0.1];
	\end{tikzpicture}}=(-\ii)\int_{0}^{1} \rmd \alpha_{m} \;\alpha_{m}^{q_m-1} \, \raisebox{-11pt}{\begin{tikzpicture}
		\node[right] at (0.5,0) {\small{$\omega_{j}\!\!+\!\alpha_{m}\omega_{m}$}};
		\draw[fill=black!10] (0,0) circle [radius=0.5];
		\path[fill=black] (0.5,0) circle[radius=0.1];
	\end{tikzpicture}}
\end{equation}
When there are several vertices adjacent to $m$, then the right-hand side will be a summation of several terms. For example,
\begin{equation}\label{eq:exampletwoadj}
\begin{aligned}
   \raisebox{-11pt}{\begin{tikzpicture}[scale=0.75]
		\coordinate (X1) at (0,0);
		\coordinate (X2) at (1.5,0);
        \coordinate (X3) at (3,0);
		\node[below] at (X1) {\small{$2$}};
		\node[below] at (X2) {\small{$1$}};
        \node[below] at (X3) {\small{$3$}};
		\draw[line width=1.5pt,->] (X1)--(0.85,0);
        \draw[line width=1.5pt] (0.7,0)--(X2);
        \draw[line width=1.5pt,->] (X3)--(2.2,0);
        \draw[line width=1.5pt] (2.5,0)--(X2);
		\path[fill=black] (X1) circle[radius=0.11];
		\path[fill=black] (X2) circle[radius=0.11];
        \path[fill=black] (X3) circle[radius=0.11];
\end{tikzpicture}}&=(-\ii)\int_{0}^{1}\rmd \alpha_{1} \, \alpha_{1}^{q_1-1}\left(\raisebox{-11pt}{\begin{tikzpicture}[scale=0.75]
		\coordinate (X1) at (0,0);
		\coordinate (X2) at (1.5,0);
        \coordinate (X3) at (3,0);
		\node[below] at (X1) {\small{$2$}};
		\node[below] at (X2) {\small{$3+1\alpha_{1}$}};
		\draw[line width=1.5pt,->] (X1)--(0.85,0);
        \draw[line width=1.5pt] (0.7,0)--(X2);
		\path[fill=black] (X1) circle[radius=0.11];
		\path[fill=black] (X2) circle[radius=0.11];
\end{tikzpicture}}+\raisebox{-11pt}{\begin{tikzpicture}[scale=0.75]
		\coordinate (X1) at (0,0);
		\coordinate (X2) at (1.5,0);
        \coordinate (X3) at (3,0);
		\node[below] at (X2) {\small{$2+1\alpha_{1}$}};
        \node[below] at (X3) {\small{$3$}};
        \draw[line width=1.5pt,->] (X3)--(2.2,0);
        \draw[line width=1.5pt] (2.5,0)--(X2);
		\path[fill=black] (X2) circle[radius=0.11];
        \path[fill=black] (X3) circle[radius=0.11];
\end{tikzpicture}}\right)\\
&=\raisebox{-11pt}{\begin{tikzpicture}[scale=0.75]
		\coordinate (X1) at (0,0);
		\coordinate (X2) at (1.5,0);
        \coordinate (X3) at (3,0);
		\node[below] at (X1) {\small{$1$}};
		\node[below] at (X2) {\small{$3$}};
        \node[below] at (X3) {\small{$2$}};
		\draw[line width=1.5pt,->] (X2)--(0.7,0);
        \draw[line width=1.5pt] (0.75,0)--(X1);
        \draw[line width=1.5pt,->] (X3)--(2.2,0);
        \draw[line width=1.5pt] (2.5,0)--(X2);
		\path[fill=black] (X1) circle[radius=0.11];
		\path[fill=black] (X2) circle[radius=0.11];
        \path[fill=black] (X3) circle[radius=0.11];
\end{tikzpicture}}+\raisebox{-11pt}{\begin{tikzpicture}[scale=0.75]
		\coordinate (X1) at (0,0);
		\coordinate (X2) at (1.5,0);
        \coordinate (X3) at (3,0);
		\node[below] at (X1) {\small{$1$}};
		\node[below] at (X2) {\small{$2$}};
        \node[below] at (X3) {\small{$3$}};
		\draw[line width=1.5pt,->] (X2)--(0.7,0);
        \draw[line width=1.5pt] (0.75,0)--(X1);
        \draw[line width=1.5pt,->] (X3)--(2.2,0);
        \draw[line width=1.5pt] (2.5,0)--(X2);
		\path[fill=black] (X1) circle[radius=0.11];
		\path[fill=black] (X2) circle[radius=0.11];
        \path[fill=black] (X3) circle[radius=0.11];
\end{tikzpicture}}
\end{aligned}.
\end{equation}
Here $\omega_{i}$ has been abbreviated to $i$ and the corresponding rule for twist exponents should be the same as before. The second line shows that this recursive solution actually incorporates the time order decomposition $\theta_{1,2}\theta_{1,3}=\theta_{1,2}\theta_{2,3}+\theta_{1,3}\theta_{3,2}$.

At last, let us briefly comment on what will happen if we choose to integrate a ``source'' vertex first. In this case, if we directly perform the soft limit of LHS, divergence will emerge\footnote{This can be seen easily by removing $\hat{\mathbf{T}}_{\hat{\mathcal{N}}_{e}}$ and replacing $\tau_{j}$ by $-\infty$ in \eqref{eq:leftlimit}.}. So we need to isolate this divergence by time decomposition (e.g. $\theta_{i,j}=1-\theta_{j,i}$) first and the remaining terms will be finite. Then \eqref{eq:Trecursion} applies.

Finally, we can perform the same analysis for $\tilde{\mathbf{T}}_{\mathcal{N}_{c}}$ and this iteration will end at the diagram with all the vertices contracted:
\begin{equation}\label{eq:root}
    \mathbf{T}^{q_{12\ldots n}}_{\bullet}(\omega_{\bullet})=-\ii\,(\ii \, \omega_{\bullet})^{-q_{1,2,\ldots,n}}\Gamma(q_{12\ldots n}),
\end{equation}
where $\omega_{\bullet}$ means the total energy on this contracted vertex, it will be a polynomial of $\alpha$ with positive coefficients $\omega$. If there are no internal ``sinks'' in a tree-level basic time integral, which is called a ``family tree" in \cite{Fan:2024iek}, then we can always choose to integrate a leaf out first in every step, and the solution will be a single term, in which case $\omega_{\bullet}$ is just the polynomial $P(\alpha,\omega)$ in the representation \eqref{eq:euler}. However, in the general case, the solution will be a summation of such integrals just like what \eqref{eq:exampletwoadj} indicates, corresponding to the so-called ``family tree decomposition" in \cite{Fan:2024iek}. Now let us first show how $P(\alpha,\omega)$ can be written for family trees in several examples. The first example is a three-site chain:
\begin{equation}\label{eq:threesitechain}
\begin{aligned}
   \raisebox{-11pt}{\begin{tikzpicture}[scale=0.75]
		\coordinate (X1) at (0,0);
		\coordinate (X2) at (1.5,0);
        \coordinate (X3) at (3,0);
		\node[below] at (X1) {\small{$1$}};
		\node[below] at (X2) {\small{$2$}};
        \node[below] at (X3) {\small{$3$}};
		\draw[line width=1.5pt,->] (X1)--(0.9,0);
        \draw[line width=1.5pt] (0.75,0)--(X2);
        \draw[line width=1.5pt,->] (X2)--(2.4,0);
        \draw[line width=1.5pt] (2,0)--(X3);
		\path[fill=black] (X1) circle[radius=0.11];
		\path[fill=black] (X2) circle[radius=0.11];
        \path[fill=black] (X3) circle[radius=0.11];
\end{tikzpicture}}&=(-\ii)\int_{0}^{1}\rmd \alpha_{3} \, \alpha_{3}^{q_3-1}\raisebox{-11pt}{\begin{tikzpicture}[scale=0.75]
		\coordinate (X1) at (0,0);
		\coordinate (X2) at (1.5,0);
		\node[below] at (X1) {\small{$1$}};
        \node[below] at (X2) {\small{$2+3\alpha_{3}$}};
        \draw[line width=1.5pt,->] (X1)--(0.9,0);
        \draw[line width=1.5pt] (0.75,0)--(X2);
		\path[fill=black] (X1) circle[radius=0.11];
        \path[fill=black] (X2) circle[radius=0.11];
\end{tikzpicture}}\\
&=(-\ii)^2\int_{0}^{1}\rmd \alpha_{3} \, \alpha_{3}^{q_3-1}\int_{0}^{1}\rmd \alpha_{2} \, \alpha_{2}^{q_{23}-1}\raisebox{-11pt}{\begin{tikzpicture}[scale=0.75]
		\coordinate (X1) at (0,0);
		\path[fill=black] (X1) circle[radius=0.11];
        \node[below] at (X1) {\small{$1+(2+3\alpha_{3})\alpha_{2}$}};
\end{tikzpicture}}\\
&=(-\ii)^{3+q_{123}}\!\!\!\int_{0}^{1}\rmd \alpha_{3} \, \alpha_{3}^{q_3-1}\!\!\!\int_{0}^{1}\rmd \alpha_{2} \, \alpha_{2}^{q_{23}-1}\!\left[\omega_{1}\!+\!(\omega_{2}\!+\!\omega_{3}\alpha_{3})\alpha_{2}\right]^{-q_{123}}\Gamma(q_{123})
\end{aligned}
\end{equation}
And the second example is a four-site star:
\begin{equation}\label{eq:foursitestar}
\begin{aligned}
   \raisebox{-25pt}{\begin{tikzpicture}[scale=0.6]
		\coordinate (X4) at (0,0);
		\coordinate (X1) at (0,1.5);
        \coordinate (X2) at (1.3,-0.75);
        \coordinate (X3) at (-1.3,-0.75);
		\node[above] at (X1) {\small{$2$}};
		\node[below] at (X2) {\small{$3$}};
        \node[below] at (X3) {\small{$4$}};
        \node[below] at (X4) {\small{$1$}};
		\draw[line width=1.5pt,->] (X4)--(0,1);
        \draw[line width=1.5pt] (0,0.5)--(X1);
        \draw[line width=1.5pt,->] (X4)--(0.87,-0.5);
        \draw[line width=1.5pt] (0.65,-0.375)--(X2);
        \draw[line width=1.5pt,->] (X4)--(-0.87,-0.5);
        \draw[line width=1.5pt] (-0.65,-0.375)--(X3);
		\path[fill=black] (X1) circle[radius=0.11];
		\path[fill=black] (X2) circle[radius=0.11];
        \path[fill=black] (X3) circle[radius=0.11];
        \path[fill=black] (X4) circle[radius=0.11];
\end{tikzpicture}}&=(-\ii)\int_{0}^{1}\rmd \alpha_{2} \, \alpha_{2}^{q_2-1}\raisebox{-11pt}{\begin{tikzpicture}[scale=0.75]
		\coordinate (X1) at (0,0);
		\coordinate (X2) at (1.5,0);
        \coordinate (X3) at (3,0);
		\node[below] at (X1) {\small{$4$}};
        \node[above] at (X2) {\small{$1\!+\!2\alpha_{2}$}};
        \node[below] at (X3) {\small{$3$}};
        \draw[line width=1.5pt,->] (X2)--(0.4,0);
        \draw[line width=1.5pt] (1,0)--(X1);
        \draw[line width=1.5pt,->] (X2)--(2.6,0);
        \draw[line width=1.5pt] (2,0)--(X3);
		\path[fill=black] (X1) circle[radius=0.11];
        \path[fill=black] (X2) circle[radius=0.11];
        \path[fill=black] (X3) circle[radius=0.11];
\end{tikzpicture}}\\
&=(-\ii)^2\int_{0}^{1}\rmd \alpha_{2} \, \alpha_{2}^{q_2-1}\int_{0}^{1}\rmd \alpha_{3} \, \alpha_{3}^{q_{3}-1}\raisebox{-11pt}{\begin{tikzpicture}[scale=0.75]
		\coordinate (X1) at (0,0);
		\coordinate (X2) at (2,0);
		\node[below] at (X1) {\small{$4$}};
        \node[below] at (X2) {\small{$1\!+\!2\alpha_{2}\!+\!3\alpha_{3}$}};
        \draw[line width=1.5pt,->] (X2)--(0.9,0);
        \draw[line width=1.5pt] (1.1,0)--(X1);
		\path[fill=black] (X1) circle[radius=0.11];
        \path[fill=black] (X2) circle[radius=0.11];
\end{tikzpicture}}\\[1em]
&=(-\ii)^{4+q_{1234}}\int_{0}^{1}\rmd \alpha_{2} \, \alpha_{2}^{q_2-1}\int_{0}^{1}\rmd \alpha_{3} \, \alpha_{3}^{q_{3}-1}\int_{0}^{1}\rmd \alpha_{4} \, \alpha_{4}^{q_{4}-1} \\
& \hspace{10em}\times \left[\omega_{1}+\omega_{2}\alpha_{2}+\omega_{3}\alpha_{3}+\omega_{4}\alpha_{4}\right]^{-q_{1234}}\Gamma(q_{1234})
\end{aligned}
\end{equation}
We summarize the general rules for the final solution of an family tree as 
\begin{enumerate}
    \item There is an overall constant factor $(-\ii)^{n+q_{12\cdots n}}\Gamma(q_{12\cdots n})$.
    \item For every vertex $i$ except for the root (the root has been integrated out using \eqref{eq:root}, we name it vertex $1$), there is an integration:
    \begin{equation}
        \int_{0}^{1}\frac{\rmd\alpha_{i}}{\alpha_{i}}\alpha_{i}^{\nu_{i}(q)}
    \end{equation}
    where $\nu_{i}(q)$ sums $q_{i}$ and all the $q_{j}$ if there is a path connecting vertex $i$ and $j$ with vertex $i$ as the source. For example, in \eqref{eq:threesitechain}, $\nu_{3}(q)=q_{3}$ and $\nu_{2}(q)=q_{23}$.
    \item The graph polynomial $P(\omega,\alpha)$ can be determined as
    \begin{equation}\label{eq:Pdef}
P(\omega,\alpha)=\sum_{i=1}^{n}\omega_{i} \, \alpha^{\mathbf{v}_{i}}, \quad \alpha^{\mathbf{v}_{i}}=\alpha_{1}^{\mathbf{v}_{i,1}}\cdots\alpha_{n}^{\mathbf{v}_{i,n}}
    \end{equation}
    where $\mathbf{v}_{i}$ is an $n$-dimension vector that indicates the exponents of $\alpha$. For example, in \eqref{eq:threesitechain}, $\mathbf{v}_{3}=(0,1,1)$ and $\mathbf{v}_{2}=(0,1,0)$. Since root has been integrated out, corresponding $\mathbf{v}_{i,1}$ is always 0 in the above examples, and $\mathbf{v}_{i,j}$ will be set to 1 if vertex $j$ is on the path from root to vertex $i$. 
    \item The final result will be
        \begin{equation}
        \begin{aligned}
            \mathbf{T}^{q_1,q_2,\ldots,q_n}_{\mathcal{N}}(\omega_1,\omega_2,\ldots,\omega_n)=(-\ii)^{n+q_{12\cdots n}}\Gamma(q_{12\cdots n})\int_{0}^{1}\prod_{i=2}^{n}\frac{\rmd\alpha_{i}}{\alpha_{i}}\alpha_{i}^{\nu(q_{i})}P(\omega,\alpha)^{-q_{12\cdots n}} 
        \end{aligned}
        \end{equation}
        where $\mathcal{N}$ is an family tree.
\end{enumerate}
Using this formula we can directly write down the expression for a five-site tree:
\begin{equation}\label{eq:5tisadp}
\begin{aligned}
    \raisebox{-2.25em}{\begin{tikzpicture}[scale=0.25]
				\coordinate (X1) at (3,4);
				\coordinate (X2) at (0,0);
				\coordinate (X3) at (4.5,2);
				\coordinate (X4) at (3,0);
				\coordinate (X5) at (6,0);
				\node[above] at (X1) {\small{$1$}};
				\node[below] at (X2) {\small{$2$}};
				\node[above right] at (X3) {\small{$3$}};
				\node[below] at (X4) {\small{$4$}};
				\node[below] at (X5) {\small{$5$}};
				\draw[line width=1.5pt,->] (X1)--(1.2,1.6);
				\draw[line width=1.5pt] (1.5,2)--(X2);
				\draw[line width=1.5pt,->] (X1)--(4.05,2.6);
				\draw[line width=1.5pt] (3.75,3)--(X3);
				\draw[line width=1.5pt,->] (X3)--(3.6,0.8);
				\draw[line width=1.5pt] (3.75,1)--(X4);
				\draw[line width=1.5pt,->] (X3)--(5.4,0.8);
				\draw[line width=1.5pt] (5.25,1)--(X5);
				\path[fill=black] (X1) circle[radius=0.32];
				\path[fill=black] (X2) circle[radius=0.32];
				\path[fill=black] (X3) circle[radius=0.32];
				\path[fill=black] (X4) circle[radius=0.32];
				\path[fill=black] (X5) circle[radius=0.32];
		\end{tikzpicture}}
    \!\!\!\!
    =&(-\ii)^{5+q_{12345}}\!\int_{0}^{1}\rmd \alpha_5\,\alpha_5^{q_5-1}\!\int_{0}^{1}\rmd \alpha_4\,\alpha_4^{q_4-1}\!\int_{0}^{1}\rmd \alpha_2\,\alpha_2^{q_2-1}\!\int_{0}^{1}\rmd \alpha_3\,\alpha_3^{q_{345}-1}\\
    & \hspace{9em} \times \left[\omega_{1}+\omega_{2}\alpha_{2}+\left(\omega_{3}+\omega_{4}\alpha_{4}+\omega_{5}\alpha_{5}\right)\alpha_{3}\right]^{-q_{12345}}\Gamma(q_{12345})
\end{aligned}
\end{equation}

At last, for cases other than family trees, we can always decompose them into the summation of family trees and products of lower-point integrals first following family tree decomposition in \cite{Fan:2024iek} by the property of theta function $\theta_{i,j}=1-\theta_{j,i}$. Then $P(\omega,\alpha)$ can be written for each term:
\begin{equation}\label{eq:Trecursivesol}
\begin{aligned}
    \mathbf{T}^{q_1,q_2,\ldots,q_n}_{\mathcal{N}}(\omega_1,\omega_2,\ldots,\omega_n)=&\sum_{P(\omega,\alpha)}(-\ii)^{n+q_{12\cdots n}}\Gamma(q_{12\cdots n})\int_{0}^{1}\prod_{i=2}^{n}\frac{\rmd\alpha_{i}}{\alpha_{i}}\alpha_{i}^{\nu(q_{i})}P(\omega,\alpha)^{-q_{12\cdots n}} \\
    &+\text{products of lower-point integrals}.
\end{aligned}
\end{equation}
Note that for tree graphs, once the root has been chosen, we can always decompose it into one term in the first line so that the summation doesn't exist. However, for loop structure, this summation usually exists. For example, the one-loop box graph can be written as a summation of two chains
\begin{equation}\label{eq:boxadp}
\begin{aligned}
    \raisebox{-2em}{\begin{tikzpicture}
        \coordinate (X1) at (0,1);
		\coordinate (X2) at (0,0);
        \coordinate (X3) at (1,1);
        \coordinate (X4) at (1,0);
		\node[above left] at (X1) {\small{$1$}};
        \node[below left] at (X2) {\small{$2$}};
        \node[above right] at (X3) {\small{$3$}};
        \node[below right] at (X4) {\small{$4$}};
		\draw[line width=1.5pt,->] (X1)--(0,0.4);
        \draw[line width=1.5pt] (0,0.5)--(X2);
        \draw[line width=1.5pt,->] (X3)--(1,0.4);
        \draw[line width=1.5pt] (1,0.5)--(X4);
        \draw[line width=1.5pt,->] (X1)--(0.6,1);
        \draw[line width=1.5pt] (0.5,1)--(X3);
        \draw[line width=1.5pt,->] (X2)--(0.6,0);
        \draw[line width=1.5pt] (0.5,0)--(X4);
		\path[fill=black] (X1) circle[radius=0.1];
		\path[fill=black] (X2) circle[radius=0.1];
        \path[fill=black] (X3) circle[radius=0.1];
		\path[fill=black] (X4) circle[radius=0.1];
    \end{tikzpicture}}
&=-\ii\int_{0}^{1}\rmd \alpha_{4} \, \alpha_{4}^{q_4-1}\left(\raisebox{-2em}{\begin{tikzpicture}
        \coordinate (X1) at (0,1);
		\coordinate (X2) at (0,0);
        \coordinate (X3) at (1,1);
        \coordinate (X4) at (1,0);
		\node[above left] at (X1) {\small{$1$}};
        \node[below left] at (X2) {\small{$2$}};
        \node[below right] at (X4) {\small{$3+4\alpha_{4}$}};
		\draw[line width=1.5pt,->] (X1)--(0,0.4);
        \draw[line width=1.5pt] (0,0.5)--(X2);
        \draw[line width=1.5pt,->] (X2)--(0.6,0);
        \draw[line width=1.5pt] (0.5,0)--(X4);
        \draw[line width=1.5pt,->,black!30] (X1)--(0.5+0.1,0.5-0.1);
        \draw[line width=1.5pt,black!30] (0.5,0.5)--(X4);
		\path[fill=black] (X1) circle[radius=0.1];
		\path[fill=black] (X2) circle[radius=0.1];
		\path[fill=black] (X4) circle[radius=0.1];
    \end{tikzpicture}}+\raisebox{-2em}{\begin{tikzpicture}
        \coordinate (X1) at (0,1);
		\coordinate (X2) at (0,0);
        \coordinate (X3) at (1,1);
        \coordinate (X4) at (1,0);
		\node[above left] at (X1) {\small{$1$}};
        \node[above right] at (X3) {\small{$3$}};
        \node[below right] at (X4) {\small{$2+4\alpha_{4}$}};
        \draw[line width=1.5pt,->] (X3)--(1,0.4);
        \draw[line width=1.5pt] (1,0.5)--(X4);
        \draw[line width=1.5pt,->] (X1)--(0.6,1);
        \draw[line width=1.5pt] (0.5,1)--(X3);
        \draw[line width=1.5pt,->,black!30] (X1)--(0.5+0.1,0.5-0.1);
        \draw[line width=1.5pt,black!30] (0.5,0.5)--(X4);
		\path[fill=black] (X1) circle[radius=0.1];
        \path[fill=black] (X3) circle[radius=0.1];
		\path[fill=black] (X4) circle[radius=0.1];
    \end{tikzpicture}}\right)\\
    &=\raisebox{-1em}{\begin{tikzpicture}[scale=0.8]
			\coordinate (X1) at (0,0);
			\coordinate (X2) at (1,0);
			\coordinate (X3) at (2,0);
			\coordinate (X4) at (3,0);
			\node[below] at (X1) {\small{$1$}};
			\node[below] at (X2) {\small{$2$}};
			\node[below] at (X3) {\small{$3$}};
			\node[below] at (X4) {\small{$4$}};
			\draw[line width=1.5pt,->] (X1)--(0.6,0);
			\draw[line width=1.5pt] (0.5,0)--(X2);
			\draw[line width=1.5pt,->] (X2)--(1.6,0);
			\draw[line width=1.5pt] (1.5,0)--(X3);
			\draw[line width=1.5pt,->] (X3)--(2.6,0);
			\draw[line width=1.5pt] (2.5,0)--(X4);
			\path[fill=black] (X1) circle[radius=0.1];
			\path[fill=black] (X2) circle[radius=0.1];
			\path[fill=black] (X3) circle[radius=0.1];
			\path[fill=black] (X4) circle[radius=0.1];
	\end{tikzpicture}}+\raisebox{-1em}{\begin{tikzpicture}[scale=0.8]
			\coordinate (X1) at (0,0);
			\coordinate (X2) at (1,0);
			\coordinate (X3) at (2,0);
			\coordinate (X4) at (3,0);
			\node[below] at (X1) {\small{$1$}};
			\node[below] at (X2) {\small{$3$}};
			\node[below] at (X3) {\small{$2$}};
			\node[below] at (X4) {\small{$4$}};
			\draw[line width=1.5pt,->] (X1)--(0.6,0);
			\draw[line width=1.5pt] (0.5,0)--(X2);
			\draw[line width=1.5pt,->] (X2)--(1.6,0);
			\draw[line width=1.5pt] (1.5,0)--(X3);
			\draw[line width=1.5pt,->] (X3)--(2.6,0);
			\draw[line width=1.5pt] (2.5,0)--(X4);
			\path[fill=black] (X1) circle[radius=0.1];
			\path[fill=black] (X2) circle[radius=0.1];
			\path[fill=black] (X3) circle[radius=0.1];
			\path[fill=black] (X4) circle[radius=0.1];
	\end{tikzpicture}}\\[3pt]
\end{aligned}
\end{equation}
In general, the differentiation of loop integrands works as an operation that opens the loop, so the loop integrands can be viewed as the growth of trees.
Compare the second line of \eqref{eq:boxadp} with \eqref{eq:exampletwoadj}, the box diagram can also be seen as nontrivial growth of the tree 
$\raisebox{-1em}{\begin{tikzpicture}[scale=0.4]
		\coordinate (X1) at (0,0);
		\coordinate (X2) at (1.5,0);
        \coordinate (X3) at (3,0);
		\node[below] at (X1) {\small{$2$}};
		\node[below] at (X2) {\small{$4$}};
        \node[below] at (X3) {\small{$3$}};
		\draw[line width=1.5pt,->] (X1)--(0.85,0);
        \draw[line width=1.5pt] (0.7,0)--(X2);
        \draw[line width=1.5pt,->] (X3)--(2.2,0);
        \draw[line width=1.5pt] (2.5,0)--(X2);
		\path[fill=black] (X1) circle[radius=0.2];
		\path[fill=black] (X2) circle[radius=0.2];
        \path[fill=black] (X3) circle[radius=0.2];
	\end{tikzpicture}}$. 
In summary, we stress that this recursion \eqref{eq:Trecursion} can be applied to arbitrary types of directed graphs beyond trees. It is also interesting to note that this recursive solution can be viewed as a dual representation of time integral representation which trades the theta functions for a structured polynomial $P(\omega,\alpha)$.

\subsection{Analytic properties of the recursive solution}
Now we study some analytic properties of the recursive solution derived in the last section. First, we will check that it indeed reproduces the series expansion in \cite{Fan:2024iek}. Then we give some comments on the tools to study them.


\paragraph{The series expansion and the method of region.} Here we will show that how we can perform the series expansion in the above representation and reproduce results in \cite{Fan:2024iek}. The first example is a two-site chain:
\begin{equation}\label{eq:twositechain}
    \mathbf{T}_{+}^{q_{1},q_{2}}(\omega_{1},\omega_{2})=\raisebox{-1em}{\begin{tikzpicture}[scale=0.4]
		\coordinate (X1) at (0,0);
		\coordinate (X2) at (3,0);
		\node[below] at (X1) {\small{$1$}};
		\node[below] at (X2) {\small{$2$}};
		\draw[line width=1.5pt,->] (X1)--(2,0);
        \draw[line width=1.5pt] (1,0)--(X2);
		\path[fill=black] (X1) circle[radius=0.2];
		\path[fill=black] (X2) circle[radius=0.2];
	\end{tikzpicture}}=(-\ii)^{2+q_{12}}\Gamma(q_{12})\int_{0}^{1}\rmd\alpha_{2}\alpha_{2}^{q_{2}-1}(\omega_{1}+\omega_{2}\alpha_{2})^{-q_{12}},
\end{equation}
where the subscript $+$ indicates the time flow on this chain is unidirectional. Supposing we are considering the region $\omega_{2}<\omega_{1}$, then we can expand the integrand in series as 
\begin{equation}
\begin{aligned}
    \mathbf{T}_{+}^{q_{1},q_{2}}(\omega_{1},\omega_{2})&=\frac{(-\ii)^{2}}{(\ii \, \omega_{1})^{q_{12}}}\int_{0}^{1}\rmd\alpha_{2}\alpha_{2}^{q_{2}-1}\sum_{n=0}^{\infty}\frac{\Gamma(q_{12})\Gamma(1-q_{12})}{\Gamma(1-q_{12}-n)n!}\left(\frac{\omega_{2}}{\omega_{1}}\right)^{n}\alpha_{2}^{n} \\
    &\hspace{-3em} =-\frac{1}{(\ii \, \omega_{1})^{q_{12}}}\sum_{n=0}^{\infty}\frac{\Gamma(q_{12})\Gamma(1-q_{12})}{(n+q_{2})\Gamma(1-q_{12}-n)n!}\left(\frac{\omega_{2}}{\omega_{1}}\right)^{n} \\
    &\hspace{-3em} =-\frac{1}{(\ii \, \omega_{1})^{q_{12}}}\sum_{n=0}^{\infty}\frac{1}{n!}\frac{\Gamma(q_{12}+n)}{(n+q_{2})}\left(-\frac{\omega_{2}}{\omega_{1}}\right)^{n}\equiv-\frac{1}{(\ii\,\omega_{1})^{q_{12}}}{}_2\mathcal{F}_1\left[
  \begin{matrix}
    q_2,q_{12} \\
    q_2+1
  \end{matrix}
  \bigg| -\frac{\omega_2}{\omega_1}\right].
\end{aligned}
\end{equation}
The last line is derived by using $\displaystyle \frac{\Gamma(q_{12})\Gamma(1-q_{12})}{\Gamma(1-q_{12}-n)}=\Gamma(q_{12}+n)(-1)^{n}$ and the definition of dressed version of hypergeometric functions:
\begin{equation}
    {}_2\mathcal{F}_1\left[
  \begin{matrix}
    a,b \\
    c
  \end{matrix}
  \bigg| z\right]\equiv \sum_{n=0}^{\infty}\frac{\Gamma(a+n)\Gamma(b+n)}{\Gamma(c+n)}\frac{z^{n}}{n!} \, .
\end{equation}
For another region $\omega_{1}\sim \omega_{2}$, we can expand by extracting an overall factor $(\ii \, \omega_{12})^{-q_{12}}$:
\begin{equation}
\begin{aligned}
    \mathbf{T}_{+}^{q_{1},q_{2}}(\omega_{1},\omega_{2})&=\frac{(-\ii)^{2}}{(\ii \, \omega_{12})^{q_{12}}}\int_{0}^{1}\rmd\alpha_{2} \, \alpha_{2}^{q_{2}-1}\sum_{n=0}^{\infty}\frac{\Gamma(q_{12})\Gamma(1-q_{12})}{\Gamma(1-q_{12}-n)n!}\left(-\frac{\omega_{2}}{\omega_{12}}\right)^{n}(1-\alpha_{2})^{n} \\
    &\hspace{-3em}=-\frac{1}{(\ii \, \omega_{12})^{q_{12}}}\sum_{n=0}^{\infty}\frac{\Gamma(q_{12})\Gamma(1-q_{12})\Gamma(q_{2})}{\Gamma(1-q_{12}-n)\Gamma(1+q_{2}+n)}\left(-\frac{\omega_{2}}{\omega_{12}}\right)^{n} \\
    &\hspace{-3em}=\!-\frac{1}{(\ii \, \omega_{12})^{q_{12}}}\sum_{n=0}^{\infty}\frac{\Gamma(q_{12}+n)\Gamma(q_{2})}{\Gamma(1+q_2+n)}\left(\frac{\omega_{2}}{\omega_{12}}\right)^{n}\!\!\equiv\!-\frac{\Gamma(q_{2})}{(\ii \, \omega_{12})^{q_{12}}}{}_2\mathcal{F}_1\left[
  \begin{matrix}
    1,q_{12} \\
    q_2\!+\!1
  \end{matrix}
  \bigg| \frac{\omega_2}{\omega_{12}}\right].
\end{aligned}
\end{equation}
This is the method of region for such integrals. In general, we could choose the root $\omega_{1}$ to be maximal among $\omega_{i}$'s and normalize the integral with $(\ii \, \omega_{1})^{q_{12\cdots n}}$ or choose to normalize with the total energy $(\ii \, \omega_{12\cdots n})^{q_{12\cdots n}}$. In both cases, the general series expansion formula for the family trees (Eq.~(32) and (33) in \cite{Fan:2024iek}) can be derived from the recursive solution. We present the complete proof in Appendix~\ref{app:series}.

\paragraph{Evolution of the recursive solution.} Since the recursive solution is derived from differential equations, it is natural to study the flow from one kinematic point (the boundary) to its neighborhood. Supposing expression (or value) of the basic time integral on one kinematic point $\omega:=(\omega_{1},\cdots,\omega_{n})$ has already been known, then we deform $\omega_{k}$ to $\omega_{k}^{\prime}\sim \omega_{k}$, that is, $\omega_{k}^{\prime}-\omega_{k}$ is small compared to $\omega_{k}$. Instead of calculating on this new kinematic point again, we want to evolve from the known expression through the differential equation. Let us start with the recursion relation:
\begin{equation}\label{eq:flow}
    \begin{aligned}
        \mathbf{T}(\omega_{k}^{\prime})&=\frac{(-\ii)}{{\omega_{k}^{\prime}}^{q_{k}}}\int_{0}^{\omega_{k}^{\prime}}\rmd t_{k}t_{k}^{q_{k}-1}\tilde{\mathbf{T}}(t_{k}) \\ 
        &=-\ii\left(\frac{\omega_{k}}{\omega^{\prime}_{k}}\right)^{q_{k}}\frac{1}{\omega_{k}^{q_{k}}}\left(\int_{\omega_{k}}^{\omega^{\prime}_{k}}\rmd t_{k}t_{k}^{q_{k}-1}\tilde{\mathbf{T}}(t_{k})+\int^{\omega_{k}}_{0}\rmd t_{k}t_{k}^{q_{k}-1}\tilde{\mathbf{T}}(t_{k})\right) \\
        &=(1+\delta)^{-q_{k}}\left(-\ii\int_{1}^{1+\delta}\rmd\alpha_{k}\alpha_{k}^{q_{k}-1}\tilde{\mathbf{T}}(\omega_{k}\alpha_{k})+\mathbf{T}(\omega_{k})\right),
    \end{aligned}
\end{equation}
where we have defined $\delta=(\omega^{\prime}_{k}-\omega_{k})/\omega_{k}$. It tells us besides an overall rescaling factor $(1+\delta)^{-q_{k}}$, we need to study a shift that amounts to changing the original integration region for $\alpha_{k}$ from $[0,1]$ to $[1,1+\delta]$. This necessitates the study of this kind of integral around its boundary. Let us take the two-site chain as a simple example. We deform $\omega_{2}$ to $\omega_{2}^{\prime}=\omega_{2}(1+\delta)$, then by \eqref{eq:flow}, we have
\begin{equation}
    \mathbf{T}_{+}^{q_{1},q_{2}}(\omega_{1},\omega_{2}^{\prime})=(1+\delta)^{-q_{2}}\mathbf{T}_{+}^{q_{1},q_{2}}(\omega_{1},\omega_{2}) + (1+\delta)^{-q_{2}}\left.\mathbf{T}_{+}^{q_{1},q_{2}}(\omega_{1},\omega_{2})\right|_{[0,1]\to [1,1\!+\!\delta]},
\end{equation}
where the subscript $[0,1]\to [1,1\!+\!\delta]$ means replacing the integral domain of $\alpha_{2}$ in \eqref{eq:twositechain} from $[0,1]$ to $[1,\delta]$. Then we directly perform the series expansion in the integrand level
\begin{equation}
\begin{aligned}
    &\left.\mathbf{T}_{+}^{q_{1},q_{2}}(\omega_{1},\omega_{2})\right|_{[0,1]\to [1,1\!+\!\delta]}=(-\ii)^{2+q_{12}}\Gamma(q_{12})(\omega_{1}+\omega_{2})^{-q_{12}}\Delta,\\
    &\Delta=(\omega_{1}+\omega_{2})^{q_{12}}\int_{0}^{\delta}\rmd\alpha_{2}(1+\alpha_{2})^{q_{2}-1}(\omega_{1}+\omega_{2}+\omega_{2}\alpha_{2})^{-q_{12}}\\
    &=\delta+\frac{\delta^2}{2!}(q_{2}-rq_{12}-1)+\frac{\delta^{3}}{3!}\left[(q_{2}-1)(q_{2}-2)-2q_{12}(q_{2}-1)r+q_{12}(q_{12}+1)r^2\right]+\mathcal{O}(\delta^4),
\end{aligned}
\end{equation}
where $r=\omega_{2}/(\omega_{1}+\omega_{2})$. The advantage of evolution only shows when the boundary value is already known (but usually not that easy to calculate) and we want to study the behavior close to this boundary value. 

\paragraph{Advanced tools for recursive solution.} As we mentioned before, many tools developed for Euler integrals can be applied to our solution. Here we only mention several of them: the coaction structure like those presented in \cite{Abreu:2019wzk}, the divergence structure under the de-Sitter limit ($q_{i}\to 0$) \cite{Salvatori:2024nva} and the analytic continuation and adjacent relations (relations among integrals with $q_{i}$ shifted) of hypergeometric systems \cite{Feng:2024xio}.
We leave the more advanced research of this kind of integrals to future works.

\section{The system of differential equations for cosmological amplitudes}\label{sec:DE}

In this section, we derive the closed differential system for the cosmological amplitudes of any graph type. We will first derive the differential equation system for general basic time integrals to gain some intuition. Though we have solved them in the last section by recursion (which can be viewed as choosing a specific integration path), some analytic properties are only manifest in the differential equation system. Especially, when we take the de Sitter limit $q_{i}\to 0$, this system can be solved as the canonical differential equation (CDE) system which is well studied in the Feynman integral literature. Then we describe how to generally obtain the canonical differential system for cosmological amplitudes which are combinations of basic time integrals. We find consistency with results in \cite{Arkani-Hamed:2023kig}, and our method applies to general cosmological amplitudes including those with loops furthermore.

\subsection{Differential equations for any basic time integrals}

To derive differential equations for any basic time integrals, we further take the total differential of the contracted terms on the RHS of \eqref{eq:dT} and eventually arrive at a closed system of differential equations. Specifically, after recursively doing this $n-1$ times ($n$ is the number of vertices in the graph), we arrive at the same one-site integral as in \eqref{eq:root}
\begin{equation}
    \mathbf{T}_{\bullet}^{q_{12\ldots n}}(\omega_{12\ldots n})=(-\ii)\int_{-\infty}^0{\rm d}\tau (-\tau)^{q_{12\cdots n}-1}e^{\ii\,\omega_{12\cdots n}\tau}{=}\frac{(-\ii)}{(\ii\,\omega_{12\ldots n})^{q_{12\ldots n}}}\mathrm{\Gamma}(q_{12\ldots n}),
\end{equation}
satisfying
\begin{equation}
	\rmd\mathbf{T}_{\bullet}^{q_{12\ldots n}}(\omega_{12\ldots n})=-q_{12\ldots n}\,\rmd\!\log\omega_{12\ldots n}\,\mathbf{T}_{\bullet}^{q_{12\ldots n}}(\omega_{12\ldots n}),
\end{equation}
where no new integral appears, thus the system closes.  In general, a time integral may correspond to a disconnected graph which is a product of several connected parts. For simplicity, we only study time integrals corresponding to connected directed graphs in this subsection, since the differential of disconnected ones can be obtained by the Leibniz rule. 

Every time we take the total differential of a basic time integral, lower-site integrals will be introduced with some edge contracted until there is no edge left (e.g. the one-site integral). We collect these lower-site integrals in each step and act total differential on them in the next step. A lower-site integral is called at $j$-th level if it appears at the $j$-th step during the procedure. For a general graph with $e$ edges, the integral basis consists of $(e{-}j)$-site integrals at $j$-th level, $j=0,1,2,\ldots,e$, and the number of integrals in each level reads $\text{C}_e^j$, {\it i.e.} one has to choose $j$ from $e$ edges to contract. Thus, the closed system naively consists of $2^{e}$ basis integrals in total. This counting is true for tree graphs. However, for graphs with loops, contraction of edges may lead to an integral with contradictory time order, so the number of basis integrals is expected to be no more than $2^{e}$.

After the general discussion, let us consider some simple examples.  Firstly, for the two-site basic time integral
\begin{equation}
    \mathbf{T}_{\theta_{1,2}}^{q_1,q_2}= \twositechain.
\end{equation}
take differential of it following \eqref{eq:dT}, we have
\begin{equation}
\begin{aligned}
    \rmd \mathbf{T}_{\theta_{1,2}}^{q_1,q_2}(\omega_1,\omega_2) & = \rmd \! \log \omega_1 \ \left( -q_1 \twositechain + \ii \onesiteT \right)\\
    &  \hspace{1em} + \rmd \! \log \omega_2 \ \left( -q_2 \twositechain - \ii \onesiteT \right)\\
    & =  \rmd \! \log \omega_1 \ \left( -q_1 \mathbf{T}_{\theta_{1,2}}^{q_1,q_2}(\omega_1,\omega_2) + \ii \, \mathbf{T}_{\bullet}^{q_{1,2}}(\omega_{1,2}) \right)\\
    &  \hspace{1em} + \rmd \! \log \omega_2 \ \left( -q_2 \mathbf{T}_{\theta_{1,2}}^{q_1,q_2}(\omega_1,\omega_2) - \ii \,  \mathbf{T}_{\bullet}^{q_{1,2}}(\omega_{1,2}) \right).
\end{aligned}
\end{equation}
Only one new term appears, whose differential is proportional to itself
\begin{equation}
    \rmd \mathbf{T}_{\bullet}^{q_{1,2}}(\omega_{1,2}) = - (q_1 + q_2) \,  \rmd \! \log (\omega_1 + \omega_2) \,  \mathbf{T}_{\bullet}^{q_{1,2}}(\omega_{1,2}).
\end{equation}
Hence the differential equation system closes, and we have obtained
\begin{equation}
    \rmd 
    \begin{pmatrix}
        \mathbf{T}_{\theta_{1,2}}^{q_1,q_2}(\omega_1,\omega_2)\\
        \mathbf{T}_{\bullet}^{q_{1,2}}(\omega_{1,2})
    \end{pmatrix} = 
    \begin{pmatrix}
        -q_1 \rmd \! \log \omega_1-q_2 \rmd \! \log \omega_2 & \ii \, (\rmd \! \log \omega_1-\rmd \! \log \omega_2) \\
        0 & -q_{1,2} \, \rmd \! \log (\omega_{1,2})
    \end{pmatrix}
    \begin{pmatrix}
        \mathbf{T}_{\theta_{1,2}}^{q_1,q_2}(\omega_1,\omega_2)\\
        \mathbf{T}_{\bullet}^{q_{1,2}}(\omega_{1,2})
    \end{pmatrix}
\end{equation}
If we further normalize $\mathbf{T}_{\bullet}^{q_{1,2}}(\omega_{1,2})$ by $ \ii (q_1+q_2) $, the above differential equation ends up like
\begin{equation}
    \rmd 
    \begin{pmatrix}
        \mathbf{T}_{\theta_{1,2}}^{q_1,q_2}(\omega_1,\omega_2)\\
        \frac{\mathbf{T}_{\bullet}^{q_{1,2}}(\omega_{1,2})}{\ii (q_1+q_2)}
    \end{pmatrix} = 
    \begin{pmatrix}
        -q_1 \rmd \! \log \omega_1-q_2 \rmd \! \log \omega_2 & -q_{1,2} \, (\rmd\!\log \omega_1-\rmd\!\log \omega_2) \\
        0 & -q_{1,2} \, \rmd \! \log (\omega_{1,2})
    \end{pmatrix}
    \begin{pmatrix}
        \mathbf{T}_{\theta_{1,2}}^{q_1,q_2}(\omega_1,\omega_2)\\
       \frac{\mathbf{T}_{\bullet}^{q_{1,2}}(\omega_{1,2})}{\ii (q_1+q_2)}
    \end{pmatrix},
\end{equation}
One sees that if we take $q_1=q_2=\epsilon$, then $\epsilon$ factors out from the matrix, and the differential equation is in canonical form like in \cite{Arkani-Hamed:2023kig}.

A similar discussion also applies to higher-site integrals. As an illustration, for the following three-site basic time integral
\begin{equation}
    \mathbf{T}^{q_1,q_2,q_3}_{\theta_{1,2},\theta_{3,2}} (\omega_1,\omega_2,\omega_3) = \raisebox{-11pt}{\begin{tikzpicture}
		\coordinate (X1) at (0,0);
		\coordinate (X2) at (1.5,0);
        \coordinate (X3) at (3,0);
		\node[below] at (X1) {\small{$\omega_1,q_1$}};
		\node[below] at (X2) {\small{$\omega_2,q_2$}};
        \node[below] at (X3) {\small{$\omega_3,q_3$}};
		\draw[line width=1.5pt,->] (X1)--(0.85,0);
        \draw[line width=1.5pt] (0.7,0)--(X2);
        \draw[line width=1.5pt,->] (X3)--(2.2,0);
        \draw[line width=1.5pt] (2.5,0)--(X2);
		\path[fill=black] (X1) circle[radius=0.1];
		\path[fill=black] (X2) circle[radius=0.1];
        \path[fill=black] (X3) circle[radius=0.1];
\end{tikzpicture}}.
\end{equation}
The differential matrix can also be deduced by applying \eqref{eq:dT} 
\begin{equation}
    \begin{pmatrix}
        -q_1 \ell_1 - q_2 \ell_2 - q_3 \ell_3 &  \ii (\ell_1-\ell_2) & \ii (\ell_3-\ell_2) & 0 \\
        0 & -q_{1,2} \ell_{1,2}-q_3 \ell_3 & 0 & \ii (\ell_3-\ell_{1,2})\\
        0 & 0 & -q_1 \ell_1 - q_{2,3} \ell_{2,3} & \ii (\ell_1 - \ell_{2,3} )\\
        0 & 0 & 0 & -q_{1,2,3} \ell_{1,2,3}
    \end{pmatrix},
\end{equation}
where we have defined
\begin{equation}
    \begin{array}{lll}
        \ell_1 = \rmd \! \log \omega_1 & \quad \ell_2 = \rmd \!\log \omega_2 & \quad  \ell_3 = \rmd \!\log \omega_3 \\
        \ell_{1,2} = \rmd \!\log \omega_{1,2} & \quad \ell_{2,3} = \rmd \!\log \omega_{2,3} & \quad \ell_{1,2,3} = \rmd \! \log \omega_{1,2,3}
    \end{array}
\end{equation}
and the basis integrals are introduced as
\begin{equation}
    \begin{aligned}
        \mathbf{T}^{q_{1,2},q_3}(\omega_{1,2},\omega_3) & = & \raisebox{-11pt}{\begin{tikzpicture}
		\coordinate (X2) at (1.5,0);
        \coordinate (X3) at (3,0);
		\node[below] at (X2) {\small{$\omega_{1,2},q_{1,2}$}};
        \node[below] at (X3) {\small{$\omega_3,q_3$}};
        \draw[line width=1.5pt,->] (X3)--(2.2,0);
        \draw[line width=1.5pt] (2.5,0)--(X2);
		\path[fill=black] (X2) circle[radius=0.1];
        \path[fill=black] (X3) circle[radius=0.1];
\end{tikzpicture}},\\
        \mathbf{T}^{q_{1},q_{2,3}}(\omega_{1},\omega_{2,3}) & = & \raisebox{-11pt}{\begin{tikzpicture}
		\coordinate (X1) at (0,0);
		\coordinate (X2) at (1.5,0);
		\node[below] at (X1) {\small{$\omega_1,q_1$}};
		\node[below] at (X2) {\small{$\omega_{2,3},q_{2,3}$}};
		\draw[line width=1.5pt,->] (X1)--(0.85,0);
        \draw[line width=1.5pt] (0.7,0)--(X2);
		\path[fill=black] (X1) circle[radius=0.1];
		\path[fill=black] (X2) circle[radius=0.1];
\end{tikzpicture}},\\
        \mathbf{T}^{q_{1,2,3}}(\omega_{1,2,3}) & = & \raisebox{-11pt}{\begin{tikzpicture}
		\coordinate (X1) at (0,0);
		\node[below] at (X1) {\small{$\omega_{1,2,3},q_{1,2,3}$}};
		\path[fill=black] (X1) circle[radius=0.1];
\end{tikzpicture}}.
    \end{aligned}
\end{equation}
It can also be brought into the canonical form after proper normalization by $q_i$'s. Explicitly, if we normalize them as 
\begin{equation}
    \begin{pmatrix}
         \mathbf{T}^{q_1,q_2,q_3}_{\theta_{1,2},\theta_{3,2}}(\omega_1,\omega_2,\omega_3)\\
         \mathbf{T}^{q_{1,2},q_3}(\omega_{1,2},\omega_3)\\
         \mathbf{T}^{q_{1},q_{2,3}}(\omega_1,\omega_{2,3})\\
         \mathbf{T}^{q_{1,2,3}}(\omega_{1,2,3})
    \end{pmatrix} \rightarrow \begin{pmatrix}
        \mathbf{T}^{q_1,q_2,q_3}_{\theta_{1,2},\theta_{3,2}}(\omega_1,\omega_2,\omega_3)\\
         -\ii \mathbf{T}^{q_{1,2},q_3}(\omega_{1,2},\omega_3)/q_3\\
         -\ii \mathbf{T}^{q_{1},q_{2,3}}(\omega_1,\omega_{2,3})/q_1\\
         -\mathbf{T}^{q_{1,2,3}}(\omega_{1,2,3})/(q_1 q_3)
    \end{pmatrix}
\end{equation}
the differential matrix will read
\begin{equation}
    \begin{pmatrix}
        -q_1 \ell_1 - q_2 \ell_2 - q_3 \ell_3 &  -q_3 (\ell_1-\ell_2) & q_1 (\ell_2-\ell_3) & 0 \\
        0 & -q_{1,2} \ell_{1,2}-q_3 \ell_3 & 0 & q_1(-\ell_3+\ell_{1,2})\\
        0 & 0 & -q_1 \ell_1 - q_{2,3} \ell_{2,3} & -q_3 (\ell_1 - \ell_{2,3} )\\
        0 & 0 & 0 & -q_{1,2,3} \ell_{1,2,3}
    \end{pmatrix}.
\end{equation}
When we take $ q_1=q_2=q_3=\epsilon $ above, $\epsilon$ factors out and the matrix is in canonical form. 

This procedure also applies to loop graphs. If we further take the differential of the integrals on the RHS of \eqref{eq:1l4pt}, we will have a 4-level differential system with $\{1,4,4,1\}$ basis integral(s) at each level. These integrals are depicted as follows,
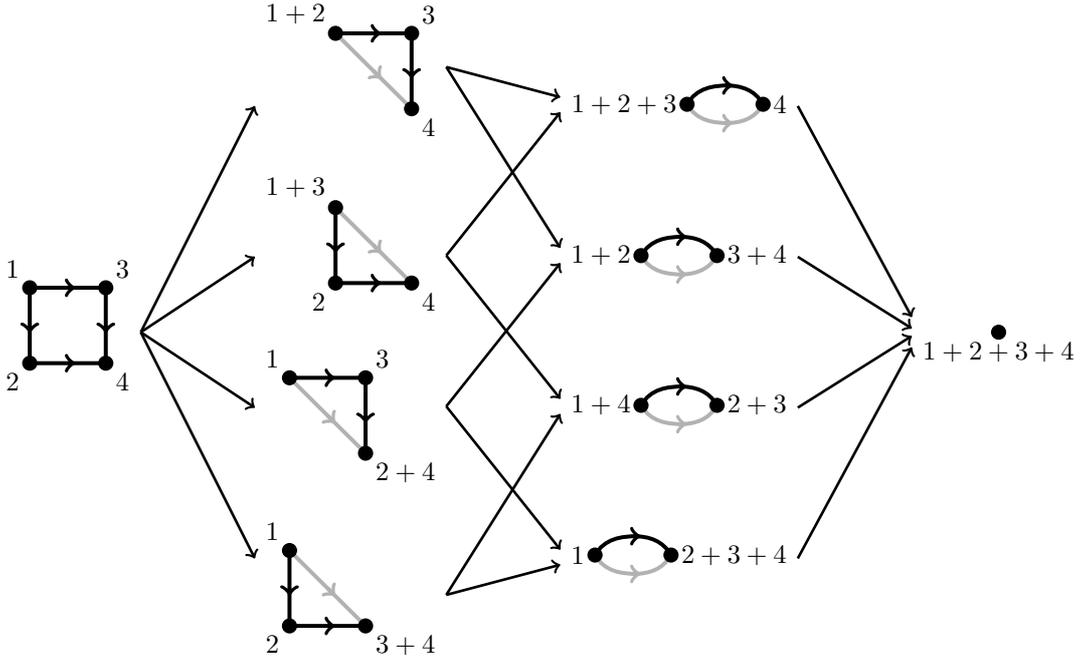
\begin{equation}\nonumber
	\raisebox{-2em}{\begin{tikzpicture}
		\coordinate (X1) at (0,1);
		\coordinate (X2) at (0,0);
		\coordinate (X3) at (1,1);
		\coordinate (X4) at (1,0);
		\node[above left] at (X1) {\small{$1$}};
		\node[below left] at (X2) {\small{$2$}};
		\node[above right] at (X3) {\small{$3$}};
		\node[below right] at (X4) {\small{$4$}};
		\draw[line width=1.5pt,->] (X1)--(0,0.4);
		\draw[line width=1.5pt] (0,0.5)--(X2);
		\draw[line width=1.5pt,->] (X3)--(1,0.4);
		\draw[line width=1.5pt] (1,0.5)--(X4);
		\draw[line width=1.5pt,->] (X1)--(0.6,1);
		\draw[line width=1.5pt] (0.5,1)--(X3);
		\draw[line width=1.5pt,->] (X2)--(0.6,0);
		\draw[line width=1.5pt] (0.5,0)--(X4);
		\path[fill=black] (X1) circle[radius=0.1];
		\path[fill=black] (X2) circle[radius=0.1];
		\path[fill=black] (X3) circle[radius=0.1];
		\path[fill=black] (X4) circle[radius=0.1];
	\end{tikzpicture}}
\raisebox{-7.5em}{\begin{tikzpicture}
	\draw[line width=1pt,->] (0,0)--(1.5,3);
	\draw[line width=1pt,->] (0,0)--(1.5,1);
	\draw[line width=1pt,->] (0,0)--(1.5,-1);
	\draw[line width=1pt,->] (0,0)--(1.5,-3);
\end{tikzpicture}}
\begin{aligned}
	\begin{tikzpicture}
		\coordinate (X1) at (0,1);
		\coordinate (X2) at (0,0);
		\coordinate (X3) at (1,1);
		\coordinate (X4) at (1,0);
		\node[above left] at (X1) {\small{$1+2$}};
		\node[above right] at (X3) {\small{$3$}};
		\node[below right] at (X4) {\small{$4$}};
		\draw[line width=1.5pt,->] (X3)--(1,0.4);
		\draw[line width=1.5pt] (1,0.5)--(X4);
		\draw[line width=1.5pt,->] (X1)--(0.6,1);
		\draw[line width=1.5pt] (0.5,1)--(X3);
		\draw[line width=1.5pt,->,black!30] (X1)--(0.5+0.1,0.5-0.1);
		\draw[line width=1.5pt,black!30] (0.5,0.5)--(X4);
		\path[fill=black] (X1) circle[radius=0.1];
		\path[fill=black] (X3) circle[radius=0.1];
		\path[fill=black] (X4) circle[radius=0.1];
	\end{tikzpicture}\\
\begin{tikzpicture}
	\coordinate (X1) at (0,1);
	\coordinate (X2) at (0,0);
	\coordinate (X3) at (1,1);
	\coordinate (X4) at (1,0);
	\node[above left] at (X1) {\small{$1+3$}};
	\node[below left] at (X2) {\small{$2$}};
	\node[below right] at (X4) {\small{$4$}};
	\draw[line width=1.5pt,->] (X1)--(0,0.4);
	\draw[line width=1.5pt] (0,0.5)--(X2);
	\draw[line width=1.5pt,->] (X2)--(0.6,0);
	\draw[line width=1.5pt] (0.5,0)--(X4);
	\draw[line width=1.5pt,->,black!30] (X1)--(0.5+0.1,0.5-0.1);
	\draw[line width=1.5pt,black!30] (0.5,0.5)--(X4);
	\path[fill=black] (X1) circle[radius=0.1];
	\path[fill=black] (X2) circle[radius=0.1];
	\path[fill=black] (X4) circle[radius=0.1];
\end{tikzpicture}\\
\begin{tikzpicture}
	\coordinate (X1) at (0,1);
	\coordinate (X2) at (0,0);
	\coordinate (X3) at (1,1);
	\coordinate (X4) at (1,0);
	\node[above left] at (X1) {\small{$1$}};
	\node[above right] at (X3) {\small{$3$}};
	\node[below right] at (X4) {\small{$2+4$}};
	\draw[line width=1.5pt,->] (X3)--(1,0.4);
	\draw[line width=1.5pt] (1,0.5)--(X4);
	\draw[line width=1.5pt,->] (X1)--(0.6,1);
	\draw[line width=1.5pt] (0.5,1)--(X3);
	\draw[line width=1.5pt,->,black!30] (X1)--(0.5+0.1,0.5-0.1);
	\draw[line width=1.5pt,black!30] (0.5,0.5)--(X4);
	\path[fill=black] (X1) circle[radius=0.1];
	\path[fill=black] (X3) circle[radius=0.1];
	\path[fill=black] (X4) circle[radius=0.1];
\end{tikzpicture}\\
\begin{tikzpicture}
	\coordinate (X1) at (0,1);
	\coordinate (X2) at (0,0);
	\coordinate (X3) at (1,1);
	\coordinate (X4) at (1,0);
	\node[above left] at (X1) {\small{$1$}};
	\node[below left] at (X2) {\small{$2$}};
	\node[below right] at (X4) {\small{$3+4$}};
	\draw[line width=1.5pt,->] (X1)--(0,0.4);
	\draw[line width=1.5pt] (0,0.5)--(X2);
	\draw[line width=1.5pt,->] (X2)--(0.6,0);
	\draw[line width=1.5pt] (0.5,0)--(X4);
	\draw[line width=1.5pt,->,black!30] (X1)--(0.5+0.1,0.5-0.1);
	\draw[line width=1.5pt,black!30] (0.5,0.5)--(X4);
	\path[fill=black] (X1) circle[radius=0.1];
	\path[fill=black] (X2) circle[radius=0.1];
	\path[fill=black] (X4) circle[radius=0.1];
\end{tikzpicture}
\end{aligned}
\raisebox{-8.75em}{\begin{tikzpicture}
		\draw[line width=1pt,->] (0,3.5)--(1.5,3.1);
		\draw[line width=1pt,->] (0,3.5)--(1.5,1.1);
		\draw[line width=1pt,->] (0,1)--(1.5,-0.9);
		\draw[line width=1pt,->] (0,1)--(1.5,2.9);
		\draw[line width=1pt,->] (0,-1)--(1.5,0.9);
		\draw[line width=1pt,->] (0,-1)--(1.5,-2.9);
		\draw[line width=1pt,->] (0,-3.5)--(1.5,-1.1);
		\draw[line width=1pt,->] (0,-3.5)--(1.5,-3.1);
\end{tikzpicture}}
\begin{aligned}
	\begin{tikzpicture}
		\node[left] at (0,0) {\small{$1+2+3$}};
		\node[right] at (1,0) {\small{$4$}};
		\draw [line width=1.5pt,->] (0,0) to[out=60,in=-180] (0.6,0.25);
		\draw [line width=1.5pt] (0.5,0.25) to[out=0,in=120] (1,0);
		\draw [line width=1.5pt,->,black!30] (0,0) to[out=-60,in=-180] (0.6,-0.25);
		\draw [line width=1.5pt,black!30] (0.5,-0.25) to[out=0,in=-120] (1,0);
		\path[fill=black] (0,0) circle[radius=0.1];
		\path[fill=black] (1,0) circle[radius=0.1];
	\end{tikzpicture}\\[3em]
\begin{tikzpicture}
	\node[left] at (0,0) {\small{$1+2$}};
	\node[right] at (1,0) {\small{$3+4$}};
	\draw [line width=1.5pt,->] (0,0) to[out=60,in=-180] (0.6,0.25);
	\draw [line width=1.5pt] (0.5,0.25) to[out=0,in=120] (1,0);
	\draw [line width=1.5pt,->,black!30] (0,0) to[out=-60,in=-180] (0.6,-0.25);
	\draw [line width=1.5pt,black!30] (0.5,-0.25) to[out=0,in=-120] (1,0);
	\path[fill=black] (0,0) circle[radius=0.1];
	\path[fill=black] (1,0) circle[radius=0.1];
\end{tikzpicture}\\[3em]
\begin{tikzpicture}
	\node[left] at (0,0) {\small{$1+4$}};
	\node[right] at (1,0) {\small{$2+3$}};
	\draw [line width=1.5pt,->] (0,0) to[out=60,in=-180] (0.6,0.25);
	\draw [line width=1.5pt] (0.5,0.25) to[out=0,in=120] (1,0);
	\draw [line width=1.5pt,->,black!30] (0,0) to[out=-60,in=-180] (0.6,-0.25);
	\draw [line width=1.5pt,black!30] (0.5,-0.25) to[out=0,in=-120] (1,0);
	\path[fill=black] (0,0) circle[radius=0.1];
	\path[fill=black] (1,0) circle[radius=0.1];
\end{tikzpicture}\\[3em]
\begin{tikzpicture}
	\node[left] at (0,0) {\small{$1$}};
	\node[right] at (1,0) {\small{$2+3+4$}};
	\draw [line width=1.5pt,->] (0,0) to[out=60,in=-180] (0.6,0.25);
	\draw [line width=1.5pt] (0.5,0.25) to[out=0,in=120] (1,0);
	\draw [line width=1.5pt,->,black!30] (0,0) to[out=-60,in=-180] (0.6,-0.25);
	\draw [line width=1.5pt,black!30] (0.5,-0.25) to[out=0,in=-120] (1,0);
	\path[fill=black] (0,0) circle[radius=0.1];
	\path[fill=black] (1,0) circle[radius=0.1];
\end{tikzpicture}
\end{aligned}
\raisebox{-7.5em}{\begin{tikzpicture}
	\draw[line width=1pt,->] (0,3)--(1.5,0.2);
	\draw[line width=1pt,->] (0,1)--(1.5,0.05);
	\draw[line width=1pt,->] (0,-1)--(1.5,-0.05);
	\draw[line width=1pt,->] (0,-3)--(1.5,-0.2);
\end{tikzpicture}}
\raisebox{-1em}{\begin{tikzpicture}
	\node[below] at (0,0) {\small{$1+2+3+4$}};
	\path[fill=black] (0,0) circle[radius=0.1];
\end{tikzpicture}}
\end{equation}
where the arrows between each level indicate which integrals appear in the total differential of the last level.

\subsection{The complete system of differential equations for cosmological amplitudes}\label{sec:wfDE}

By recursively using \eqref{eq:dT} and the Leibniz rule, we can derive the differential equation system for any wavefunction coefficients or correlators. Let us generally describe as an algorithm how the canonical differential equations for wavefunction/correlator are computed. As stated in Sec.\ref{sec:decomp}, a wavefunction $\psi$ can be decomposed into a combination of (products of) basic time integrals. The total differential of the wavefunction is easy to compute by using \eqref{eq:dT}. Schematically, we have 
\begin{equation}
    \rmd \psi = \rmd \sum_{i}  \mathbf{T}_{\mathcal{N}_i} = \sum_j \rmd \!\log (\omega_j) \ \mathbf{B}_j
\end{equation}
where on the RHS we collect the $\rmd \! \log$'s and denote their coefficients as $\mathbf{B}_j$, which are linear combinations of basic time integrals including ones with contracted edges. This is the 1st level of the differential equation system. Letters $\omega_j$ appearing at this level are the {\it physical singularities} of the form $ \sum_i X_i + \sum_j Y_j $, as well as  {\it folded singularities} which can be obtained by replacing some of ``$+Y_j$" by ``$-Y_j$" in physical ones. Then we collect the $\mathbf{B}_j$'s associated with folded singularities, along with the original wavefunction we get a set of integrals $ \{ \psi, \mathbf{B}_j \} $. All other $\mathbf{B}_j$'s corresponding to physical singularities are expected to be linear combinations of these integrals. We then take differential of $ \mathbf{B}_j $'s to get the 2nd level total differentials
\begin{equation}
    \rmd \mathbf{B}_j = \sum_k \rmd\! \log (\omega_{j,k}) \ \mathbf{C}_{j,k}
\end{equation}
where new folded singularities appear, and we extend the integral basis by their coefficients, and express other coefficients as a linear combination of  $ \{ \psi, \mathbf{B}_j, \mathbf{C}_{j,k} \} $. If these integrals are not complete, we further extend them by coefficients of the newly appearing letters at the last level. Recursively, we will have a closed differential equation system. Then we are going to normalize every integral in the basis by the product of $ (-q_i) $'s, where the $q_i$'s are those that show up as coefficients in this basis integral. For example, we normalize the integral
\begin{equation}
    \ii \, (q_1+q_2) \, q_3 \raisebox{-11pt}{\begin{tikzpicture}
		\coordinate (X1) at (0,0);
		\coordinate (X2) at (2,0);
		\node[below] at (X1) {\tiny{$X_1+X_2+Y_2,q_{1,2}$}};
		\node[below] at (X2) {\tiny{$X_3-Y,q_3$}};
		\draw[line width=1.5pt] (X1)--(1.1,0);
        \draw[line width=1.5pt,->] (X2)--(0.9,0);
		\path[fill=black] (X1) circle[radius=0.1];
		\path[fill=black] (X2) circle[radius=0.1];
\end{tikzpicture}} + \ii \, q_3 \raisebox{-11pt}{\begin{tikzpicture}
		\coordinate (X1) at (0,0);
		\node[below] at (X1) {\tiny{$X_1+X_2+X_3,\ q_{1,2,3}$}};
		\path[fill=black] (X1) circle[radius=0.1];
\end{tikzpicture}}
\end{equation}
by $-1/(q_1 q_2 q_3)$. This integral appears as a basis integral of the three-site chain at the 4th level.

Since for individual basic time integral, all possible new graphs produced by taking differential can be obtained by contracting some of the edges of the original graph. Thus, for a general wavefunction with $e$ edges, there are 4 different choices for each edge: 2 different time orders, 1 dash line, 1 contracting. So the number of time integrals appearing in the closed differential equation system is (at most) $4^{e}$, which coincides with that in \cite{Arkani-Hamed:2023kig}. This number can be reduced when the diagram contains loops, where the contraction of edges may lead to a vanishing graph with an incompatible time order. In the rest part of this subsection, we will give some explicit examples. 


\paragraph{Two-site chain wavefunction/correlator} The two-site chain wavefunction in terms of time integrals is written as
\begin{equation}
    \psi_{\text{2-chain}} = \raisebox{-11pt}{\begin{tikzpicture}
		\coordinate (X1) at (0,0);
		\coordinate (X2) at (2,0);
		\node[below] at (X1) {\tiny{$X_1+Y,q_1$}};
		\node[below] at (X2) {\tiny{$X_2-Y,q_2$}};
		\draw[line width=1.5pt,->] (X1)--(1.1,0);
        \draw[line width=1.5pt] (0.9,0)--(X2);
		\path[fill=black] (X1) circle[radius=0.1];
		\path[fill=black] (X2) circle[radius=0.1];
\end{tikzpicture}}+\raisebox{-11pt}{\begin{tikzpicture}
		\coordinate (X1) at (0,0);
		\coordinate (X2) at (2,0);
		\node[below] at (X1) {\tiny{$X_1-Y,q_1$}};
		\node[below] at (X2) {\tiny{$X_2+Y,q_2$}};
		\draw[line width=1.5pt] (X1)--(1.1,0);
        \draw[line width=1.5pt,->] (X2)--(0.9,0);
		\path[fill=black] (X1) circle[radius=0.1];
		\path[fill=black] (X2) circle[radius=0.1];
\end{tikzpicture}}-\raisebox{-11pt}{\begin{tikzpicture}
		\coordinate (X1) at (0,0);
		\coordinate (X2) at (1.5,0);
		\node[below] at (X1) {\tiny{$X_1+Y,q_1$}};
		\node[below] at (X2) {\tiny{$X_2+Y,q_2$}};
		\path[fill=black] (X1) circle[radius=0.1];
		\path[fill=black] (X2) circle[radius=0.1];
\end{tikzpicture}},
\end{equation}
Since we are adding basic time integrals together and get undirected integrals, we change variables from $\omega_i$ in each directed graph to $\{X_i,Y_{i,j}\}$, and the energy and twist exponent of each site are labeled below ($Y:=Y_{1,2}$ in this example).  By taking differential of $\psi_{\text{2-chain}}$, we get
\begin{equation}\label{eq:dpsi2}
\begin{aligned}
    \rmd \psi_{\text{2-chain}} & = \rmd \! \log (X_1+Y) \left( -q_1 \raisebox{-11pt}{\begin{tikzpicture}
		\coordinate (X1) at (0,0);
		\coordinate (X2) at (2,0);
		\node[below] at (X1) {\tiny{$X_1+Y,q_1$}};
		\node[below] at (X2) {\tiny{$X_2-Y,q_2$}};
		\draw[line width=1.5pt,->] (X1)--(1.1,0);
        \draw[line width=1.5pt] (0.9,0)--(X2);
		\path[fill=black] (X1) circle[radius=0.1];
		\path[fill=black] (X2) circle[radius=0.1];
\end{tikzpicture}}+ \ii \raisebox{-11pt}{\begin{tikzpicture}
		\coordinate (X1) at (0,0);
		\node[below] at (X1) {\tiny{$X_1+X_2,\ q_{1,2}$}};
		\path[fill=black] (X1) circle[radius=0.1];
\end{tikzpicture}} + q_1 \raisebox{-11pt}{\begin{tikzpicture}
		\coordinate (X1) at (0,0);
		\coordinate (X2) at (1.5,0);
		\node[below] at (X1) {\tiny{$X_1+Y,q_1$}};
		\node[below] at (X2) {\tiny{$X_2+Y,q_2$}};
		\path[fill=black] (X1) circle[radius=0.1];
		\path[fill=black] (X2) circle[radius=0.1];
\end{tikzpicture}} \right)\\
    & +\rmd \! \log (X_2+Y) \left( -q_2 \raisebox{-11pt}{\begin{tikzpicture}
		\coordinate (X1) at (0,0);
		\coordinate (X2) at (2,0);
		\node[below] at (X1) {\tiny{$X_1-Y,q_1$}};
		\node[below] at (X2) {\tiny{$X_2+Y,q_2$}};
		\draw[line width=1.5pt] (X1)--(1.1,0);
        \draw[line width=1.5pt,->] (X2)--(0.9,0);
		\path[fill=black] (X1) circle[radius=0.1];
		\path[fill=black] (X2) circle[radius=0.1];
\end{tikzpicture}} + \ii \raisebox{-11pt}{\begin{tikzpicture}
		\coordinate (X1) at (0,0);
		\node[below] at (X1) {\tiny{$X_1+X_2,\ q_{1,2}$}};
		\path[fill=black] (X1) circle[radius=0.1];
\end{tikzpicture}}+ q_2 \raisebox{-11pt}{\begin{tikzpicture}
		\coordinate (X1) at (0,0);
		\coordinate (X2) at (1.5,0);
		\node[below] at (X1) {\tiny{$X_1+Y,q_1$}};
		\node[below] at (X2) {\tiny{$X_2+Y,q_2$}};
		\path[fill=black] (X1) circle[radius=0.1];
		\path[fill=black] (X2) circle[radius=0.1];
\end{tikzpicture}} \right)\\
    & +\rmd \! \log (X_1-Y) \left( -q_1 \raisebox{-11pt}{\begin{tikzpicture}
		\coordinate (X1) at (0,0);
		\coordinate (X2) at (2,0);
		\node[below] at (X1) {\tiny{$X_1-Y,q_1$}};
		\node[below] at (X2) {\tiny{$X_2+Y,q_2$}};
		\draw[line width=1.5pt] (X1)--(1.1,0);
        \draw[line width=1.5pt,->] (X2)--(0.9,0);
		\path[fill=black] (X1) circle[radius=0.1];
		\path[fill=black] (X2) circle[radius=0.1];
\end{tikzpicture}} - \ii \raisebox{-11pt}{\begin{tikzpicture}
		\coordinate (X1) at (0,0);
		\node[below] at (X1) {\tiny{$X_1+X_2,\ q_{1,2}$}};
		\path[fill=black] (X1) circle[radius=0.1];
\end{tikzpicture}} \right)\\
    & +\rmd \! \log (X_2-Y) \left( -q_2 \raisebox{-11pt}{\begin{tikzpicture}
		\coordinate (X1) at (0,0);
		\coordinate (X2) at (2,0);
		\node[below] at (X1) {\tiny{$X_1+Y,q_1$}};
		\node[below] at (X2) {\tiny{$X_2-Y,q_2$}};
		\draw[line width=1.5pt,->] (X1)--(1.1,0);
        \draw[line width=1.5pt] (0.9,0)--(X2);
		\path[fill=black] (X1) circle[radius=0.1];
		\path[fill=black] (X2) circle[radius=0.1];
\end{tikzpicture}} - \ii \raisebox{-11pt}{\begin{tikzpicture}
		\coordinate (X1) at (0,0);
		\node[below] at (X1) {\tiny{$X_1+X_2,\ q_{1,2}$}};
		\path[fill=black] (X1) circle[radius=0.1];
\end{tikzpicture}} \right),
\end{aligned}
\end{equation}
where letters $X_1+Y$ and $X_2+Y$ are ``physical", and $X_1-Y$ and $X_2-Y$ are the so-called ``folded" ones. If we define the coefficients of folded letters to be
\begin{equation}\label{eq:Bdef}
    \begin{aligned}
        & \textbf{B}_1 = -q_1 \raisebox{-11pt}{\begin{tikzpicture}
		\coordinate (X1) at (0,0);
		\coordinate (X2) at (2,0);
		\node[below] at (X1) {\tiny{$X_1-Y,q_1$}};
		\node[below] at (X2) {\tiny{$X_2+Y,q_2$}};
		\draw[line width=1.5pt] (X1)--(1.1,0);
        \draw[line width=1.5pt,->] (X2)--(0.9,0);
		\path[fill=black] (X1) circle[radius=0.1];
		\path[fill=black] (X2) circle[radius=0.1];
\end{tikzpicture}} - \ii \raisebox{-11pt}{\begin{tikzpicture}
		\coordinate (X1) at (0,0);
		\node[below] at (X1) {\tiny{$X_1+X_2,\ q_{1,2}$}};
		\path[fill=black] (X1) circle[radius=0.1];
\end{tikzpicture}} ,\\
        & \textbf{B}_2=  -q_2 \raisebox{-11pt}{\begin{tikzpicture}
		\coordinate (X1) at (0,0);
		\coordinate (X2) at (2,0);
		\node[below] at (X1) {\tiny{$X_1+Y,q_1$}};
		\node[below] at (X2) {\tiny{$X_2-Y,q_2$}};
		\draw[line width=1.5pt,->] (X1)--(1.1,0);
        \draw[line width=1.5pt] (0.9,0)--(X2);
		\path[fill=black] (X1) circle[radius=0.1];
		\path[fill=black] (X2) circle[radius=0.1];
\end{tikzpicture}} - \ii \raisebox{-11pt}{\begin{tikzpicture}
		\coordinate (X1) at (0,0);
		\node[below] at (X1) {\tiny{$X_1+X_2,\ q_{1,2}$}};
		\path[fill=black] (X1) circle[radius=0.1];
\end{tikzpicture}}.
    \end{aligned}
\end{equation}
Then it is easy to see \eqref{eq:dpsi2} can be recast as
\begin{equation}\label{eq:d2chain}
\begin{aligned}
    \rmd \psi_{\text{2-chain}}  = \rmd \! \log (X_1+Y) \  (-q_1 \psi_{\text{2-chain}} & -\mathbf{B}_1) + \rmd \! \log (X_2+Y) \  (-q_2 \psi_{\text{2-chain}}-\mathbf{B}_2) \\
    &  + \rmd \! \log (X_1-Y) \ \mathbf{B}_1 + \rmd \! \log (X_2-Y) \ \mathbf{B}_2.
\end{aligned}
\end{equation}
And here is the 1st level differential equation, which only contains one single differential.

To get the 2nd level differential equation, we further take differential of $ \textbf{B}_1 $ and $ \textbf{B}_2 $ and get
\begin{equation}
    \begin{aligned}
        \rmd \mathbf{B}_1 & =  \rmd \! \log (X_2+Y) \left( q_1 q_2 \raisebox{-11pt}{\begin{tikzpicture}
		\coordinate (X1) at (0,0);
		\coordinate (X2) at (2,0);
		\node[below] at (X1) {\tiny{$X_1-Y,q_1$}};
		\node[below] at (X2) {\tiny{$X_2+Y,q_2$}};
		\draw[line width=1.5pt] (X1)--(1.1,0);
        \draw[line width=1.5pt,->] (X2)--(0.9,0);
		\path[fill=black] (X1) circle[radius=0.1];
		\path[fill=black] (X2) circle[radius=0.1];
\end{tikzpicture}} - \ii \, q_1 \raisebox{-11pt}{\begin{tikzpicture}
		\coordinate (X1) at (0,0);
		\node[below] at (X1) {\tiny{$X_1+X_2,\ q_{1,2}$}};
		\path[fill=black] (X1) circle[radius=0.1];
\end{tikzpicture}} \right)\\
        & + \rmd \! \log (X_1-Y) \left( q_1^2 \raisebox{-11pt}{\begin{tikzpicture}
		\coordinate (X1) at (0,0);
		\coordinate (X2) at (2,0);
		\node[below] at (X1) {\tiny{$X_1-Y,q_1$}};
		\node[below] at (X2) {\tiny{$X_2+Y,q_2$}};
		\draw[line width=1.5pt] (X1)--(1.1,0);
        \draw[line width=1.5pt,->] (X2)--(0.9,0);
		\path[fill=black] (X1) circle[radius=0.1];
		\path[fill=black] (X2) circle[radius=0.1];
\end{tikzpicture}} + \ii \, q_1 \raisebox{-11pt}{\begin{tikzpicture}
		\coordinate (X1) at (0,0);
		\node[below] at (X1) {\tiny{$X_1+X_2,\ q_{1,2}$}};
		\path[fill=black] (X1) circle[radius=0.1];
\end{tikzpicture}} \right)\\
        & + \rmd \! \log (X_1+X_2) \left( \ii \, (q_1+q_2) \raisebox{-11pt}{\begin{tikzpicture}
		\coordinate (X1) at (0,0);
		\node[below] at (X1) {\tiny{$X_1+X_2,\ q_{1,2}$}};
		\path[fill=black] (X1) circle[radius=0.1];
\end{tikzpicture}} \right)
    \end{aligned}
\end{equation}
\begin{equation}
    \begin{aligned}
        \rmd \mathbf{B}_2 & = \rmd \! \log (X_2+Y) \left( q_1 q_2 \raisebox{-11pt}{\begin{tikzpicture}
		\coordinate (X1) at (0,0);
		\coordinate (X2) at (2,0);
		\node[below] at (X1) {\tiny{$X_1+Y,q_1$}};
		\node[below] at (X2) {\tiny{$X_2-Y,q_2$}};
		\draw[line width=1.5pt,->] (X1)--(1.1,0);
        \draw[line width=1.5pt] (0.9,0)--(X2);
		\path[fill=black] (X1) circle[radius=0.1];
		\path[fill=black] (X2) circle[radius=0.1];
\end{tikzpicture}} - \ii q_2 \raisebox{-11pt}{\begin{tikzpicture}
		\coordinate (X1) at (0,0);
		\node[below] at (X1) {\tiny{$X_1+X_2,\ q_{1,2}$}};
		\path[fill=black] (X1) circle[radius=0.1];
\end{tikzpicture}} \right)\\
        & + \rmd \! \log (X_1-Y) \left( q_2^2 \raisebox{-11pt}{\begin{tikzpicture}
		\coordinate (X1) at (0,0);
		\coordinate (X2) at (2,0);
		\node[below] at (X1) {\tiny{$X_1+Y,q_1$}};
		\node[below] at (X2) {\tiny{$X_2-Y,q_2$}};
		\draw[line width=1.5pt,->] (X1)--(1.1,0);
        \draw[line width=1.5pt] (0.9,0)--(X2);
		\path[fill=black] (X1) circle[radius=0.1];
		\path[fill=black] (X2) circle[radius=0.1];
\end{tikzpicture}} + \ii q_2 \raisebox{-11pt}{\begin{tikzpicture}
		\coordinate (X1) at (0,0);
		\node[below] at (X1) {\tiny{$X_1+X_2,\ q_{1,2}$}};
		\path[fill=black] (X1) circle[radius=0.1];
\end{tikzpicture}} \right)\\
 & + \rmd \! \log (X_1+X_2) \left( \ii (q_1+q_2) \raisebox{-11pt}{\begin{tikzpicture}
		\coordinate (X1) at (0,0);
		\node[below] at (X1) {\tiny{$X_1+X_2,\ q_{1,2}$}};
		\path[fill=black] (X1) circle[radius=0.1];
\end{tikzpicture}} \right)
    \end{aligned}
\end{equation}
where we recognize the coefficient of $ \rmd \! \log (X_1 + X_2) $ as a new basis integral 
\begin{equation}
    \mathbf{C}=\ii \, (q_1+q_2) \raisebox{-11pt}{\begin{tikzpicture}
		\coordinate (X1) at (0,0);
		\node[below] at (X1) {\tiny{$X_1+X_2,\ q_{1,2}$}};
		\path[fill=black] (X1) circle[radius=0.1];
\end{tikzpicture}}
\end{equation}
Together with basis integrals at 1st level, the 2nd level differential equation can be re-written as
\begin{equation}
    \begin{aligned}
        & \rmd \textbf{B}_1=-\rmd \! \log (X_2+Y)(q_2 \mathbf{B}_1+\mathbf{C})-\rmd \! \log (X_1-Y) (q_1 \mathbf{B}_1)+\rmd \! \log (X_1+X_2) \mathbf{C} ,\\
        & \rmd \textbf{B}_2= -\rmd \! \log (X_1+Y) (q_1 \mathbf{B}_2 + \mathbf{C})-\rmd \! \log (X_2-Y) (q_2 \mathbf{B}_2 ) + \rmd \! \log (X_1+X_2) \mathbf{C}.
    \end{aligned}
\end{equation}

Then we calculate the 3rd level differential equation, which is simply 
\begin{equation}
    \rmd \textbf{C} = -(q_1+q_2) \rmd \! \log (X_1+X_2) \ \textbf{C}.
\end{equation}
There are neither new integrals nor new letters in this differential equation, and we finally arrive at a closed differential equation system
\begin{equation}\label{eq:2chainDE}
    \rmd
    \begin{pmatrix}
     \psi_{\text{2-chain}} \\
     \textbf{B}_1\\
     \textbf{B}_2\\
     \textbf{C}
    \end{pmatrix} = 
    \begin{pmatrix}
        -q_1 \ell_1- q_2 \ell_2 & - \ell_1+ \ell_3 & - \ell_2 + \ell_4 & 0 \\
        0 & -q_1 \ell_3 - q_2 \ell_2 & 0 & -\ell_2 + \ell_5\\
        0 & 0 & -q_1 \ell_1- q_2 \ell_4 & -\ell_1+ \ell_5\\
        0 & 0 & 0 & -(q_1+q_2) \ell_5
    \end{pmatrix}
    \begin{pmatrix}
     \psi_{\text{2-chain}} \\
     \textbf{B}_1\\
     \textbf{B}_2\\
     \textbf{C}
    \end{pmatrix},
\end{equation}
where we have defined
\begin{equation}\label{eq:ldef}
\begin{aligned}
    & \ell_1=\rmd \! \log (X_1+Y), & \ell_2=\rmd \! \log (X_2+Y),\ \ \   & \ell_3=\rmd \! \log (X_1-Y),\\
    & \ell_4=\rmd \! \log (X_2-Y), & \ell_5=\rmd \! \log (X_1+X_2). & 
\end{aligned}
\end{equation}
After proper normalization of these integrals, we have
\begin{equation}\label{eq:2chainCDE}
    \rmd
    \begin{pmatrix}
     \psi_{\text{2-chain}} \\
     -\textbf{B}_1 / q_1\\
     -\textbf{B}_2/ q_2\\
     \textbf{C}/ (q_1 q_2)
    \end{pmatrix} = \begin{pmatrix}
        -q_1 \ell_1- q_2 \ell_2 & q_1 \ell_1 - q_1 \ell_3 & q_2 \ell_2 - q_2 \ell_4 & 0 \\
        0 & -q_1 \ell_3 - q_2 \ell_2 & 0 & q_2 \ell_2 - q_2 \ell_5\\
        0 & 0 & -q_1 \ell_1- q_2 \ell_4 & q_1 \ell_1 -q_1 \ell_5\\
        0 & 0 & 0 & -(q_1+q_2) \ell_5
    \end{pmatrix} \begin{pmatrix}
     \psi_{\text{2-chain}} \\
     -\textbf{B}_1 / q_1\\
     -\textbf{B}_2/ q_2\\
     \textbf{C}/ (q_1 q_2)
    \end{pmatrix}
\end{equation}
which is the same differential equation system obtained by kinematic flow when taking $q_i=\epsilon$.

For the two-site chain correlator, we have the following expansion
\begin{equation}
    \mathcal{T}_{\text{2-chain}} = \raisebox{-11pt}{\begin{tikzpicture}
		\coordinate (X1) at (0,0);
		\coordinate (X2) at (2,0);
		\node[below] at (X1) {\tiny{$X_1+Y,q_1$}};
		\node[below] at (X2) {\tiny{$X_2-Y,q_2$}};
		\draw[line width=1.5pt,->] (X1)--(1.1,0);
        \draw[line width=1.5pt] (0.9,0)--(X2);
		\path[fill=black] (X1) circle[radius=0.1];
		\path[fill=black] (X2) circle[radius=0.1];
\end{tikzpicture}}+\raisebox{-11pt}{\begin{tikzpicture}
		\coordinate (X1) at (0,0);
		\coordinate (X2) at (2,0);
		\node[below] at (X1) {\tiny{$X_1-Y,q_1$}};
		\node[below] at (X2) {\tiny{$X_2+Y,q_2$}};
		\draw[line width=1.5pt] (X1)--(1.1,0);
        \draw[line width=1.5pt,->] (X2)--(0.9,0);
		\path[fill=black] (X1) circle[radius=0.1];
		\path[fill=black] (X2) circle[radius=0.1];
\end{tikzpicture}}-\raisebox{-11pt}{\begin{tikzpicture}
		\coordinate (X1) at (0,0);
		\coordinate (X2) at (1.5,0);
		\node[below] at (X1) {\tiny{$X_1+Y,q_1$}};
		\node[below] at (X2) {\tiny{$-X_2-Y,q_2$}};
		\path[fill=black] (X1) circle[radius=0.1];
		\path[fill=black] (X2) circle[radius=0.1];
\end{tikzpicture}}+\text{c.c.}
\end{equation}
For simplicity, we overlook the complex conjugate part. After a similar calculation as in the two-site wavefunction case, the whole differential system reads the same as \eqref{eq:2chainDE} and \eqref{eq:2chainCDE}, except with $\psi_{\text{2-chain}}$ exchanged by $\mathcal{T}_{\text{2-chain}}$.

\paragraph{Two-site loop wavefunction/correlator} We can apply our method to arbitrary graphs even with loops. For the two-site bubble, we have 3 levels in the DE, and $\{1, 6, 3\}$ basis integral(s) in each level. Interestingly, for the two-site $L$-loop (banana) wavefunction, we found that there are still 3 levels of DE and \{$1,\, 2\,(2^{L+1}-1),\, (2^{L+1}-1)$\} basis integrals in each level. Explicitly, we choose the basis integrals as 
\begin{equation}
    \begin{aligned}
        \psi_{\text{banana}}&=\raisebox{-0.5em}{\begin{tikzpicture}
		\node[left] at (0,0) {\small{$1$}};
		\node[right] at (1,0) {\small{$2$}};
		\draw [line width=1.5pt] (0,0) to[out=60,in=-180] (0.6,0.25);
		\draw [line width=1.5pt] (0.5,0.25) to[out=0,in=120] (1,0);
		\draw [line width=1.5pt] (0,0) to[out=-60,in=-180] (0.5,-0.25);
		\draw [line width=1.5pt] (1,0) to[out=-120,in=0] (0.4,-0.25);
		\path[fill=black] (0,0) circle[radius=0.1];
		\path[fill=black] (1,0) circle[radius=0.1];
        \path[fill=black] (0.5,0.1) circle[radius=0.025];
		\path[fill=black] (0.5,0.0) circle[radius=0.025];
		\path[fill=black] (0.5,-0.1) circle[radius=0.025];
\end{tikzpicture}}= (2^{L+2}-1) \text{ terms},\\[1.1em]
\mathbf{B}^{(1)}(\omega_{i}^{(1)})&= -q_1 \raisebox{-11pt}{\begin{tikzpicture}
		\coordinate (X1) at (0,0);
		\coordinate (X2) at (1.5,0);
		\node[below] at (X1) {\tiny{$\omega_i^{(1)},q_1$}};
		\node[below] at (X2) {\tiny{$X_2^{+},q_2$}};
		\draw[line width=1.5pt] (X1)--(0.8,0);
        \draw[line width=1.5pt,->] (X2)--(0.65,0);
		\path[fill=black] (X1) circle[radius=0.1];
		\path[fill=black] (X2) circle[radius=0.1];
\end{tikzpicture}} - \ii \raisebox{-11pt}{\begin{tikzpicture}
		\coordinate (X1) at (0,0);
		\node[below] at (X1) {\tiny{$\omega_i^{(1)}+X_2^{+},\ q_{1,2}$}};
		\path[fill=black] (X1) circle[radius=0.1];
\end{tikzpicture}}, \quad i=1,2,\ldots,2^{L+1}-1,\\
\mathbf{B}^{(2)}(\omega_{i}^{(2)})&= -q_2 \raisebox{-11pt}{\begin{tikzpicture}
		\coordinate (X1) at (0,0);
		\coordinate (X2) at (1.5,0);
		\node[below] at (X1) {\tiny{$X_1^{+},q_1$}};
		\node[below] at (X2) {\tiny{$\omega_i^{(2)},q_2$}};
		\draw[line width=1.5pt,->] (X1)--(0.85,0);
        \draw[line width=1.5pt] (X2)--(0.7,0);
		\path[fill=black] (X1) circle[radius=0.1];
		\path[fill=black] (X2) circle[radius=0.1];
\end{tikzpicture}} - \ii \raisebox{-11pt}{\begin{tikzpicture}
		\coordinate (X1) at (0,0);
		\node[below] at (X1) {\tiny{$X_1^{+}+\omega_i^{(2)},\ q_{1,2}$}};
		\path[fill=black] (X1) circle[radius=0.1];
\end{tikzpicture}}, \quad i=1,2,\ldots,2^{L+1}-1,\\[1.1em]
\mathbf{C}\left(\omega_i\right) & = \ii (q_1+q_2) \raisebox{-11pt}{\begin{tikzpicture}
		\coordinate (X1) at (0,0);
		\node[below] at (X1) {\tiny{$\omega_i,\ q_{1,2}$}};
		\path[fill=black] (X1) circle[radius=0.1];
\end{tikzpicture}}, \quad i=1,2,\ldots,2^{L+1}-1,
    \end{aligned}
\end{equation}
where the energies are defined as:
\begin{equation}
\begin{aligned}
    &X_i^{+}=X_i+\sum_{j=1}^{L+1}Y_j, \quad i=1,2\\
    &\omega_i=X_1+X_2+2 \sum_{k \in \alpha_i} Y_k, \quad \alpha_i \subsetneq \{1,2,\ldots,2^{L+1}\}\\
    &\omega_i^{(j)}=X_j+\sum_{k \in \alpha_i} Y_k-\sum_{k \in \beta_i} Y_k, \quad \alpha_i \cup \beta_i=\{1,2,\ldots,2^{L+1}\}, \, \alpha_i \cap \beta_i=\emptyset, \, \beta_i \neq \emptyset
\end{aligned}
\end{equation}
and the DE takes the following form:
\begin{equation}
\begin{aligned}
    \rmd \psi_{\text{banana}} & = -(q_1 \, \rmd \! \log X_1^+ + q_2 \, \rmd \! \log X_2^+) \,  \psi_{\text{banana}} + \sum_{i,j} (-1)^{|\alpha_i|} \rmd \! \log \left(\frac{\omega_i^{(j)}}{X_j^+}\right)\mathbf{B}^{(j)}(\omega_i^{(j)})\\[1.1em]
    \rmd \mathbf{B}^{(1)}(\omega_i^{(1)}) & = -(q_1 \, \rmd \! \log \omega_i^{(1)}+q_2 \, \rmd \! \log X_2^+) \, \mathbf{B}^{(1)}(\omega_i^{(1)})+ \rmd \! \log \left(\frac{\omega_i^{(1)}+X_2^{+}}{X_2^+}\right) \, \mathbf{C} (\omega_i^{(1)}+X_2^{+})\\
    \rmd \mathbf{B}^{(2)}(\omega_i^{(2)}) & = -(q_1 \,  \rmd \! \log X_1^+ +q_2 \, \rmd \! \log \omega_i^{(2)}) \, \mathbf{B}^{(2)}(\omega_i^{(2)})+ \rmd \! \log \left(\frac{X_1^{+}+\omega_i^{(2)}}{X_1^+}\right) \, \mathbf{C} (X_1^{+}+\omega_i^{(2)})\\[1.1em]
    \rmd \mathbf{C}\left(\omega_i\right) & = -(q_1+q_2) \, \rmd \! \log \omega_ i  \, \mathbf{C}\left(\omega_i\right) 
\end{aligned}
\end{equation}
which can be easily normalized to become canonical. And it can be checked that when $L=0$ it goes back to \eqref{eq:2chainDE}.

For the two-site $L$-loop correlator, the structure is simpler. According to the discussion at the end of Sec.\ref{sec:decomp}, there are only $ 2\times 3 $ time integrals in the decomposition, and we only consider 3 of them, the other 3 terms are the complex conjugate. Thus, the structure of DE is expected to be very similar to tree-level ones \eqref{eq:2chainDE} and \eqref{eq:2chainCDE}. Explicitly, we choose the basis integrals as 
\begin{equation}
    \begin{aligned}
        \mathcal{T}_{\text{banana}} & =  \raisebox{-0.5em}{\begin{tikzpicture}
				\node[left] at (0,0) {\tiny{$1,+$}};
				\node[right] at (1,0) {\tiny{$2,+$}};
				\draw [line width=1.5pt,->] (0,0) to[out=60,in=-180] (0.6,0.25);
				\draw [line width=1.5pt] (0.5,0.25) to[out=0,in=120] (1,0);
				\draw [line width=1.5pt,->] (0,0) to[out=-60,in=-180] (0.6,-0.25);
				\draw [line width=1.5pt] (0.5,-0.25) to[out=0,in=-120] (1,0);
				\path[fill=black] (0,0) circle[radius=0.1];
				\path[fill=black] (1,0) circle[radius=0.1];
    \path[fill=black] (0.5,0.1) circle[radius=0.025];
				\path[fill=black] (0.5,0.0) circle[radius=0.025];
				\path[fill=black] (0.5,-0.1) circle[radius=0.025];
		\end{tikzpicture}} + \raisebox{-0.5em}{\begin{tikzpicture}
		\node[left] at (0,0) {\tiny{$1,+$}};
		\node[right] at (1,0) {\tiny{$2,+$}};
		\draw [line width=1.5pt] (0,0) to[out=60,in=-180] (0.5,0.25);
		\draw [line width=1.5pt,->] (1,0) to[out=120,in=0] (0.4,0.25);
		\draw [line width=1.5pt] (0,0) to[out=-60,in=-180] (0.5,-0.25);
		\draw [line width=1.5pt,->] (1,0) to[out=-120,in=0] (0.4,-0.25);
		\path[fill=black] (0,0) circle[radius=0.1];
		\path[fill=black] (1,0) circle[radius=0.1];
  \path[fill=black] (0.5,0.1) circle[radius=0.025];
				\path[fill=black] (0.5,0.0) circle[radius=0.025];
				\path[fill=black] (0.5,-0.1) circle[radius=0.025];
	\end{tikzpicture}}+ \raisebox{-0.5em}{\begin{tikzpicture}
	\node[left] at (0,0) {\tiny{$1,+$}};
	\node[right] at (1,0) {\tiny{$2,-$}};
	\draw [line width=1.5pt,dashed] (0,0) to[out=60,in=-180] (0.6,0.25);
	\draw [line width=1.5pt,dashed] (0.5,0.25) to[out=0,in=120] (1,0);
	\draw [line width=1.5pt,dashed] (0,0) to[out=-60,in=-180] (0.5,-0.25);
	\draw [line width=1.5pt,dashed] (1,0) to[out=-120,in=0] (0.4,-0.25);
	\path[fill=black] (0,0) circle[radius=0.1];
	\path[fill=black] (1,0) circle[radius=0.1];
 \path[fill=black] (0.5,0.1) circle[radius=0.025];
				\path[fill=black] (0.5,0.0) circle[radius=0.025];
				\path[fill=black] (0.5,-0.1) circle[radius=0.025];
\end{tikzpicture}}\\[0.5em]
\mathbf{B}_1 & = -q_1 \raisebox{-0.5em}{\begin{tikzpicture}
		\node[left] at (0,0) {\tiny{$1,+$}};
		\node[right] at (1,0) {\tiny{$2,+$}};
		\draw [line width=1.5pt] (0,0) to[out=60,in=-180] (0.5,0.25);
		\draw [line width=1.5pt,->] (1,0) to[out=120,in=0] (0.4,0.25);
		\draw [line width=1.5pt] (0,0) to[out=-60,in=-180] (0.5,-0.25);
		\draw [line width=1.5pt,->] (1,0) to[out=-120,in=0] (0.4,-0.25);
		\path[fill=black] (0,0) circle[radius=0.1];
		\path[fill=black] (1,0) circle[radius=0.1];
  \path[fill=black] (0.5,0.1) circle[radius=0.025];
				\path[fill=black] (0.5,0.0) circle[radius=0.025];
				\path[fill=black] (0.5,-0.1) circle[radius=0.025];
	\end{tikzpicture}} - \ii \raisebox{-11pt}{\begin{tikzpicture}
		\coordinate (X1) at (0,0);
		\node[below] at (X1) {\tiny{$X_1+X_2,\ q_{1,2}$}};
		\path[fill=black] (X1) circle[radius=0.1];
\end{tikzpicture}}\\[0.5em]
\mathbf{B}_2 & = -q_2 \raisebox{-0.5em}{\begin{tikzpicture}
				\node[left] at (0,0) {\tiny{$1,+$}};
				\node[right] at (1,0) {\tiny{$2,+$}};
				\draw [line width=1.5pt,->] (0,0) to[out=60,in=-180] (0.6,0.25);
				\draw [line width=1.5pt] (0.5,0.25) to[out=0,in=120] (1,0);
				\draw [line width=1.5pt,->] (0,0) to[out=-60,in=-180] (0.6,-0.25);
				\draw [line width=1.5pt] (0.5,-0.25) to[out=0,in=-120] (1,0);
				\path[fill=black] (0,0) circle[radius=0.1];
				\path[fill=black] (1,0) circle[radius=0.1];
    \path[fill=black] (0.5,0.1) circle[radius=0.025];
				\path[fill=black] (0.5,0.0) circle[radius=0.025];
				\path[fill=black] (0.5,-0.1) circle[radius=0.025];
		\end{tikzpicture}}- \ii \raisebox{-11pt}{\begin{tikzpicture}
		\coordinate (X1) at (0,0);
		\node[below] at (X1) {\tiny{$X_1+X_2,\ q_{1,2}$}};
		\path[fill=black] (X1) circle[radius=0.1];
\end{tikzpicture}}\\[0.5em]
\mathbf{C} & = \ii (q_1+q_2)\raisebox{-11pt}{\begin{tikzpicture}
		\coordinate (X1) at (0,0);
		\node[below] at (X1) {\tiny{$X_1+X_2,\ q_{1,2}$}};
		\path[fill=black] (X1) circle[radius=0.1];
\end{tikzpicture}}
    \end{aligned}
\end{equation}
then the DE takes the same form as \eqref{eq:2chainDE} and \eqref{eq:2chainCDE}, with the letters redefined as 
\begin{equation}\label{eq:ldef}
\begin{aligned}
    & \ell_1=\rmd \! \log (X_1+\sum_{j=1}^{L+1}Y_j), & \ell_2=\rmd \! \log (X_2+\sum_{j=1}^{L+1}Y_j),\ \ \   & \ell_3=\rmd \! \log (X_1-\sum_{j=1}^{L+1}Y_j),\\
    & \ell_4=\rmd \! \log (X_2-\sum_{j=1}^{L+1}Y_j), & \ell_5=\rmd \! \log (X_1+X_2). & 
\end{aligned}
\end{equation}

\paragraph{Three-site chain wavefunction} For wavefunction of three-site chain $ \psi_{\text{3-chain}} $, a new feature appears at the differential equation of the 2nd level integrals, where the integral basis is not complete after extended by the coefficients of newly appeared letters. Explicitly, there is 5 integrals $ \mathbf{B}_i $, $ i=1,2,\ldots,5 $, in the 2nd level of differential equation. Taking the differential of these 5 integrals, we would find 4 new letters appearing, whose coefficients $ \mathbf{C}_j $, $ j=1,\ldots,4 $ are considered as new members of the integral basis. But we could not express the differentials $ \rmd \mathbf{B}_i $ by $\{ \psi_{\text{3-chain}}, \mathbf{B}_1,\ldots,\mathbf{B}_5,\mathbf{C}_1,\ldots, \mathbf{C}_4 \}$. 
Then we search for new time integrals in the coefficients of previously appearing letters. We choose 
\begin{equation}
\begin{aligned}
    \mathbf{C}_5=\frac{\rmd \mathbf{B}_4}{\rmd (X_2+Y_1+Y_2)},\\
    \mathbf{C}_6=\frac{\rmd \mathbf{B}_4}{\rmd (X_1-Y_1)},\\
    \mathbf{C}_7=\frac{\rmd \mathbf{B}_5}{\rmd (X_3-Y_2)}
\end{aligned}
\end{equation}
to be the new integrals and the differential equation can be recast as (the equations are lengthy, we only present a simple one)
\begin{equation}
\begin{aligned}
\rmd \mathbf{B}_3 = \rmd \! \log (X_1+Y_1) (-q_1 \mathbf{B}_3-\mathbf{C}_3)+\rmd \! \log (X_3+Y_2) (-q_3 \mathbf{B}_3-\mathbf{C}_1)\\
+\rmd \! \log (X_1+X_2-Y_2) \mathbf{C}_3 + \rmd \! \log (X_2+X_3-Y_1) \mathbf{C}_1 - q_2 \rmd \! \log (X_2-Y_1-Y_2) \mathbf{B}_3
\end{aligned}
\end{equation}
Recursively, this procedure will truncate at the 4th level, where the differential equation system closes. Each level have $\{1,5,7,3\}$ integral(s) respectively. Similarly, normalize every time integral in the basis by product of $(-q_i)$'s they involve, we have a CDE system.

Compared to CDE in \cite{Arkani-Hamed:2023kig}, we find that our CDE system turns out to be the same, after taking $q_i \rightarrow \epsilon$ and applying a linear transformation to our basis:
\begin{equation}
    \mathbf{U}_{\text{trans}} = \left(
\begin{array}{cccccccccccccccc}
 1 & 0 & 0 & 0 & 0 & 0 & 0 & 0 & 0 & 0 & 0 & 0 & 0 & 0 & 0 & 0 \\
 0 & 1 & 0 & 0 & 0 & 0 & 0 & 0 & 0 & 0 & 0 & 0 & 0 & 0 & 0 & 0 \\
 0 & 0 & 1 & 0 & 0 & 0 & 0 & 0 & 0 & 0 & 0 & 0 & 0 & 0 & 0 & 0 \\
 0 & 0 & 0 & 0 & 0 & 1 & 0 & 0 & 0 & 0 & 0 & 0 & 0 & 0 & 0 & 0 \\
 0 & 0 & 0 & 1 & 0 & 0 & 0 & 0 & 0 & 0 & 0 & 0 & 0 & 0 & 0 & 0 \\
 0 & 0 & 0 & 0 & 1 & 0 & 0 & 0 & 0 & 0 & 0 & 0 & 0 & 0 & 0 & 0 \\
 0 & 0 & 0 & 0 & 0 & 0 & 0 & 0 & 0 & 0 & 1 & 0 & 0 & 0 & 0 & 0 \\
 0 & 0 & 0 & 0 & 0 & 0 & 0 & 0 & 0 & 1 & 0 & 0 & 0 & 0 & 0 & 0 \\
 0 & 0 & 0 & 0 & 0 & 0 & 0 & 0 & 1 & 0 & 0 & 0 & 0 & 0 & 0 & 0 \\
 0 & 0 & 0 & 0 & 0 & 0 & 0 & 0 & -1 & 0 & 0 & 1 & 0 & 0 & 0 & 0 \\
 0 & 0 & 0 & 0 & 0 & 0 & -1 & 0 & 0 & 0 & 0 & 0 & 1 & 0 & 0 & 0 \\
 0 & 0 & 0 & 0 & 0 & 0 & 1 & 0 & 0 & 0 & 0 & 0 & 0 & 0 & 0 & 0 \\
 0 & 0 & 0 & 0 & 0 & 0 & 0 & 1 & 0 & 0 & 0 & 0 & 0 & 0 & 0 & 0 \\
 0 & 0 & 0 & 0 & 0 & 0 & 0 & 0 & 0 & 0 & 0 & 0 & 0 & 0 & 0 & 1 \\
 0 & 0 & 0 & 0 & 0 & 0 & 0 & 0 & 0 & 0 & 0 & 0 & 0 & 1 & 0 & 0 \\
 0 & 0 & 0 & 0 & 0 & 0 & 0 & 0 & 0 & 0 & 0 & 0 & 0 & 0 & 1 & 0 \\
\end{array}
\right)
\end{equation}
We also calculated DEs for four-chain and four-star, and found consistency with those in \cite{Arkani-Hamed:2023kig}, once proper normalization and linear transformation are employed.

\paragraph{$n$-gon wavefunction}For more general graphs, the procedure also applies. We further calculated the DE system for \{3,4,5\}-gon, and obtain \{50, 226, 962\} basis integrals respectively. The naive counting for the number of basis integral is $4^n$, which is greater than the actual numbers by \{14, 30, 62\} respectively for \{3,4,5\}-gon. These numbers can be interpreted as the counting of all graphs with incompatible time ordering among total $4^n$ ones. For example, all incompatible time orderings of clockwise type for 3- and 4-gon are as follows:
\begin{equation}\label{eq:3gon}
\begin{gathered}
		\underbrace{\raisebox{-2em}{
}}_{4=\text{C}_4^3 \text{ terms}}
\end{gathered}
\end{equation}
where the shade means contraction of edges. Generally for 1-loop $n$-gon wavefunction, incompatible time ordering corresponds to directed cycles (clockwise and anti-clockwise). Then it suffices to consider the two $n$-cycles (such as the first term in \eqref{eq:3gon} and \eqref{eq:4gon}) and all their contractions. The counting for incompatible time ordering is then
\begin{equation}
    2 \times \sum_{k=0}^{n-1} \text{C}_n^k= 2 \,(2^n{-}1),
\end{equation}
Note that the all-edge contraction is compatible, so we exclude the $\text{C}_n^n$ term. Thus we conclude that for general $n$-gon wavefunctions,  the number of basis integrals is $[4^{n}{-}2\,(2^n{-}1)]$.

\section{Back to de Sitter: truncated differential equations and symbology}\label{sec:dS}
In previous sections, we kept the twists $q_i$ to be generic. However, we can also consider some particular choices of $q_i$, which are both interesting in physics and mathematics. In this subsection, we study the wavefunction coefficients at $q_i\to0$ limit under $\phi^3$ interaction, which corresponds to de Sitter spacetime as we have explained in review part, and all the special functions become multi-polylogarithm (MPL). We will see that it is straightforward to take the dS limit and truncate our differential equations at $q_i=0$, and we will study the MPL symbol results for some special wavefunction integrals following the truncated differential equations\footnote{It is also possible to truncate the DE system (before canonical normalization) in the dS limit. Then to calculate the symbol is simply to multiply the derivative matrix several times. Notice the derivative matrix corresponding to the last level integrals will have all zero entries in the dS limit, the symbol will thus eventually depend only on the boundary condition of the last level, which is easy to calculate. However, this procedure is quite straightforward and not very illuminating, so we suppress the discussion here.}. 

To take the limit, the main subtlety is that although the wavefunction in dS limit is a finite object, each integral \eqref{eq:Tdef} will probably diverge under $q_i\to0$. For instance, at $q_1\to0$ limit, a single-point diagram reads
\begin{equation}\label{eq:sgexp}
    \raisebox{-10pt}{\begin{tikzpicture}
		\coordinate (X1) at (0,0);
		\node[below] at (X1) {\tiny{$1$}};
		\path[fill=black] (X1) circle[radius=0.1];
\end{tikzpicture}}=\frac{(-\ii)}{(\ii\,\omega_1)^{q_1}}\Gamma(q_1)=-\frac{\ii}{q_1}+\ii\log{\omega_1}+\mathcal{O}(q_1),
\end{equation}
where we have omitted the Euler's constant $\gamma_E$ and $\pi$ appearing in the finite part. However, after combining basic time integrals to get wavefunction coefficients, all the divergence must cancel, and we will have a finite truncated differential equation system for the wavefunction integral.

From the last subsection, we see that after adding basic time integrals together, the total differential for an $n$-site $e$-edge wavefunction coefficient integral, with $X_j$ as sum of energies from bulk-to-boundary propagators \eqref{eq:vertexwf} attaching vertex $j$, and $Y_{i,j}$ as the energies along the bulk-to-bulk propagators \eqref{eq:wavefuncprop} between vertices $\{i,j\}$, can be expressed generally as\footnote{Although the following discussion is  very general, explicit examples of $\psi_n$ considered in this section should only have vertices with up to $3$ bulk-to-bulk propagators attached, corresponding to only $\phi^3$ interaction. For vertices with $3$ bulk-to-bulk propagators, no more bulk-to-boundary propagators can be attached, and $X_i{=}0$.} 
\begin{equation}
\begin{aligned}\label{eq:dpsic}
    & \rmd \psi_n(X_1,\cdots,X_n;Y_{i,j})\\
    & \hspace{3em}{=}\sum_{j=1}^n\sum_{k}\rmd \! \log{X_j^{\bf a_{j,k}}}\sum_{\mathbf{T}}\biggl(-q_j\mathbf{T}^{q_1,q_2,\ldots,q_n}(\omega_1,\ldots,\omega_{j-1},X_j^{\bf a_{j,k}},\omega_{j+1},\ldots,\omega_n)\\
    & \hspace{4em} {+}\ii\sum_{m\neq j}\mathrm{sgn}^{j,m}_\mathbf{T}\,\mathbf{T}^{q_1,q_2,\ldots,q_{j,m},\ldots,q_n}(\omega_1,\ldots,\omega_{j-1},X_j^{\bf a_{j,k}}{+}\omega_m,\omega_{j+1},\ldots,\omega_n)\biggr).
\end{aligned}
\end{equation}
The first sum is over all vertices. The second sum is over all letters $X_j^{\bf a_{j,k}}=X_j+{\bf a_{j,k}}\cdot {\bf Y}_{j}$, where ${\bf a_{j,k}}$ are vectors with $\{0,1,-1\}$ as components,  and ${\bf Y_i}=(Y_{i,1},Y_{i,2},\cdots)$ are vectors for all energies along propagators between vertex $i$ and others. Finally, the summation over $\mathbf{T}$ is to sum over all $\mathbf{T}^{q_1,q_2,\ldots,q_n}(\omega_1,\ldots,\omega_{j-1},X_j^{\bf a_{j,k}},\omega_{j+1},\omega_n)$ with the energy of the vertex $j$ being $X_j^{\bf a_{j,k}}$, together with the contact terms associated after contracting any of its edges connecting $j$. In another word, we add up the total differential for all components of basic time integrals to get the total differential of the whole $\psi_n$ and collect the same ${\rm d}\log$ forms among all these terms in \eqref{eq:dpsic}. 

Now we want to take the de Sitter limit. To make our discussion more general, we temporarily do not restrict to the special case $d=k=3$, and still assume $q_i$ to be generic twist variables. Firstly, we consider limit $q_j\to0$ and keep all $q_k$ with $k\neq j$ generic in $j$th summand. Although $q_j$ only appears in terms like the first line of \eqref{eq:dpsic} superficially, we cannot just get rid of the terms in the first line and obtain the truncated differential equation, since $\mathbf{T}$ itself is likely divergent in this limit. However, other contracted terms in the second line are finite under the limit. Therefore, to obtain the finite result of the RHS in \eqref{eq:dpsic} when $q_j\to0$, the only thing we need to do is carefully extracting the $\mathcal{O}\left(\frac{1}{q_j}\right)$ divergence of the term $\mathbf{T}^{q_1,q_2,\ldots,q_n}(\omega_1,\cdots,X_j^{\bf a_{j,k}},\cdots,\omega_n)$ in \eqref{eq:dpsic}, and we will show that $\mathbf{T}^{q_1,q_2,\ldots,q_n}(\omega_1,\cdots,X_j^{\bf a_{j,k}},\cdots,\omega_n)$ is only of $\mathcal{O}\left(\frac{1}{q_j}\right)$ divergence. Moreover, recall that $\psi_n$ itself is finite in de Sitter limit, so are its coproducts in the summation $\sum_{\mathbf{T}}$. Adding up this term with all contracted terms in the summation $\sum_{\mathbf{T}}$, due to the finiteness of ${\rm d}\psi_n$ and its coproduct, we can freely set all other $q_i\to0$ as well, and automatically get the (first-level) truncated differential equation for $\psi_n$.

Now let us compute the divergence. when $j$ is not a source, {\it i.e.} $j$ is at least later than one of the vertices it relates, we can use the recursive relation \eqref{eq:Trecursion} to extract its leading term directly: We firstly express $\mathbf{T}^{q_1,q_2,\ldots,q_n}(\omega_1,\cdots,X_j^{\bf a_{j,k}},\cdots,\omega_n)$ as a one-fold integration following \eqref{eq:Trecursion} as
\begin{align}
\mathbf{T}^{q_1,q_2,\ldots,q_n}_\mathcal{N}(&\omega_1,\ldots,X_j^{\bf a_{j,k}},\ldots,\omega_n)\nonumber\\
&=\ii\int_{0}^{1}\rmd \alpha \, \alpha^{q_k-1}\sum_{j\neq k}\mathrm{sgn}^{j,k}_\mathcal{N}\,\mathbf{T}^{q_1,q_2,\ldots,q_{j,k},\ldots,q_n}_{k}(\omega_1,\omega_2,\ldots,\omega_{j}+X_j^{\bf a_{j,k}}\alpha\ldots,\omega_n),
\end{align}
where $k$ is over all other vertices except $j$. As a function of integration variable $t$, each contact term in the integrand can be expanded as
\begin{equation} \mathbf{T}^{q_1,q_2,\ldots,q_{j,k},\ldots,q_n}_{k}(\omega_1,\ldots,\omega_{k-1},\omega_k+X_j^{\bf a_{j,k}}\alpha,\omega_{k+1},\ldots,\omega_n)=\sum_{m=0}^\infty a_{m,k}\left(X_j^{\bf a_{j,k}}\right)^{m} \, \alpha^m.
\end{equation}
 The summation start from $m=0$ since the time integral is finite for generic $q_i$ where $i\neq j$ and $a_{0,k}=\mathbf{T}^{q_1,q_2,\ldots,q_{j,k},\ldots,q_n}_{k}(\omega_1,\ldots,\omega_{k-1},\omega_k,\omega_{k+1},\ldots,\omega_n)$. Adding together all terms and  performing the integration in \eqref{eq:Trecursion}, 
\begin{equation}
    \sum_{k\neq j}\sum_{m=0}^\infty \mathrm{sgn}^{j,k}_\mathcal{N}a_{m,k}\left(X_j^{\bf a_{j,k}}\right)^{m}\,\ii\int_{0}^{1}\rmd \alpha  \, \alpha^{q_j+m-1}=\sum_{k\neq j}\sum_{m=0}^\infty\mathrm{sgn}^{j,k}_\mathcal{N}\frac{\ii \, a_{m,k}}{q_j+m}(X_j^{\bf a_{j,k}})^m,
\end{equation}
we can see that at $q_i\to 0$, only $m=0$ term contributes to the $\mathcal{O}\left(\frac{1}{q_j}\right)$ divergence and all the $m\neq0$ terms remain finite. Thus, $\mathbf{T}^{q_1,q_2,\ldots,q_n}_\mathcal{N}(\omega_1,\ldots,X_j^{\bf a_{j,k}},\ldots,\omega_n)$ is indeed of $\mathcal{O}\left(\frac{1}{q_j}\right)$ divergence, and the divergence can be obtained from summing over all $a_{0,k}$ with proper signatures, {\it i.e.} summing over all graphs where the links involving vertex $j$ and $k$ are contracted as
\begin{tcolorbox}
\begin{equation}\label{eq:gq} 
q_j \raisebox{-6em}{
}\!+\mathcal{O}(q_j),
\end{equation}
\end{tcolorbox}
The graphs $G_i$ can be either connected or disconnected directed graph. For example, if we go back to the chain graphs
\begin{equation}\label{eq:dsc1}
    \begin{aligned}
        & q_j \;\raisebox{-1.2em}{%
} +\mathcal{O}(q_j)
    \end{aligned},
\end{equation}
where in the last equation, we used $\theta_{i,j}+\theta_{j,i}=1$.

When $j$ is a source, \eqref{eq:Trecursion} can not be used directly. However, we can use the identity $\theta_{i,j}+\theta_{j,i}=1$ to rewrite it as a summation over graphs where $j$ is not a source
\begin{equation}\label{eq:ps}
	\raisebox{-4.5em}{
%
}
\end{equation}
All the $\mathcal{O}\left(\frac{1}{q_j}\right)$ divergence terms are given in \eqref{eq:gq} and \eqref{eq:sgexp}. For example, if we go to the chain graphs again, we can see that there is always no $\mathcal{O}\left(\frac{1}{q_j}\right)$ divergence when $j$ is a source,
\begin{equation}\label{eq:dsc2}
\begin{aligned}
& \begin{aligned}
q_j \, \raisebox{-1.2em}{%
}+\mathcal{O}(q_j)=\mathcal{O}(q_j)
	\end{aligned}
 \end{aligned}
\end{equation}
where in the computation, we used the fact \eqref{eq:sgexp} that the $\mathcal{O}(\frac{1}{q_j})$ divergence of the single vertex $j$ is $-\frac{\ii}{q_j}$. However, if we consider sources with more than two propagators attached, then they contribute non-trivial $O\left(\frac1{q_j}\right)$ divergence furthermore.

In summary, after taking the $\mathcal{O}(\frac{1}{q_j})$ leading divergent terms on the first line of \eqref{eq:dpsic} and sending $q_j\to0$ in $j$th summand,  basic time integrals on the right-hand side of \eqref{eq:dpsic} are of $n-1$ sites. As we have explained, after collecting independent ${\rm d}\log$-forms, each $n{-}1$ coproduct in $\sum_{\mathbf{T}}$ is finite in dS limit again, like $\psi_n$ itself. We can therefore treat the total differential equation relation as in dS spacetime, {\it i.e.} (first-level) truncated differential equation for $\psi_n$, by just simply setting other $q_k=0$ as well. Furthermore, we can take the total differential of $(n{-}1)$-site integrals recursively and finally get the truncated differential equation system of an $n$-site integral under the dS limit, from which we can get the symbol/function results very easily. 

In the following, to illustrate the truncation procedure, we discuss two kinds of special wavefunction coefficient integrals and their symbol in dS limit, {\it i.e.} n-site chain and n-gon integrals.

\paragraph{$n$-chain graphs:}
As a warm-up, by using \eqref{eq:dsc1} and \eqref{eq:dsc2}, we can obtain the truncated differential equation from \eqref{eq:d2chain} for the finite part of the $2$-site chain wavefunction directly as
\begin{equation}
\begin{aligned}
    \rmd\psi_{\text{2-chain}}&= \ii \, \rmd\!\log{\frac{X_1-Y}{X_1+Y}}\left(\raisebox{-10pt}{\begin{tikzpicture}
		\coordinate (X1) at (0,0);
		\node[below] at (X1) {\tiny{$X_2^+$}};
		\path[fill=black] (X1) circle[radius=0.1];
\end{tikzpicture}}-\raisebox{-10pt}{\begin{tikzpicture}
		\coordinate (X1) at (0,0);
		\node[below] at (X1) {\tiny{$X_{12}^{\textcolor{white}{+}}$}};
		\path[fill=black] (X1) circle[radius=0.1];
\end{tikzpicture}}\right)+\ii \, \rmd\!\log{\frac{X_2-Y}{X_2+Y}}\left(\raisebox{-10pt}{\begin{tikzpicture}
		\coordinate (X1) at (0,0);
		\node[below] at (X1) {\tiny{$X_1^+$}};
		\path[fill=black] (X1) circle[radius=0.1];
\end{tikzpicture}}-\raisebox{-10pt}{\begin{tikzpicture}
		\coordinate (X1) at (0,0);
		\node[below] at (X1) {\tiny{$X_{12}^{\textcolor{white}{+}}$}};
		\path[fill=black] (X1) circle[radius=0.1];
\end{tikzpicture}}\right)\\[1em]
    &=\rmd\!\log{\frac{X_1-Y}{X_1+Y}}\log{\frac{X_{1}+X_2}{X_2+Y}}+\rmd\!\log{\frac{X_2-Y}{X_2+Y}}\log{\frac{X_{1}+X_2}{X_1+Y}}.
\end{aligned}
\end{equation}
which exactly meets the correct symbol result for the 2-site wavefunction coefficient under the dS limit, and the function result can also be very easily obtained after integration. More importantly, we also have a diagrammatic interpretation for all coproducts in the result.

For an $n$-site chain diagram, we have $n$ energies $X_i$ on the nodes and $n{-}1$ energies $Y_{i,i{+}1}$ on the edges. After taking the $q_i\to0$ limit by \eqref{eq:dsc1} and \eqref{eq:dsc2}, total differential of an $n$-site chain graph can be organized as the following
\begin{equation}
\begin{aligned}
    \rmd \psi_{\text{n-chain}}&=-\ii \, \rmd\!\log{\frac{X_1^-}{X_1^+}}(D_1+C_1)\\
    &+\!\ii \!\sum_{j=2}^{n-1} \!\!\left(\!-\rmd\!\log{\frac{X_{j}^{--}}{X_{j}^{++}}}(D_j\!+\!C_j^{--})+\rmd\!\log{\frac{X_{j}^{+-}}{X_{j}^{++}}}(D_j\!+\!C_j^{+-})+\rmd\!\log{\frac{X_{j}^{-+}}{X_{j}^{++}}}(D_j\!+\!C_j^{-+})\right)\\
    &-\ii \, \rmd\!\log{\frac{X_n^-}{X_n^+}}(D_n+C_n)
\end{aligned}
\end{equation}
where $X_1^\pm=X_1\pm Y_{1,2}$, $X_n^\pm=X_n\pm Y_{n{-}1,n}$, $X_j^{\pm\pm}=X_j\pm Y_{j{-}1,j}\pm Y_{j,j{+}1}$, and
\begin{equation}\label{eq:CD}
    \begin{aligned}
    &D_1=-\sum_{G}\raisebox{-1em}{\begin{tikzpicture}
		\draw[fill=black!10] (0,0) circle [radius=0.5];
		\path[fill=black] (-0.5,0) circle[radius=0.1];
		\draw[line width=1.5pt] (-0.5,0)--(-1.25,0);
		\path[fill=black] (-1.25,0) circle[radius=0.1];
		\node at (0,0) {\small{$G$}};
		\node[below] at (-1.25,0) {\tiny{$X_2+Y_{1,2}$}};
	\end{tikzpicture}},\;\; \, \, \,  C_1=\sum_{G}\raisebox{-1em}{\begin{tikzpicture}
		\draw[fill=black!10] (0,0) circle [radius=0.5];
		\path[fill=black] (-0.5,0) circle[radius=0.1];
		\draw[line width=1.5pt] (-0.5,0)--(-1.25,0);
		\path[fill=black] (-1.25,0) circle[radius=0.1];
		\node at (0,0) {\small{$G$}};
		\node[below] at (-1.25,0) {\tiny{$X_1+X_2$}};
	\end{tikzpicture}},\\
  & D_n=-\sum_{G}\raisebox{-1em}{\begin{tikzpicture}
		\draw[fill=black!10] (0,0) circle [radius=0.5];
		\path[fill=black] (0.5,0) circle[radius=0.1];
		\draw[line width=1.5pt] (0.5,0)--(1.25,0);
		\path[fill=black] (1.25,0) circle[radius=0.1];
		\node at (0,0) {\small{$G$}};
		\node[below] at (1.25,0) {\tiny{$X_n+Y_{n{-}1,n}$}};
	\end{tikzpicture}},\;\;C_n=\sum_{G} \raisebox{-1em}{\begin{tikzpicture}
		\draw[fill=black!10] (0,0) circle [radius=0.5];
		\path[fill=black] (0.5,0) circle[radius=0.1];
		\draw[line width=1.5pt] (0.5,0)--(1.25,0);
		\path[fill=black] (1.25,0) circle[radius=0.1];
		\node at (0,0) {\small{$G$}};
		\node[below] at (1.25,0) {\tiny{$X_{n{-}1}+X_n$}};
	\end{tikzpicture}}, \, \, D_j=-\sum_{G_1,G_2} \raisebox{-1em}{\begin{tikzpicture};
		\draw[fill=black!10] (0,0) circle [radius=0.5];
		\draw[fill=black!10] (3,0) circle [radius=0.5];
		\path[fill=black] (0.5,0) circle[radius=0.1];
		\path[fill=black] (2.5,0) circle[radius=0.1];
		\draw[line width=1.5pt] (0.5,0)--(1.25,0);
		\draw[line width=1.5pt] (1.75,0)--(2.5,0);
		\draw[fill=black] (1.25,0) circle [radius=0.1];
		\draw[fill=black] (1.75,0) circle [radius=0.1];
		\node at (0,0) {\small{$G_1$}};
		\node at (3,0) {\small{$G_2$}};
		\node[above] at (1.25-0.05,0) {\tiny{$X_{j-1}{+}Y_{j{-}1,j}$}};
		\node[below] at (1.75+0.05,0) {\tiny{$X_{j+1}{+}Y_{j,j{+}1}$}};
	\end{tikzpicture}}\\
        &C_j^{--}=\!\!\!\sum_{G_1,G_2}\raisebox{-1em}{\begin{tikzpicture};
		\draw[fill=black!10] (0,0) circle [radius=0.5];
		\draw[fill=black!10] (3,0) circle [radius=0.5];
		\path[fill=black] (0.5,0) circle[radius=0.1];
		\path[fill=black] (2/3+0.5-0.2,0) circle[radius=0.1];
		\path[fill=black] (4/3+0.5+0.2,0) circle[radius=0.1];
		\path[fill=black] (2.5,0) circle[radius=0.1];
		\draw[line width=1.5pt] (0.5,0)--(4/3+0.5,0);
		\drawA{1.166667}{0}{1.833333}{0}
		\draw[line width=1.5pt] (0.5+4/3,0)--(2.5,0);
		\node at (0,0) {\small{$G_1$}};
		\node at (3,0) {\small{$G_2$}};
		\node[above] at (0.5+2/3,0.05) {\tiny{$X_{j-1}$}};
		\node[below] at (0.5+4/3+0.2,-0.05) {\tiny{$X_{j}{+}X_{j{+}1}$}};
	\end{tikzpicture}}+\raisebox{-1em}{\begin{tikzpicture};
		\draw[fill=black!10] (0,0) circle [radius=0.5];
		\draw[fill=black!10] (3,0) circle [radius=0.5];
		\path[fill=black] (0.5,0) circle[radius=0.1];
		\path[fill=black] (2/3+0.5-0.2,0) circle[radius=0.1];
		\path[fill=black] (4/3+0.5+0.2,0) circle[radius=0.1];
		\path[fill=black] (2.5,0) circle[radius=0.1];
		\draw[line width=1.5pt] (0.5,0)--(4/3+0.5,0);
		\drawA{1.833333}{0}{1.166667}{0}
		\draw[line width=1.5pt] (0.5+4/3,0)--(2.5,0);
		\node at (0,0) {\small{$G_1$}};
		\node at (3,0) {\small{$G_2$}};
		\node[above] at (0.5+2/3,0.05) {\tiny{$X_{j-1}{+}X_{j}$}};
		\node[below] at (0.5+4/3+0.2,-0.05) {\tiny{$X_{j+1}$}};
	\end{tikzpicture}}\\
        &C_j^{+-}=\!\!\!\sum_{G_1,G_2}\raisebox{-1em}{\begin{tikzpicture};
		\draw[fill=black!10] (0,0) circle [radius=0.5];
		\draw[fill=black!10] (3,0) circle [radius=0.5];
		\path[fill=black] (0.5,0) circle[radius=0.1];
		\path[fill=black] (2.5,0) circle[radius=0.1];
		\draw[line width=1.5pt] (0.5,0)--(1.25,0);
		\draw[line width=1.5pt] (1.75,0)--(2.5,0);
		\draw[fill=black] (1.25,0) circle [radius=0.1];
		\draw[fill=black] (1.75,0) circle [radius=0.1];
		\node at (0,0) {\small{$G_1$}};
		\node at (3,0) {\small{$G_2$}};
		\node[above] at (1.25-0.05,0) {\tiny{$X_{j-1}{+}Y_{j{-}1,j}$}};
		\node[below] at (1.75+0.05,0) {\tiny{$X_{j}{+}X_{j{+}1}{+}Y_{j{-}1,j}$}};
	\end{tikzpicture}}+\raisebox{-1em}{\begin{tikzpicture};
		\draw[fill=black!10] (0,0) circle [radius=0.5];
		\draw[fill=black!10] (3,0) circle [radius=0.5];
		\path[fill=black] (0.5,0) circle[radius=0.1];
		\path[fill=black] (2/3+0.5-0.2,0) circle[radius=0.1];
		\path[fill=black] (4/3+0.5+0.2,0) circle[radius=0.1];
		\path[fill=black] (2.5,0) circle[radius=0.1];
		\draw[line width=1.5pt] (0.5,0)--(4/3+0.5,0);
		\drawA{1.833333}{0}{1.166667}{0}
		\draw[line width=1.5pt] (0.5+4/3,0)--(2.5,0);
		\node at (0,0) {\small{$G_1$}};
		\node at (3,0) {\small{$G_2$}};
		\node[above] at (0.5+2/3,0.05) {\tiny{$X_{j{-}1}{+}X_{j}$}};
		\node[below] at (0.5+4/3,-0.05) {\tiny{$X_{j+1}$}};
	\end{tikzpicture}}-\raisebox{-1em}{\begin{tikzpicture};
		\draw[fill=black!10] (0,0) circle [radius=0.5];
		\draw[fill=black!10] (3,0) circle [radius=0.5];
		\path[fill=black] (0.5,0) circle[radius=0.1];
		\path[fill=black] (2/3+0.5-0.2,0) circle[radius=0.1];
		\path[fill=black] (4/3+0.5+0.2,0) circle[radius=0.1];
		\path[fill=black] (2.5,0) circle[radius=0.1];
		\draw[line width=1.5pt] (0.5,0)--(4/3+0.5,0);
		\drawA{1.833333}{0}{1.166667}{0}
		\draw[line width=1.5pt] (0.5+4/3,0)--(2.5,0);
		\node at (0,0) {\small{$G_1$}};
		\node at (3,0) {\small{$G_2$}};
		\node[above] at (0.5+2/3,0.05) {\tiny{$X_{j-1}$}};
		\node[below] at (0.5+4/3+0.2,-0.05) {\tiny{$X_{j}{+}X_{j{+}1}$}};
	\end{tikzpicture}}\\
        &C_j^{-+}=\!\!\!\sum_{G_1,G_2}\raisebox{-1em}{\begin{tikzpicture};
		\draw[fill=black!10] (0,0) circle [radius=0.5];
		\draw[fill=black!10] (3,0) circle [radius=0.5];
		\path[fill=black] (0.5,0) circle[radius=0.1];
		\path[fill=black] (2.5,0) circle[radius=0.1];
		\draw[line width=1.5pt] (0.5,0)--(1.25,0);
		\draw[line width=1.5pt] (1.75,0)--(2.5,0);
		\draw[fill=black] (1.25,0) circle [radius=0.1];
		\draw[fill=black] (1.75,0) circle [radius=0.1];
		\node at (0,0) {\small{$G_1$}};
		\node at (3,0) {\small{$G_2$}};
		\node[above] at (1.25-0.05,0.05) {\tiny{$X_{j{-}1}{+}X_{j}{+}Y_{j,j{+}1}$}};
		\node[below] at (1.75+0.05,0) {\tiny{$X_{j+1}{+}Y_{j,j{+}1}$}};
	\end{tikzpicture}}+\raisebox{-1em}{\begin{tikzpicture};
		\draw[fill=black!10] (0,0) circle [radius=0.5];
		\draw[fill=black!10] (3,0) circle [radius=0.5];
		\path[fill=black] (0.5,0) circle[radius=0.1];
		\path[fill=black] (2/3+0.5-0.2,0) circle[radius=0.1];
		\path[fill=black] (4/3+0.5+0.2,0) circle[radius=0.1];
		\path[fill=black] (2.5,0) circle[radius=0.1];
		\draw[line width=1.5pt] (0.5,0)--(4/3+0.5,0);
		\drawA{1.166667}{0}{1.833333}{0}
		\draw[line width=1.5pt] (0.5+4/3,0)--(2.5,0);
		\node at (0,0) {\small{$G_1$}};
		\node at (3,0) {\small{$G_2$}};
		\node[above] at (0.5+2/3,0.05) {\tiny{$X_{j-1}$}};
		\node[below] at (0.5+4/3+0.2,-0.05) {\tiny{$X_{j}{+}X_{j{+}1}$}};
	\end{tikzpicture}}-\raisebox{-1em}{\begin{tikzpicture};
		\draw[fill=black!10] (0,0) circle [radius=0.5];
		\draw[fill=black!10] (3,0) circle [radius=0.5];
		\path[fill=black] (0.5,0) circle[radius=0.1];
		\path[fill=black] (2/3+0.5-0.2,0) circle[radius=0.1];
		\path[fill=black] (4/3+0.5+0.2,0) circle[radius=0.1];
		\path[fill=black] (2.5,0) circle[radius=0.1];
		\draw[line width=1.5pt] (0.5,0)--(4/3+0.5,0);
		\drawA{1.166667}{0}{1.833333}{0}
		\draw[line width=1.5pt] (0.5+4/3,0)--(2.5,0);
		\node at (0,0) {\small{$G_1$}};
		\node at (3,0) {\small{$G_2$}};
		\node[above] at (0.5+2/3,0.05) {\tiny{$X_{j{-}1}{+}X_j$}};
		\node[below] at (0.5+4/3+0.2,-0.05) {\tiny{$X_{j+1}$}};
	\end{tikzpicture}}
    \end{aligned},
\end{equation}
In the expression, $D_j$ are contributed from the $\mathcal{O}\left(\frac{1}{q_j}\right)$ divergent terms, 
and $C_j^{a,b}$ are contributed from the contact terms. 
Notice that in \eqref{eq:CD}, these blocks are related to some shifted $(n-1)$-site chain integrals as
\begin{equation}\label{eq:DCpsi}
    \begin{aligned}
        &D_1=-\left.\psi_{n-1}\right|_{X_2\to X_2+Y_{1,2}},\;\;
        C_1=\left.\psi_{n-1}\right|_{X_2\to X_{1}{+}X_2}\\
        &D_n=-\left.\psi_{n{-}1}\right|_{X_{n-1}\to X_{n{-}1}+Y_{n{-}1,n}},\;\;
        C_n=\left.\psi_{n-1}\right|_{X_{n-1}\to X_{n{-}1}{+}X_{n}}\\
        &C_j^{--}-C_j^{+-}=\left.\psi_{n-1}\right|_{X_{j+1}\to X_{j}{+}X_{j{+}1}},\;\;
        C_j^{--}-C_j^{-+}=\left.\psi_{n-1}\right|_{X_{j-1}\to X_{j-1}{+}X_{j}}\\
        &D_j+C_j^{--}=\left.\psi_{n-1}\right|_{X_{j{-}1}\to \tilde{X}_{j{-}1},X_{j}\to \tilde{X}_{j{+}1},Y_{j{-}1,j}\to \tilde{Y}_{j{-}1,j{+}1},X_i\to X_{i{+}1},\ \text{for}\ i>j }.
    \end{aligned}
\end{equation}
with the definition
\begin{align}\label{tildeXY}    
\tilde{X}_{j{-}1}{=}X_{j{-}1}{+}\frac12(X_j&{+}Y_{j{-}1,j}{-}Y_{j,j{+}1}),\ \tilde{X}_{j{+}1}{=}X_{j{+}1}{+}\frac12(X_j{-}Y_{j{-}1,j}{+}Y_{j,j{+}1})\nonumber\\
&\tilde{Y}_{j{-}1,j{+}1}=\frac12({-}X_j{+}Y_{j{-}1,j}{+}Y_{j,j{+}1})
\end{align}
Therefore, all the coproducts can be converted to shifted $n{-}1$ site chains and graphically we have the following expression
\begin{equation}
\begin{aligned}
\rmd\psi_{\text{n-chain}}&
=-\ii \, \rmd\!\log{\frac{X_1^-}{X_1^+}}\left({-}\raisebox{-1em}{\begin{tikzpicture}
		\draw[line width=1.5pt] (0,0)--(1,0);
		\draw[line width=1.5pt] (1,0)--(1.5,0);
		\draw[line width=1.5pt] (2,0)--(2.5,0);
		\draw[fill=black] (0,0) circle [radius=0.1];
		\draw[fill=black] (1,0) circle [radius=0.1];
		\draw[fill=black] (2.5,0) circle [radius=0.1];
		\node[below] at (0,0) {\tiny{$X_2+Y_1$}};
		\node[below] at (1,-0.01) {\tiny{$3$}};
		\node[below] at (2.5,-0.01) {\tiny{$n$}};
		\draw[fill=black] (1.5+0.5/4,0) circle [radius=0.025];
		\draw[fill=black] (1.5+2*0.5/4,0) circle [radius=0.025];
		\draw[fill=black] (1.5+3*0.5/4,0) circle [radius=0.025];
	\end{tikzpicture}}+\raisebox{-1em}{\begin{tikzpicture}
		\draw[line width=1.5pt] (0,0)--(1,0);
		\draw[line width=1.5pt] (1,0)--(1.5,0);
		\draw[line width=1.5pt] (2,0)--(2.5,0);
		\draw[fill=black] (0,0) circle [radius=0.1];
		\draw[fill=black] (1,0) circle [radius=0.1];
		\draw[fill=black] (2.5,0) circle [radius=0.1];
		\node[below] at (0,0) {\tiny{$X_{1}{+}X_{2}$}};
		\node[below] at (1,-0.01) {\tiny{$3$}};
		\node[below] at (2.5,-0.01) {\tiny{$n$}};
		\draw[fill=black] (1.5+0.5/4,0) circle [radius=0.025];
		\draw[fill=black] (1.5+2*0.5/4,0) circle [radius=0.025];
		\draw[fill=black] (1.5+3*0.5/4,0) circle [radius=0.025];
	\end{tikzpicture}}\right)\\[0.75em]
&+\ii \sum_{j=2}^{n-1}\left( \rmd\!\log{\frac{X_j^{+-}X_j^{-+}}{X_j^{++}X_j^{--}}} \raisebox{-1em}{\begin{tikzpicture}[scale=1.3]
		\draw[line width=1.5pt] (0,0)--(0.5,0);
		\draw[line width=1.5pt] (1,0)--(3,0);
		\draw[line width=1.5pt] (3.5,0)--(4,0);
		\draw[fill=black] (0,0) circle [radius=0.075];
		\draw[fill=black] (1.5-0.2,0) circle [radius=0.075];
		\draw[fill=black] (2.5+0.2,0) circle [radius=0.075];
		\draw[fill=black] (4,0) circle [radius=0.075];
		\draw[fill=black] (0.5+0.5/4,0) circle [radius=0.02];
		\draw[fill=black] (0.5+2*0.5/4,0) circle [radius=0.02];
		\draw[fill=black] (0.5+3*0.5/4,0) circle [radius=0.02];
		\draw[fill=black] (3+0.5/4,0) circle [radius=0.02];
		\draw[fill=black] (3+2*0.5/4,0) circle [radius=0.02];
		\draw[fill=black] (3+3*0.5/4,0) circle [radius=0.02];
		\node[below] at (0,-0.01) {\tiny{$1$}};
		\node[below] at (4,-0.01) {\tiny{$n$}};
		\node[below] at (1.5-0.25,-0.01) {\tiny{$\tilde{X}_{j-1}$}};
		\node[below] at (2.5+0.25,-0.01) {\tiny{$\tilde{X}_{j+1}$}};
		\node[above] at (2,0.01) {\tiny{$\tilde{Y}_{j{-}1,j{+}1}$}};
	\end{tikzpicture}} \right. \\
&\left.-\rmd\!\log{\frac{X_j^{+-}}{X_j^{++}}}\raisebox{-1em}{\begin{tikzpicture}
		\draw[line width=1.5pt] (0,0)--(0.5,0);
		\draw[line width=1.5pt] (1,0)--(2,0);
		\draw[line width=1.5pt] (2.5,0)--(3,0);
		\draw[fill=black] (0,0) circle [radius=0.1];
		\draw[fill=black] (1.5,0) circle [radius=0.1];
		\draw[fill=black] (3,0) circle [radius=0.1];
		\draw[fill=black] (0.5+0.5/4,0) circle [radius=0.025];
		\draw[fill=black] (0.5+2*0.5/4,0) circle [radius=0.025];
		\draw[fill=black] (0.5+3*0.5/4,0) circle [radius=0.025];
		\draw[fill=black] (2+0.5/4,0) circle [radius=0.025];
		\draw[fill=black] (2+2*0.5/4,0) circle [radius=0.025];
		\draw[fill=black] (2+3*0.5/4,0) circle [radius=0.025];
		\node[below] at (0,-0.01) {\tiny{$1$}};
		\node[below] at (3,-0.01) {\tiny{$n$}};
		\node[below] at (1.5,-0.01) {\tiny{$X_{j}{+}X_{j{+}1}$}};
		\node[above] at (1.1,0) {\tiny{$Y_{j{-}1,j}$}};
		\node[above] at (2.1,0) {\tiny{$Y_{j{+}1,j{+}2}$}};
	\end{tikzpicture}}-\rmd\!\log{\frac{X_j^{-+}}{X_j^{++}}}\raisebox{-1em}{\begin{tikzpicture}
		\draw[line width=1.5pt] (0,0)--(0.5,0);
		\draw[line width=1.5pt] (1,0)--(2,0);
		\draw[line width=1.5pt] (2.5,0)--(3,0);
		\draw[fill=black] (0,0) circle [radius=0.1];
		\draw[fill=black] (1.5,0) circle [radius=0.1];
		\draw[fill=black] (3,0) circle [radius=0.1];
		\draw[fill=black] (0.5+0.5/4,0) circle [radius=0.025];
		\draw[fill=black] (0.5+2*0.5/4,0) circle [radius=0.025];
		\draw[fill=black] (0.5+3*0.5/4,0) circle [radius=0.025];
		\draw[fill=black] (2+0.5/4,0) circle [radius=0.025];
		\draw[fill=black] (2+2*0.5/4,0) circle [radius=0.025];
		\draw[fill=black] (2+3*0.5/4,0) circle [radius=0.025];
		\node[below] at (0,-0.01) {\tiny{$1$}};
		\node[below] at (3,-0.01) {\tiny{$n$}};
		\node[below] at (1.5,-0.01) {\tiny{$X_{j{-}1}{+}X_j$}};
		\node[above] at (1.1,0) {\tiny{$Y_{j{-}2,j{-}1}$}};
		\node[above] at (2.1,0) {\tiny{$Y_{j,j{+}1}$}};
	\end{tikzpicture}}\right)\\[0.75em]
& -\ii \, \rmd\!\log{\frac{X_n^-}{X_n^+}} \left({-}\raisebox{-1em}{\begin{tikzpicture}
		\draw[line width=1.5pt] (0,0)--(-1,0);
		\draw[line width=1.5pt] (-1,0)--(-1.5,0);
		\draw[line width=1.5pt] (-2,0)--(-2.5,0);
		\draw[fill=black] (0,0) circle [radius=0.1];
		\draw[fill=black] (-1,0) circle [radius=0.1];
		\draw[fill=black] (-2.5,0) circle [radius=0.1];
		\node[above] at (0,0) {\tiny{$X_{n{-}1}{+}Y_{n{-}1,n}$}};
		\node[below] at (-1,0) {\tiny{$n-2$}};
		\node[below] at (-2.5,0) {\tiny{$1$}};
		\draw[fill=black] (-1.5-0.5/4,0) circle [radius=0.025];
		\draw[fill=black] (-1.5-2*0.5/4,0) circle [radius=0.025];
		\draw[fill=black] (-1.5-3*0.5/4,0) circle [radius=0.025];
	\end{tikzpicture}} + \raisebox{-1em}{\begin{tikzpicture}
		\draw[line width=1.5pt] (0,0)--(-1,0);
		\draw[line width=1.5pt] (-1,0)--(-1.5,0);
		\draw[line width=1.5pt] (-2,0)--(-2.5,0);
		\draw[fill=black] (0,0) circle [radius=0.1];
		\draw[fill=black] (-1,0) circle [radius=0.1];
		\draw[fill=black] (-2.5,0) circle [radius=0.1];
		\node[below] at (0,0) {\fontsize{6pt}{0}{$X_{n-1}{+}X_n$}};
		\node[below] at (-1,0) {\tiny{$n-2$}};
		\node[below] at (-2.5,0) {\tiny{$1$}};
		\draw[fill=black] (-1.5-0.5/4,0) circle [radius=0.025];
		\draw[fill=black] (-1.5-2*0.5/4,0) circle [radius=0.025];
		\draw[fill=black] (-1.5-3*0.5/4,0) circle [radius=0.025];
	\end{tikzpicture}}\right),
\end{aligned}
\end{equation}
where each graph is a shifted $(n{-}1)$-point chain. This turns out to be a very efficient recursive expression to generate symbols of chain graphs, compared with \cite{Hillman:2019wgh} or direct integration. Take the $3$-site chain as an example, the differential of the finite part is
\begin{equation}
\begin{aligned}
    &\rmd\psi_{\text{3-chain}}(X_1,X_2,X_3,Y_{1,2},Y_{2,3})\\
    =&-\ii \, \rmd\!\log{\frac{X_{1}^-}{X_1^+}}\left({-}\psi_{2}(X_2{+}Y_{1,2},X_3,Y_{2,3})+\psi_2(X_{1}{+}X_{2},X_3,Y_{2,3})\right)\\
+&\ii \, \Bigg(\rmd\!\log{\frac{X_{2}^{+-}X_{2}^{-+}}{X_2^{++}X_2^{--}}}\psi_2(\tilde{X}_1,\tilde{X}_3,\tilde{Y}_{1,3}){-}\rmd\!\log{\frac{X_{2}^{+-}}{X_2^{++}}\psi_2(X_1,X_{2}{+}X_{3},Y_{1,2})} {-}\rmd\!\log{\frac{X_{2}^{-+}}{X_2^{++}}\psi_2(X_{1}{+}X_{2},X_{3},Y_{2,3})}\Bigg)\\
-&\ii \, \rmd\!\log{\frac{X_{3}^-}{X_3^+}}\left({-}\psi_2(X_1,X_2{+}Y_{2,3},Y_{1,2})+\psi_2(X_{1},X_{2}{+}X_{3},Y_{1,2})\right),
\end{aligned}
\end{equation}
where $\psi_2(A,B,C)$ denotes the $2$-site chain wavefunction where the external energies on the two vertices are $A,B$, and the energy of the propagator is $C$. 

\paragraph{$n$-gon graphs}
Similarly, we can also consider the dS limit of $n$-gon graphs at one-loop in \eqref{eq:ngon}. 
The truncated differential equation of its wavefunction $\psi_{\text{n-gon}}$ can be obtained from \eqref{eq:dsc1} and \eqref{eq:dsc2}
\begin{equation}\label{eq:dgonDC}
\rmd\psi_{\text{n-gon}}=\ii\sum_{j=1}^n\sum_{a,b=\pm}\rmd\!\log{X_j^{a,b}}\left(D_j^{a,b}+C_j^{a,b}\right),
\end{equation}
where  $X_j^{a,b}=X_j+a Y_j+b Y_{j-1}$ (we use the shorthand $Y_i:=Y_{i,i{+}1}$ and $Y_n:=Y_{n,1}$ ). $D_j^{a,b}$ comes from the $\mathcal{O}\left(\frac{1}{q_j}\right)$ divergent terms with relations $D_{j}^{+-}=D_{j}^{-+}=-D_{j}^{++}=-D_{j}^{--}\equiv D_j$, and $C_j^{a,b}$ in \eqref{eq:dgonDC} are contributed from the contact terms in \eqref{eq:dpsic}. Similar to the $n$-site chain case \eqref{eq:DCpsi}, where now every node is valency-$2$, the coproduct of each letter can be written as some shifted $(n-1)$-gon wavefunction. After a similar computation as the chain cases, the truncated differential equation for an n-gon wavefunction integral can be written as 
\begin{equation}
\begin{aligned}
\rmd\psi_{\text{n-gon}}
= & \ii \, \sum_{j=1}^{n} \left( \rmd\!\log{\frac{X_j^{+-}X_j^{-+}}{X_j^{++}X_j^{--}}}\raisebox{-2em}{\begin{tikzpicture}[scale=0.6]
		\draw[line width=1.5pt] (0,0) circle [radius=1.25];
		\draw[fill=black] (0.803485, 0.957556) circle [radius=0.1];
		\draw[fill=black] (-0.803485, 0.957556) circle [radius=0.1];
		\draw[fill=black] (1.14681, 0.803007) circle [radius=0.03];
		\draw[fill=black] (1.31557, 0.478828) circle [radius=0.03];
		\draw[fill=black] (1.39467, 0.122018) circle [radius=0.03];
		\draw[fill=black] (-1.14681, 0.803007) circle [radius=0.03];
		\draw[fill=black] (-1.31557, 0.478828) circle [radius=0.03];
		\draw[fill=black] (-1.39467, 0.122018) circle [radius=0.03];
		\node[above left] at (-0.803485, 0.957556) {\tiny{$\tilde{X}_{j{-}1}$}};
		\node[above right] at (0.803485, 0.957556) {\tiny{$\tilde{X}_{j{+}1}$}};
		\node[above] at (0,1.25) {\tiny{$\tilde{Y}_{j{-}1,j{+}1}$}};
	\end{tikzpicture}}-\rmd\!\log{\frac{X_j^{+-}}{X_j^{++}}}\raisebox{-2em}{\begin{tikzpicture}[scale=0.6]
		\draw[line width=1.5pt] (0,0) circle [radius=1.25];
		\draw[fill=black] (0.803485, 0.957556) circle [radius=0.1];
		\draw[fill=black] (-0.803485, 0.957556) circle [radius=0.1];
		\draw[fill=black] (1.14681, 0.803007) circle [radius=0.03];
		\draw[fill=black] (1.31557, 0.478828) circle [radius=0.03];
		\draw[fill=black] (1.39467, 0.122018) circle [radius=0.03];
		\draw[fill=black] (-1.14681, 0.803007) circle [radius=0.03];
		\draw[fill=black] (-1.31557, 0.478828) circle [radius=0.03];
		\draw[fill=black] (-1.39467, 0.122018) circle [radius=0.03];
		\node[above left] at (-0.803485, 0.957556) {\tiny{$X_{j{-}1}$}};
		\node[above right] at (0.803485, 0.957556) {\tiny{$X_{j}{+}X_{j{+}1}$}};
	\end{tikzpicture}}\right. \\
  & \hspace{17em}\left. -\rmd\!\log{\frac{X_j^{-+}}{X_j^{++}}}\raisebox{-2em}{\begin{tikzpicture}[scale=0.6]
		\draw[line width=1.5pt] (0,0) circle [radius=1.25];
		\draw[fill=black] (0.803485, 0.957556) circle [radius=0.1];
		\draw[fill=black] (-0.803485, 0.957556) circle [radius=0.1];
		\draw[fill=black] (1.14681, 0.803007) circle [radius=0.03];
		\draw[fill=black] (1.31557, 0.478828) circle [radius=0.03];
		\draw[fill=black] (1.39467, 0.122018) circle [radius=0.03];
		\draw[fill=black] (-1.14681, 0.803007) circle [radius=0.03];
		\draw[fill=black] (-1.31557, 0.478828) circle [radius=0.03];
		\draw[fill=black] (-1.39467, 0.122018) circle [radius=0.03];
		\node[above left] at (-0.803485, 0.957556) {\tiny{$X_{j{-}1}{+}X_{j}$}};
		\node[above right] at (0.803485, 0.957556) {\tiny{$X_{j{+}1}$}};
	\end{tikzpicture}} \right).
 \end{aligned}
\end{equation}
where $\tilde{X}_{j{-}1}$, $\tilde{X}_{j{+}1}$ and $\tilde{Y}_{j{-}1,j{+}1}$ follows \eqref{tildeXY}. As a quick example, we present the total differential of $\psi_{\text{3-gon}}$ as
\begin{align}
\rmd\psi_{\text{3-gon}}(X_1,X_2,X_3,Y_1,Y_2,Y_3)&=-\ii\ {\rm d}\log\frac{X_1^{+-}X_2^{-+}}{X_1^{++}X_2^{++}}\psi_{\text{2-gon}} (X_1{+}X_2,X_3,Y_3,Y_2)\nonumber\\
&+\ii\  {\rm d}\log\frac{X_1^{+-}X_1^{-+}}{X_1^{++}X_1^{--}}\psi_{\text{2-gon}}(\tilde{X}_0,\tilde{X}_2,\tilde{Y}_{0,2},Y_2)+\text{cyclic}.
\end{align}
Note that here we should identify index $0$ as $3$ after applying \eqref{tildeXY}.

\section{Conclusions and outlook}

In this paper, we have revisited the computation of ``cosmological amplitudes" in an important class of toy models, namely conformal scalars with time-dependent interactions in the power-law FRW universe. Starting from the time-integral representation of wavefunction coefficients and correlators, there is a canonical decomposition into simpler building blocks, and we have proposed a method for deriving the complete system of differential equations based on DE for each building block. We have also obtained recursive solutions in terms of Euler-Mellin integrals or series representation of the resulting generalized hypergeometric functions, and studied how to truncate such differential equations back to the dS case, where they simplify to MPL functions. 

Let us end by listing just a few directions for future investigations.
\begin{itemize}
\item From our DEs, we still need to finish the interesting graphical/combinatorial exercise to obtain the canonical DEs for any tree-level wavefunction/correlator (the slightly generalized version with $n$ twists) as we have done in examples, and also to extend these results to loop integrands; indeed it would be very satisfying to have a complete set of graphic rules for any graph (tree or loops).  
\item As already initiated in~\cite{Arkani-Hamed:2023bsv,Arkani-Hamed:2023kig}, it would be very interesting to study the DEs (and their solutions) for the full wavefunction/correlator when combining all graphs together,  {\it e.g.} in a tr$\phi^3$ model. Since the latter are again expressed as linear combinations of our building blocks, we can derive the resulting DEs and even the solutions. 
\item It would be highly desirable to study the solutions of our DEs more carefully, both in terms of series expansion and Euler-Mellin integrals (both as definitions of certain generalized hypergeometric functions). How can we classify such functions and are there Hopf algebraic structures for such functions similar to those studied in Feynman integrals~\cite{Abreu:2019wzk}? It would also be interesting to study further the symbology and structure of MPL functions when expanded in powers of $q_i$'s. Another interesting thing is to study the integration-by-part (IBP) relations satisfied by the solutions of general $q_{i}$'s case via (relative) twisted cohomology~\cite{Aomoto:2011ggg,Mastrolia:2018uzb,Caron-Huot:2021iev,Caron-Huot:2021xqj,De:2023xue} or methods developed for GKZ systems~\cite{Chestnov:2022alh,Feng:2024xio} and even not only restricted to toy models we studied here~\cite{Chen:2023iix}.
\item Finally, we have only considered loop integrands and it would be interesting to explore integrated loop amplitudes (see~\cite{Benincasa:2024lxe,Benincasa:2024rfw} for interesting steps in this direction). Could we derive DEs for integrated amplitudes instead? What types of functions we would encounter after performing loop integrations (even in the limit with $q_i \to 0$)? 
\end{itemize}  

\section*{Acknowledgement}
We thank Nima Arkani-Hamed, Daniel Baumann, Zhong-zhi Xianyu for inspiring discussions. The work of S.H. has been supported by the National Natural Science Foundation of China under Grant No. 12225510, 11935013, 12047503, 12247103, and by the New Cornerstone Science Foundation through the XPLORER PRIZE.

\appendix

\section{The soft limit of time integrals}\label{app:boundary}
In this section, we calculate the soft limit $\omega_{m}\to 0$ of time integrals $\mathbf{T}^{q_{1},\ldots,q_{n}}_{\mathcal{N}}(\omega_{1},\ldots,\omega_{n})$ supposing that $m$ is not a ``source'' and prove by induction that it will evaluate to \eqref{eq:rightlimit} presented in Sec.~\ref{sec:recursivesol}.

Let us start from the two-site chain:
\begin{equation}
\begin{aligned}
    \lim_{\omega_{m\to 0}}\raisebox{-1em}{\begin{tikzpicture}[scale=0.3]
		\coordinate (X1) at (0,0);
		\coordinate (X2) at (3,0);
		\node[below] at (X1) {\small{$j$}};
		\node[below] at (X2) {\small{$m$}};
		\draw[line width=1.5pt,->] (X1)--(2,0);
        \draw[line width=1.5pt] (1,0)--(X2);
		\path[fill=black] (X1) circle[radius=0.2];
		\path[fill=black] (X2) circle[radius=0.2];
	\end{tikzpicture}}\!\!&=(-\ii)^2\!\!\int_{-\infty}^{0}\!\!\!\rmd\tau_{j}(-\tau_{j})^{q_{j}\!-\!1}e^{\ii\omega_{j}\tau_{j}}\!\!\int_{\tau_{j}}^{0}\!\!\rmd\tau_{m}(-\tau_{m})^{q_{m}\!-\!1}\!\! \\
     &=\frac{(-\ii)^2}{q_{m}}\!\!\int_{-\infty}^{0}\!\!\rmd\tau_{j}(-\tau_{j})^{q_{j,m}-1}e^{\ii\omega_{j}\tau_{j}}\!\!=\frac{(-\ii)}{q_{m}}\!\!\raisebox{-1em}{\begin{tikzpicture}[scale=0.3]
		\coordinate (X1) at (0,0);
        \node[below] at (X1) {\small{$q_{j,m},\omega_{j}$}};
		\path[fill=black] (X1) circle[radius=0.2];
	\end{tikzpicture}}.
\end{aligned}
\end{equation}
We first prove by induction that for $m$ as a ``sink'' which means $\tau_{m}$ is later than all the adjacent vertices, the soft limit will be
\begin{equation}\label{eq:sinkboundary}
    \lim_{\omega_{m}\to 0}\mathbf{T}^{q_{1},\ldots,q_{n}}_{\mathcal{N}_{n}}(\omega_{1},\ldots,\omega_{n})=\frac{(-\ii)}{q_{m}}\tilde{\mathbf{T}}_{\mathcal{N}_{n-1}}(\omega_{m}=0).
\end{equation}
$\mathcal{N}_{n}$ indicates there are $n$ different parts directly connected to $m$ which we call an $n$-connection structure. $\hat{\mathbf{T}}_{\mathcal{N}_{n-1}}$ indicates the summation of time integrals with vertex $m$ contracted with some adjacent one which results in an $(n-1)$-connection structure.
Supposing for an $(n-1)$-connection configuration we have,
\begin{equation}
    \lim_{\omega_{m}\to 0}\mathbf{T}^{q_{1},\ldots,q_{n-1}}_{\mathcal{N}_{n-1}}(\omega_{1},\ldots,\omega_{n-1})=\frac{(-\ii)}{q_{m}}\tilde{\mathbf{T}}_{\mathcal{N}_{n-2}}(\omega_{m}=0).
\end{equation}
Now we add an additional part $\mathcal{A}$ which connects to $m$ through vertex $n$ and the time arrow is from $n$ to $m$, that is, there is a theta function $\theta(\tau_{m}-\tau_{n})$ in the integrand. Note that, this additional part $\mathcal{A}$ influences original time integral \textit{only} through this theta function. Then we can decompose this new integral into two parts according to whether $\tau_{n}$ is later than the lastest time in the original integral or not, that is, adding the following decomposition into the integrand
\begin{equation}\label{eq:decompose}
    1=\theta(\tau_{n}-\mathrm{max}(\tau_{1},\ldots,\tau_{n-1}))+\theta(\mathrm{max}(\tau_{1},\ldots,\tau_{n-1})-\tau_{n})
\end{equation}
The first part of \eqref{eq:decompose} restricts the integral of $\tau_{m}$ to
\begin{equation}\label{eq:temp}
\begin{aligned}
    \lim_{\omega_{m}\to 0}\mathbf{T}^{q_{1},\ldots,q_{n}}_{\mathcal{N}_{n}}(\omega_{1},\ldots,\omega_{n})&=\hat{\mathcal{T}}\int_{\tau_{n}}^{0}\rmd\tau_{m}(-\tau_{m})^{q_{m}-1}\\
    &=\frac{-\ii}{q_{m}}\mathbf{T}^{q_{1},\ldots,\hat{q}_{m},\ldots,q_{m}+q_{n}}_{\mathcal{N}_{n-1}}(\omega_{1},\ldots,\hat{\omega}_{m},\ldots,\omega_{n}),
\end{aligned}
\end{equation}
where $\hat{\mathcal{T}}$ means the integration which is irrelevant to $\tau_{m}$. $\hat{q}_{m}$ and $\hat{\omega}_{m}$ means they are missing. And we note that $-\ii$ arises from the different normalization of $\mathbf{T}_{\mathcal{N}_{n}}$ and $\mathbf{T}_{\mathcal{N}_{n-1}}$. The RHS of \eqref{eq:temp} is just the $(n-1)$-connection diagram with $m$ contracted with $n$ under the limit $\omega_{m}\to 0$. And it is easy to know that the second part of \eqref{eq:decompose} equals to directly attaching $\mathcal{A}$ to $\tilde{\mathbf{T}}_{\mathcal{N}_{n-2}}$ since in this case $\tau_{n}$ does not influence the integration of $\tau_{m}$. Add the two parts together, we get \eqref{eq:sinkboundary}. This finishes the induction.

Now we want to relax the condition that $m$ is a ``sink'' and allow the time arrows incident to $m$ to flip. Following the notations before, we call the $n$-th part connecting to $m$ as $\mathcal{A}_{n}$, $n$-connection structure as $\mathcal{N}_{n}$ and the remaining $(n-1)$-connection part without $\mathcal{A}_{n}$ as $\mathcal{N}_{n-1}$. Then we define a flip action as reversing the time arrow from $\mathcal{A}_{n}$ to $m$. Then we show that \eqref{eq:sinkboundary} holds if we apply the flip action to both sides by reduction. We know from the identity for ``sink'' proved before that the following two identity $B_{1}$ and $B_{2}$ holds:
\begin{equation}
    \begin{aligned}
        B_{1}:&\quad\lim_{\omega_{m}\to 0}\raisebox{-1.2em}{\begin{tikzpicture}
		\node[below,xshift=0.2cm] at (0.6,0) {\small{$m$}};
        \node[below,xshift=-0.1cm] at (1.7,0) {\small{$n$}};
		\draw[fill=black!10] (0,0) circle [radius=0.6];
        \draw[fill=black!10] (2.1,0) circle [radius=0.4];
		\path[fill=black] (0.6,0) circle[radius=0.1];
		\path[fill=black] (1.7,0) circle[radius=0.1];
        \node at (0,0) {\small{$\mathcal{N}_{n-1}$}};
        \node at (2.1,0) {\small{$\mathcal{A}_{n}$}};
	\end{tikzpicture}}=
     \raisebox{-1.2em}{\begin{tikzpicture}
		\node[below,xshift=0.2cm] at (0.6,0) {\small{$m$}};
        \node[below,xshift=-0.1cm] at (1.7,0) {\small{$n$}};
		\draw[fill=black!10] (0,0) circle [radius=0.6];
        \draw[fill=black!10] (2.1,0) circle [radius=0.4];
		\path[fill=black] (0.6,0) circle[radius=0.1];
		\path[fill=black] (1.7,0) circle[radius=0.1];
        \node at (0,0) {\small{$\mathcal{N}_{n-2}$}};
        \node at (2.1,0) {\small{$\mathcal{A}_{n}$}};
	\end{tikzpicture}}, \\
        B_{2}:&\quad\lim_{\omega_{m}\to 0}\raisebox{-1.2em}{\begin{tikzpicture}
		\node[below,xshift=0.2cm] at (0.6,0) {\small{$m$}};
        \node[below,xshift=-0.1cm] at (1.7,0) {\small{$n$}};
		\draw[fill=black!10] (0,0) circle [radius=0.6];
        \draw[fill=black!10] (2.1,0) circle [radius=0.4];
		\path[fill=black] (0.6,0) circle[radius=0.1];
		\draw[line width=1.5pt,->] (1.7,0)--(1.1,0);
		\draw[line width=1.5pt] (0.6,0)--(1.7,0);
		\path[fill=black] (1.7,0) circle[radius=0.1];
        \node at (0,0) {\small{$\mathcal{N}_{n-1}$}};
        \node at (2.1,0) {\small{$\mathcal{A}_{n}$}};
	\end{tikzpicture}}=\frac{-\ii}{q_{m}}
     \raisebox{-1.2em}{\begin{tikzpicture}
    		\draw[fill=black!10] (0,0) circle [radius=0.6];
            \draw[fill=black!10] (1,0) circle [radius=0.4];
    		\path[fill=black] (0.6,0) circle[radius=0.1];
            \node at (0,0) {\small{$\mathcal{N}_{n-1}$}};
            \node at (1,0) {\small{$\mathcal{A}_{n}$}};
    	\end{tikzpicture}}\,\, +\,\,
     \raisebox{-1.2em}{\begin{tikzpicture}
		\node[below,xshift=0.2cm] at (0.6,0) {\small{$m$}};
        \node[below,xshift=-0.1cm] at (1.7,0) {\small{$n$}};
		\draw[fill=black!10] (0,0) circle [radius=0.6];
        \draw[fill=black!10] (2.1,0) circle [radius=0.4];
		\path[fill=black] (0.6,0) circle[radius=0.1];
		\draw[line width=1.5pt,->] (1.7,0)--(1.1,0);
		\draw[line width=1.5pt] (0.6,0)--(1.7,0);
		\path[fill=black] (1.7,0) circle[radius=0.1];
        \node at (0,0) {\small{$\mathcal{N}_{n-2}$}};
        \node at (2.1,0) {\small{$\mathcal{A}_{n}$}};
	\end{tikzpicture}}
    \end{aligned}
\end{equation}
Then subtracting $B_{1}$ with $B_{2}$, we directly get
\begin{equation}\label{eq:B3}
    B_{3}:\quad\lim_{\omega_{m}\to 0}\raisebox{-1.2em}{\begin{tikzpicture}
		\node[below,xshift=0.2cm] at (0.6,0) {\small{$m$}};
        \node[below,xshift=-0.1cm] at (1.7,0) {\small{$n$}};
		\draw[fill=black!10] (0,0) circle [radius=0.6];
        \draw[fill=black!10] (2.1,0) circle [radius=0.4];
		\path[fill=black] (0.6,0) circle[radius=0.1];
		\draw[line width=1.5pt,->] (0.6,0)--(1.3,0);
		\draw[line width=1.5pt] (0.6,0)--(1.7,0);
		\path[fill=black] (1.7,0) circle[radius=0.1];
        \node at (0,0) {\small{$\mathcal{N}_{n-1}$}};
        \node at (2.1,0) {\small{$\mathcal{A}_{n}$}};
	\end{tikzpicture}}=-\frac{-\ii}{q_{m}}
     \raisebox{-1.2em}{\begin{tikzpicture}
    		\draw[fill=black!10] (0,0) circle [radius=0.6];
            \draw[fill=black!10] (1,0) circle [radius=0.4];
    		\path[fill=black] (0.6,0) circle[radius=0.1];
            \node at (0,0) {\small{$\mathcal{N}_{n-1}$}};
            \node at (1,0) {\small{$\mathcal{A}_{n}$}};
    	\end{tikzpicture}}\,\, +\,\,
     \raisebox{-1.2em}{\begin{tikzpicture}
		\node[below,xshift=0.2cm] at (0.6,0) {\small{$m$}};
        \node[below,xshift=-0.1cm] at (1.7,0) {\small{$n$}};
		\draw[fill=black!10] (0,0) circle [radius=0.6];
        \draw[fill=black!10] (2.1,0) circle [radius=0.4];
		\path[fill=black] (0.6,0) circle[radius=0.1];
		\draw[line width=1.5pt,->] (0.6,0)--(1.3,0);
		\draw[line width=1.5pt] (0.6,0)--(1.7,0);
		\path[fill=black] (1.7,0) circle[radius=0.1];
        \node at (0,0) {\small{$\mathcal{N}_{n-2}$}};
        \node at (2.1,0) {\small{$\mathcal{A}_{n}$}};
	\end{tikzpicture}}
\end{equation}
Note that $\omega_{m}$ in the RHS of the above identities are all taken to 0 automatically. Above we only act one flip action to the ``sink'' diagram for $m$. Then we can perform this flip several times to get a diagram with more arrow incident to $m$ reversed. But we need to keep in mind that in $\mathcal{N}_{n-1}$, there must be at least one arrow flowing into $m$ for the induction assumption to hold.
Then we finish our induction and conclude that
\begin{equation}
    \lim_{\omega_{m}\to 0}\mathbf{T}^{q_{1},\ldots,q_{n}}_{\mathcal{N}_{n}}(\omega_{1},\ldots,\omega_{n})=\frac{(-\ii)}{q_{m}}\tilde{\mathbf{T}}_{\mathcal{N}_{n-1}}(\omega_{m}=0), \text{ for $m$ not a ``source''}.
\end{equation}
The definition of $\tilde{\mathbf{T}}_{\mathcal{N}_{n-1}}$ will account for the minus sign before the contracted term when the arrows are flipped like the first term in the right-hand side of \eqref{eq:B3} and this matches the sign before our definition of contracted diagrams in \eqref{eq:tildeTdef}.

\section{Series expansion from the recursive solutions}\label{app:series}
In this section, we derive the two general series representations of the family tree presented in \cite{Fan:2024iek} from the recursive solution in Sec.~\ref{sec:recursivesol}. The two series representations are equivalent to each other but normalized with different energy factors.
Let us start from the series expansion for $(N-1)$-site family tree\footnote{In the following, $\mathcal{N}$ means a family tree and vertex $1$ is always chosen as root as before.} nomralized with $\omega_{1}$ (Eq.~(32) in \cite{Fan:2024iek}),
\begin{equation}
    \mathbf{T}^{q_1,\ldots,q_{N-1}}_{\mathcal{N}}(\omega_1,\ldots,\omega_{N-1})=\frac{(-\ii)^{N-1}}{(\ii \, \omega_{1})^{q_{1\cdots (N-1)}}}\!\!\!\!\!\sum^{\infty}_{n_{2},\cdots,n_{N\!-\!1}=0}\!\!\!\!\!\!\!\!\Gamma\left(q_{1\cdots(N-1)}+n_{2\cdots (N-1)}\right)\!\prod_{j=2}^{N-1}\frac{(-\omega_{j}/\omega_{1})^{n_{j}}}{(\tilde{q}_{j}+\tilde{n}_{j})n_{j}!},
\end{equation}
we can add a leaf $N$ to vertex $k$ in this family tree\footnote{Here we remind the reader that $\tilde{q}_{j}$ and $\tilde{n}_{j}$ means the summation of all descendants of vertex $j$.}. The recursion relation \eqref{eq:Trecursion} indicates that
\begin{equation}\label{eq:srecursion}
    \begin{aligned}
        \mathbf{T}^{q_1,\ldots,q_{N}}_{\mathcal{N}}(\omega_1,\ldots,\omega_{N})=&\frac{(-\ii)^{N}}{(\ii\omega_{1})^{q_{12\cdots N}}}\sum_{\substack{n_{2},\cdots,n_{k}^{\prime},\\ \cdots,n_{N-1}=0}}^{\infty}\Gamma\left(q_{1\cdots N}+n_{2\cdots (N-1)}\right)\prod_{j=2,j\ne k}^{N-1}\frac{(-\omega_{j}/\omega_{1})^{n_{j}}}{(\tilde{q}_{j}+\tilde{n}_{j})n_{j}!} \\
        &\times \frac{1}{\tilde{q}_{k}+\tilde{n}^{\prime}_{k}}\int_{0}^{1}\rmd\alpha_{N}\alpha_{N}^{q_{N}-1}\left(-\frac{\omega_{k}+\omega_{N}\alpha_{N}}{\omega_{1}}\right)^{n_{k}^{\prime}}\frac{1}{n_{k}^{\prime}!},
    \end{aligned}
\end{equation}
where the definition of $\tilde{q}_{j}$ has been adapted to the $N$-site case, but the definition of $\tilde{n}_{j}$ still remains the same as $(N-1)$-site case currently. We focus on the second line of \eqref{eq:srecursion},
\begin{equation}
    \begin{aligned}
        \int_{0}^{1}\rmd\alpha_{N}\alpha_{N}^{q_{N}-1}\!\!\left(-\frac{\omega_{k}+\omega_{N}\alpha_{N}}{\omega_{1}}\right)^{n_{k}^{\prime}}\!\!\frac{1}{n_{k}^{\prime}!}&=\left(-\frac{\omega_{k}}{\omega_{1}}\right)^{n_{k}^{\prime}}\!\!\int_{0}^{1}\rmd\alpha_{N}\sum_{n_{N}=0}^{n_{k}^{\prime}}\alpha_{N}^{n_{N}+q_{N}-1}\left(\frac{\omega_{N}}{\omega_{j}}\right)^{n_{N}}\frac{C_{n_{k}^{\prime}}^{n_{N}}}{n_{k}^{\prime}!}, \\
        &=\sum_{n_{N}=0}^{n_{k}^{\prime}}\frac{1}{q_{N}+n_{N}}\left(-\frac{\omega_{k}}{\omega_{1}}\right)^{n_{k}^{\prime}-n_{N}}\left(-\frac{\omega_{N}}{\omega_{1}}\right)^{n_{N}}\frac{C_{n_{k}^{\prime}}^{n_{N}}}{n_{k}^{\prime}!},
    \end{aligned}
\end{equation}
where $C_{n_{k}^{\prime}}^{n_{N}}$ is the binomial and since $n_{k}^{\prime}$ is an integer, above expansion is just a binomial expansion. Then after a redefinition of $n_{k}\equiv n_{k}^{\prime}-n_{N}$, above equation becomes
\begin{equation}
    \int_{0}^{1}\rmd\alpha_{N}\alpha_{N}^{q_{N}-1}\!\!\left(-\frac{\omega_{k}+\omega_{N}\alpha_{N}}{\omega_{1}}\right)^{n_{k}^{\prime}}\!\!\frac{1}{n_{k}^{\prime}!}=\left(-\frac{\omega_{k}}{\omega_{1}}\right)^{n_{k}}\frac{1}{n_{k}!}\sum_{n_{N}=0}\frac{1}{q_{N}+n_{N}}\left(-\frac{\omega_{N}}{\omega_{1}}\right)^{n_{N}}\frac{1}{n_{N}!}.
\end{equation}
Putting it back into \eqref{eq:srecursion} and replacing every $n_{k}^{\prime}$ with $n_{k}+n_{N}$, we immediately finish the reduction:
\begin{equation}
    \mathbf{T}^{q_1,\ldots,q_{N}}_{\mathcal{N}}(\omega_1,\ldots,\omega_{N})=\frac{(-\ii)^{N}}{(\ii \, \omega_{1})^{q_{1\cdots N}}}\!\!\!\!\!\sum^{\infty}_{n_{2},\cdots,n_{N}=0}\!\!\!\!\!\!\!\!\Gamma\left(q_{1\cdots N}+n_{2\cdots N}\right)\!\prod_{j=2}^{N}\frac{(-\omega_{j}/\omega_{1})^{n_{j}}}{(\tilde{q}_{j}+\tilde{n}_{j})n_{j}!}.
\end{equation}

For series representation normalized with $(\ii\omega_{12\cdots n})^{q_{12\cdots n}}$, it is not suitable to prove by recursion relation. Instead, it is straightforward to derive from the recursive solution \eqref{eq:Trecursivesol}. Since $P(\omega,\alpha)$ is already organized according to the family tree structure, we can easily prove by induction that $P(\omega,\alpha)$ can be organized as 
\begin{equation}
    P(\omega,\alpha)=\omega_{12\cdots n}-\sum_{j=2}^{N}\tilde{\omega}_{j}(1-\alpha_{j})\alpha^{\tilde{\mathbf{v}}_{j}},
\end{equation}
where $\tilde{\omega}_{j}$ is the summation of energies of all descendants including vertex $j$.  $\tilde{\mathbf{v}}_{j}$ is defined basically the same as $\mathbf{v}_{j}$ under \eqref{eq:Pdef} except that $\tilde{\mathbf{v}}_{j,j}$ is also set to 0 rather than 1 since we have decomposed $\alpha_{j}$ to $1-(1-\alpha_{j})$. Then we series expand $P(\omega,\alpha)^{-q_{12\cdots n}}$ from leaves to root step by step and arrive at
\begin{equation}
    \begin{aligned}
        \mathbf{T}^{q_1,\ldots,q_{N}}_{\mathcal{N}}(\omega_1,\ldots,\omega_{N})&=\frac{(-\ii)^{N}\Gamma(q_{12\cdots N})}{(\ii \, \omega_{12\cdots N})^{q_{12\cdots N}}}\sum_{n_{2},\cdots,n_{N}=0}^{\infty}\frac{\Gamma(1-q_{12\cdots N})}{\Gamma(1-q_{12\cdots N}-n_{2\cdots N})} \\
        &\hspace{10em}\times \prod_{j=2}^{N}\frac{\Gamma(\tilde{q}_{j}+\tilde{n}_{j}-n_{j})}{\Gamma(\tilde{q}_{j}+\tilde{n}_{j}+1)}\left(-\frac{\tilde{\omega}_{j}}{\omega_{12\cdots N}}\right)^{n_{j}}.
    \end{aligned}
\end{equation}
We note that the Gamma function in the second line all arise from \[\int_{0}^{1}\rmd\alpha_{j}\alpha_{j}^{\tilde{q}_{j}+\tilde{n}_{j}-n_{j}-1}(1-\alpha_{j})^{n_{j}}.\]
Using the relation\[\frac{\Gamma(a)\Gamma(1-a)}{\Gamma(1-a-n)}=\Gamma(a+n)(-1)^{n}, \]we can finally bring the above equation into the following form (Eq.~(33) in \cite{Fan:2024iek}):
\begin{equation}
    \begin{aligned}
        \mathbf{T}^{q_1,\ldots,q_{N}}_{\mathcal{N}}(\omega_1,\ldots,\omega_{N})&=\frac{(-\ii)^{N}}{(\ii \, \omega_{12\cdots N})^{q_{12\cdots N}}}\sum_{n_{2},\cdots,n_{N}=0}^{\infty}\Gamma\left(q_{12\cdots N}+n_{2\cdots N}\right)\\
        &\hspace{10em}\times\prod_{j=2}^{N}\frac{\Gamma(-\tilde{q}_{j}-\tilde{n}_{j})}{\Gamma(-\tilde{q}_{j}-\tilde{n}_{j}+n_{j}+1)}\left(-\frac{\tilde{\omega}_{j}}{\omega_{12\cdots N}}\right)^{n_{j}}.
    \end{aligned}
\end{equation}

\bibliographystyle{JHEP}
\bibliography{ref.bib}

\end{document}